\documentclass[11pt,a4paper]{article} 
\usepackage{jheppub, hyperref}
\usepackage{array}
\newcolumntype{L}[1]{>{\raggedright\let\newline\\\arraybackslash\hspace{0pt}}m{#1}}
\newcolumntype{C}[1]{>{\centering\let\newline\\\arraybackslash\hspace{0pt}}m{#1}}
\newcolumntype{R}[1]{>{\raggedleft\let\newline\\\arraybackslash\hspace{0pt}}m{#1}}
\usepackage{float}
\usepackage{graphicx}
\usepackage{epsfig}
\usepackage{color}

\newcommand*{\be}{\begin{equation}}
\newcommand*{\ee}{\end{equation}}
\newcommand*{\bea}{\begin{eqnarray}}
\newcommand*{\eea}{\end{eqnarray}}

\newcommand{\comment}[1]{}

\newcommand{\cref}[1]{Chapter~\ref{c.#1}}

\def\beq{\begin{equation}}
\def\eeq{\end{equation}}
\def\bea{\begin{eqnarray}}
\def\eea{\end{eqnarray}}
\def\ba{\begin{array}}
	\def\ea{\end{array}}
\def\bi{\begin{itemize}}
	\def\ei{\end{itemize}}
\def\be{\begin{enumerate}}
	\def\ee{\end{enumerate}}
\def\bc{\begin{center}}
	\def\ec{\end{center}}
\def\bt{\begin{table}}
	\def\et{\end{table}}
\def\btb{\begin{tabular}}
	\def\etb{\end{tabular}}

	\def\lsim{\raise0.3ex\hbox{$\;<$\kern-0.75em\raise-1.1ex\hbox{$\sim\;$}}}
	\def\gsim{\raise0.3ex\hbox{$\;>$\kern-0.75em\raise-1.1ex\hbox{$\sim\;$}}}

\title{Composite Higgs revealed in Higgs pair photo-production at future colliders}

\author[1]{A.~Bharucha,}
\author[2]{G.~Cacciapaglia,}
\author[2]{A.~Deandrea,}
\author[3]{N.~Gaur,}
\author[4,5]{D.~Harada,}
\author[2,6]{F.~Mahmoudi,}
\author[7,8]{K. Sridhar}
\affiliation[1]{Aix Marseille Univ, Universit\'e de Toulon, CNRS, CPT, Marseille, France}
\affiliation[2]{Universit\'e de Lyon, Universit\'e Claude Bernard Lyon 1, CNRS/IN2P3, \\
Institut de Physique des 2 Infinis de Lyon, UMR 5822, F-69622, Villeurbanne, France}
\affiliation[3]{Department of Physics, Dyal Singh College (University of Delhi), Lodi Road, New Delhi, 110003, India}
\affiliation[4]{KEK Theory Center, Institute of Particle and Nuclear Studies, KEK, 1-1 Oho, Tsukuba, Ibaraki 305-0801, Japan}
\affiliation[5]{Rudjer Boskovic Institute, Division of Theoretical Physics, Bijenicka cesta 54, 10000 Zagreb, Croatia}
\affiliation[6]{Theoretical Physics Department, CERN, CH-1211 Geneva 23, Switzerland}
\affiliation[7]{Department of Theoretical Physics, Tata Institute of Fundamental Research, Homi Bhabha Road, Colaba, Mumbai 400005, India}
\affiliation[8]{KREA University, Sri City, Andhra Pradesh-517646, India}
\abstract{
The next generation electron-positron colliders are designed for precision studies of the Standard Model and its extensions, in particular in the Higgs sector. We consider the potential for discovery of composite Higgs models in Higgs pair production through photon collisions. This process is loop-generated, thus it provides access to all Higgs couplings and can show new physics effects in polarized and unpolarized cross-sections starting at relatively low collider energies. It is, therefore, relevant for all electron-positron colliders planned or in preparation. Sizeable deviations from the Standard Model predictions are present in a general class of composite Higgs models, as couplings of one or more Higgs bosons to fermions, or fermionic and scalar resonances, modify the destructive interference present in the Standard Model. 
In particular, large effects are due to the new quartic coupling of the Higgs to tops and to the presence of a light scalar resonance.
}

\emailAdd{aoife.bharucha@cpt.univ-mrs.fr}
\emailAdd{g.cacciapaglia@ipnl.in2p3.fr}
\emailAdd{deandrea@ipnl.in2p3.fr}
\emailAdd{naveengaur@dsc.du.ac.in}
\emailAdd{dharada@post.kek.jp}
\emailAdd{nazila@cern.ch}
\emailAdd{sridhar@theory.tifr.res.in}

\begin{document}

\begin{flushright}
  CERN-TH-2020-212,\\
  KEK-TH-2285,\\
  RBI-ThPhys-2020-49,\\
  TIFR/TH/20-50
\end{flushright}

\maketitle

\section{Introduction \label{section:intro}}

After the discovery of the Higgs boson at the Large Hadron Collider (LHC)~\cite{Aad:2012tfa, Chatrchyan:2012xdj, Aad:2015zhl}, the question of the origin of the scalar sector has become a central focus for both theorists and experimentalists.
Model building efforts and specific searches aim at discovering features that can shed light on the fundamental mechanism behind the Higgs sector. 
A popular extension of the Standard Model (SM) consists of replacing the Higgs sector by a new strong-interaction at the electroweak scale, giving rise to electroweak symmetry breaking of a dynamical origin~\cite{Weinberg:1975gm,Dimopoulos:1979es,Eichten:1979ah}. In this framework, the relative lightness of the Higgs boson can be explained by the pseudo Nambu-Goldstone Boson (pNGB) nature of this particle, which stems from the broken global symmetry~\cite{Kaplan:1983fs}. In recent years detailed models were proposed based on a holographic description~\cite{Contino:2003ve,Agashe:2004rs, Agashe:2005dk, Contino:2006qr}, or based on an underlying gauge-fermion theory, where the global symmetry is broken by the bilinear condensate of techni-fermions~\cite{Peskin:1980gc,Ryttov:2008xe,Galloway:2010bp,Cacciapaglia:2014uja}. The underlying theories are designed to feature a vacuum alignment that does not break the SM gauge symmetry and a Higgs doublet in the pNGB sector (for a review see \cite{Cacciapaglia:2020kgq}), contrary to the old-school Technicolor theories that break the electroweak symmetry at the condensation scale without a Higgs boson~\cite{Weinberg:1975gm}. 
The exploration of these Beyond the Standard Model (BSM) scenarios is an active research subject both at present and for future colliders.

However, BSM searches at the LHC have not yet led to a discovery of new particles or new phenomena, a sign that BSM physics is subtler and/or fainter than what was originally expected. Possible search strategies at colliders include increasing the luminosity, increasing the center-of-mass energy, colliding other types of particles, or searching for new physics in more complex configurations.
The electron-positron collider option provides rather clean experimental conditions in comparison to the LHC, where the quantum chromodynamics (QCD) background is intense and hard to master. This is an asset for high precision measurements, at the price of a reduced center of mass energy. Among the different future electron-positron colliders there are both circular designs, as for example the FCC-ee at CERN \cite{Abada:2019zxq,Abada:2019lih} and the CEPC in China \cite{CEPCStudyGroup:2018rmc,CEPCStudyGroup:2018ghi}, and linear ones, such as the International Linear Collider (ILC) \cite{Baer:2013cma}. All proposals allow the exploration, to high precision, of the energy domain from around the $Z$ boson mass up to the TeV scale. In particular, the first runs at the $Z$ pole and at the $W^+ W^-$ threshold will allow for high-accuracy measurements of the electroweak sector, surpassing those of the LEP experiments, while runs at a Higgs factory stage of future colliders at  $\sqrt{s} \sim 240-250$ GeV and at the $t\bar{t}$ threshold and above will allow for the study of the Higgs production, both singly and in pairs. It is the pair production that will be the main focus of this work.

Compton back-scattering of laser photons on electrons at linear colliders allows the production of high energy photons. These  photon beams can reach energies close to those of the initial electrons. The photon collider is therefore a compelling option for the ILC \cite{Ginzburg:2019yws}. The possibility of measuring the triple Higgs coupling via the process $\gamma \gamma \to hh$ has been widely discussed in the literature,  as the sensitivity of this channel to the Higgs self-coupling is maximal at  the  threshold $2 m_h$ and greater than the sensitivities achieved directly at $e^+e^-$ colliders in processes such as $e^+e^- \to Zhh$ and $e^+e^-\to \nu {\bar{\nu}} hh$ in a wide range of center of mass energies \cite{Jikia:1992zw,Belusevic:2004pz}.  The $\gamma \gamma \to hh$ process~\cite{Jikia:1992mt} is loop mediated and goes through $t$ and $W^{\pm}$ loops, see for example 
\cite{Asakawa:2008se, Asakawa:2010xj, Kawada:2012uy} for details. In the SM these loops have destructive interference, resulting in small values of the cross-section, which reaches a maximum of roughly $0.5$~fb. In BSM scenarios  
there can be a substantial change in the production cross-sections  as the cancellation between top quark and $W$ boson loops can be spoilt~ \cite{Asakawa:2008se, Asakawa:2010xj, Kawada:2012uy}. Compared to other di-Higgs channels at tree level, e.g.~the production in association with a pair of neutral or charged leptons, the photon fusion channel is of more interest because it provides access to more couplings of the Higgs, which enter in the loops \cite{Asakawa:2010xj} . This is particularly attractive in BSM scenarios, where the top quark plays an important role due to its large mass .


The $\gamma \gamma \to hh$ process is an excellent probe of Composite Higgs models (CHMs): in fact, all models in this class feature modifications of the Higgs couplings to gauge bosons and fermions (in particular the top quark), which can spoil the cancellations that occur in the SM. The importance of the new coupling of two Higgses to the top (generated by pNGB non-linearity) has already been highlighted for the gluon-fusion process at the LHC \cite{Grober:2010yv,Contino:2012xk}. Composite resonances, if relatively light, can also provide additional contributions.
This in turn has effects both on the cross-section and on the helicity distributions, allowing models based on the composite Higgs idea to be further constrained (or discovered). In this paper we therefore study the (polarised) cross-section for the $\gamma\gamma\to hh$ process at $e^+e^-$ colliders, comparing predictions in the SM to those in CHMs, with the aim to understand the sensitivity of this process to the modified couplings and possible new particles. In minimal CHMs, where only the modifications of the pNGB Higgs are included, we focus on the interplay between the couplings to fermions and gauge bosons, involving both one and two Higgses.
We also explore models with light resonances, in particular models with a heavy scalar below the TeV and with top partners. In both cases, the presence of these states is required by the generation of the top mass and the softening of the constraints from electroweak precision on the compositeness scale.

In the next section we summarise the details of the calculation of di-Higgs production via photon fusion in the SM. In Sec.~\ref{section:2} we discuss minimal CHMs, where only modifications of the SM Higgs couplings are included, with the exception of a coupling of two Higgses to tops, which is peculiar to the non-linear pNGB couplings. 
As a next step in Sec.~\ref{section:4} we include a second singlet scalar, a resonance of the composite sector, discuss the implications for the model and analyse how this affects the cross section. While the couplings of this scalar are analogous of those of the Higgs to fermions and gauge bosons, we note that there is a derivative coupling to two Higgs bosons relevant to our process.
In Sec.~\ref{section:5} we add top partners to the model, and study the consequences at the numerical level for cross-sections. We conclude in Sec.~\ref{section:conclusion}, and consider the future outlook.

\section{Di-Higgs production in photon fusion \label{section:1}}

\begin{table}[tb]
\begin{center}
\begin{tabular}{|c|c|}
\hline
Diagrams & Amplitude \\
\hline
\begin{minipage}{0.6\textwidth} \begin{center}
 \epsfig{file=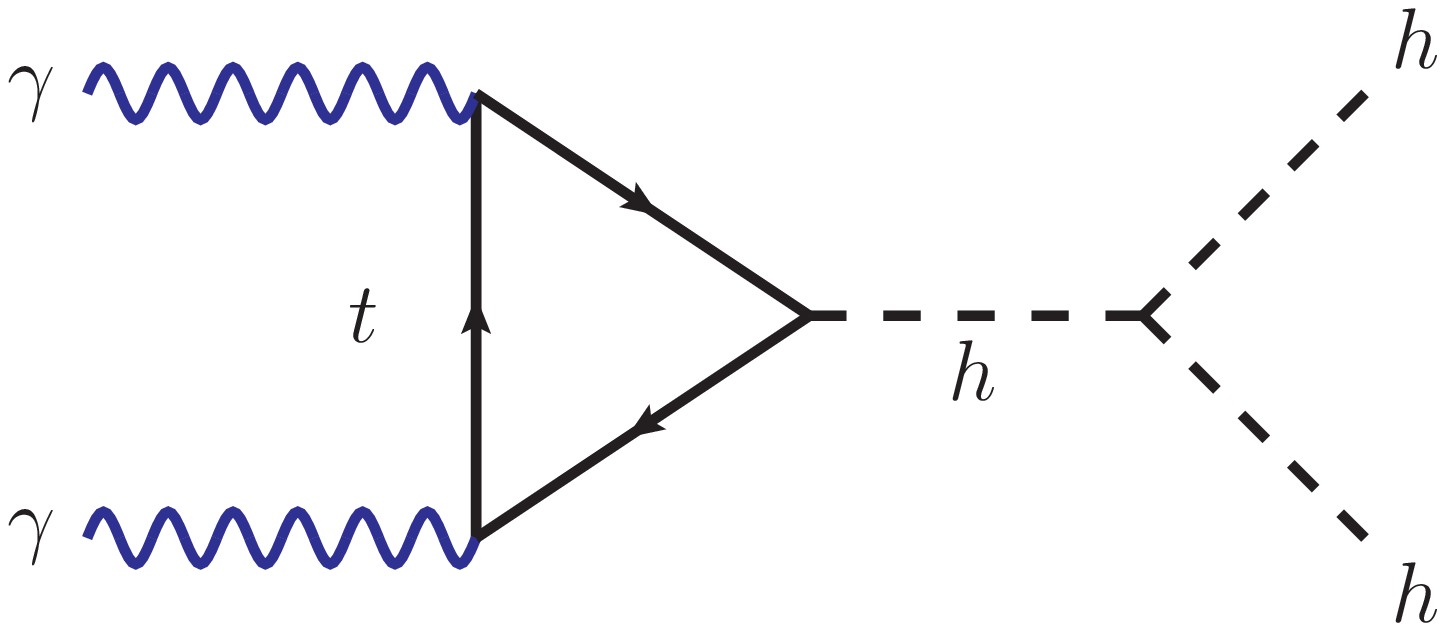,width=0.49\textwidth}  \epsfig{file=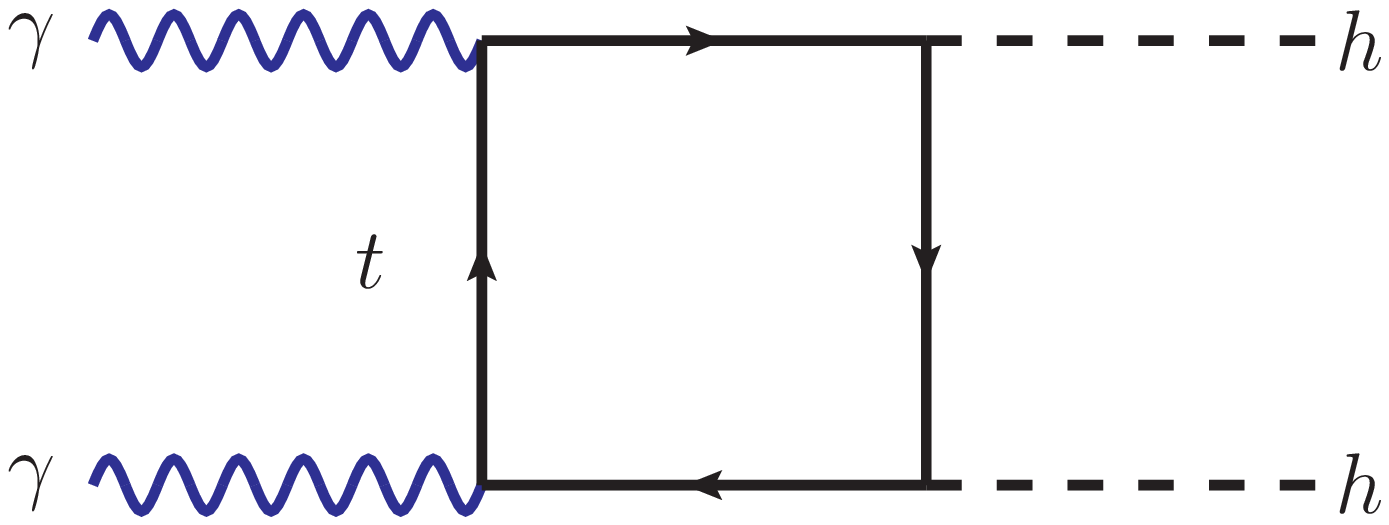,width=0.49\textwidth} \end{center} \end{minipage}    & $\mathcal{M}_{c_f}$ \\ \hline
\begin{minipage}{0.6\textwidth} \begin{center} \epsfig{file=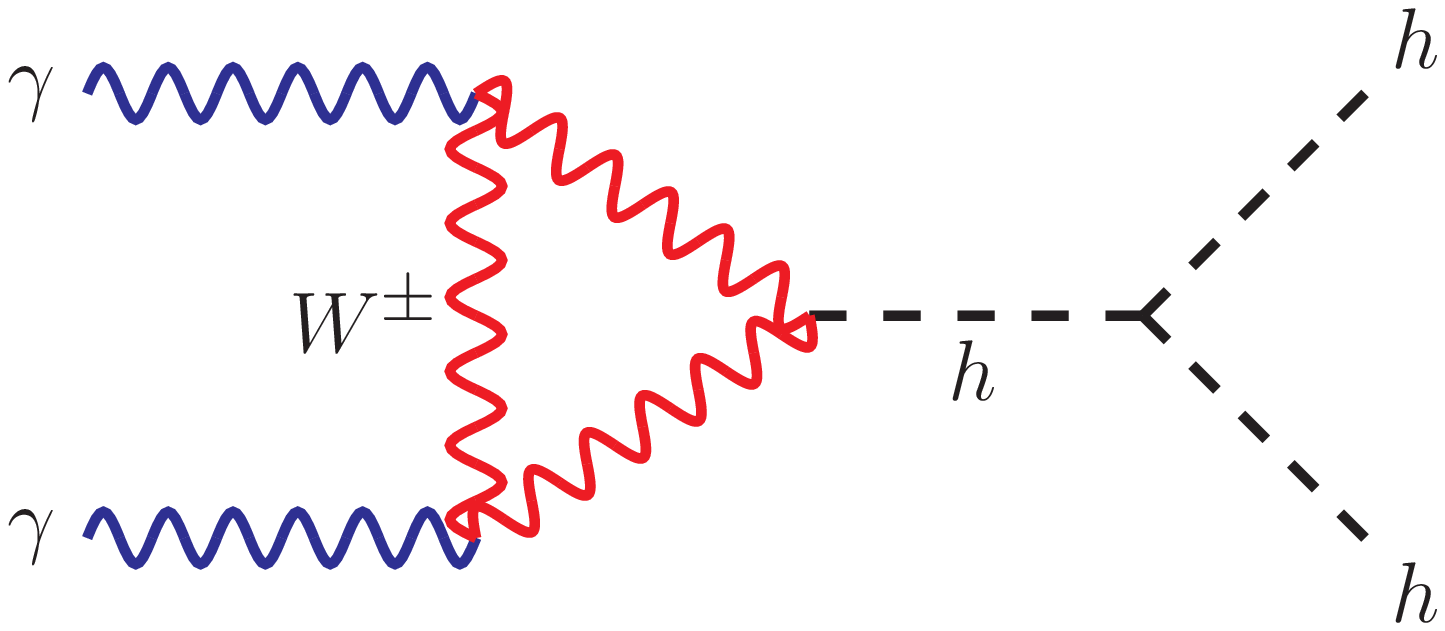,width=0.49\textwidth}  \epsfig{file=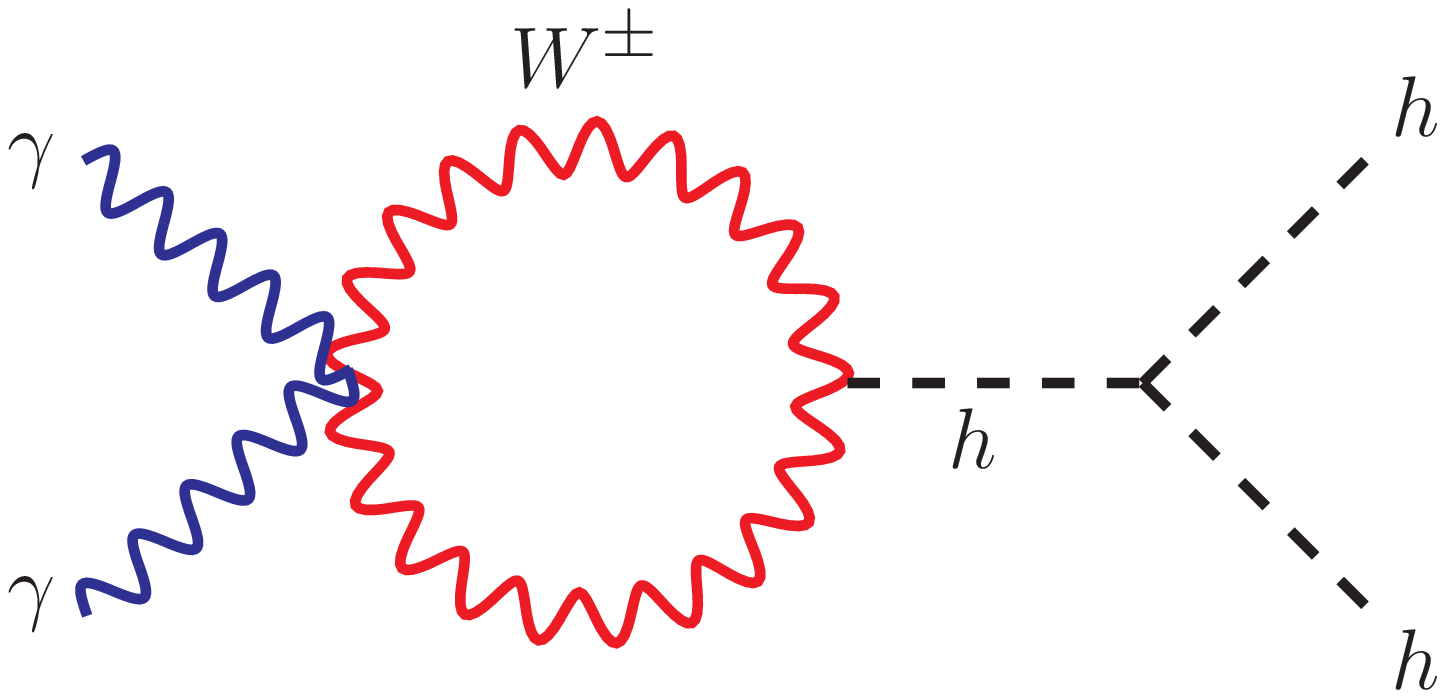,width=0.49\textwidth} \epsfig{file=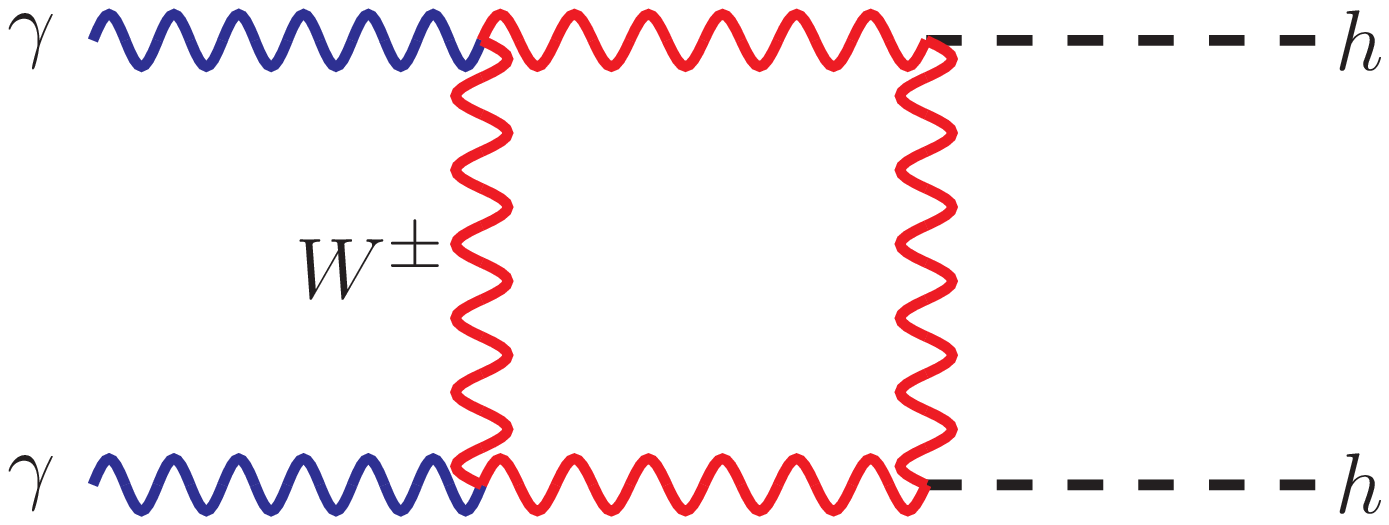,width=0.49\textwidth}  \epsfig{file=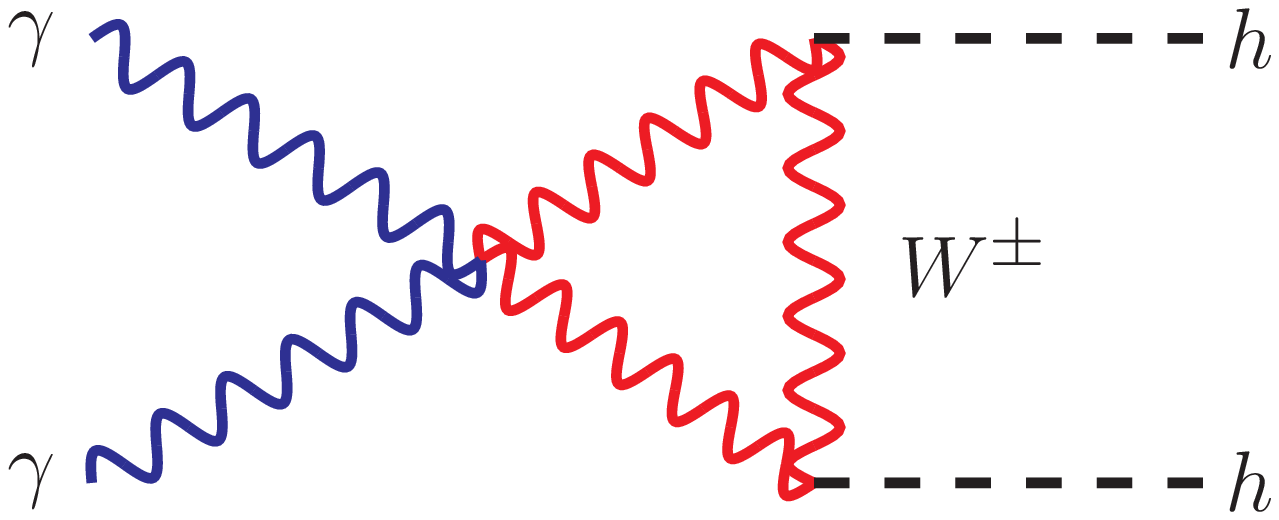,width=0.49\textwidth} \end{center} \end{minipage}& $\mathcal{M}_{c_v}$ \\ \hline
\begin{minipage}{0.6\textwidth} \begin{center} \epsfig{file=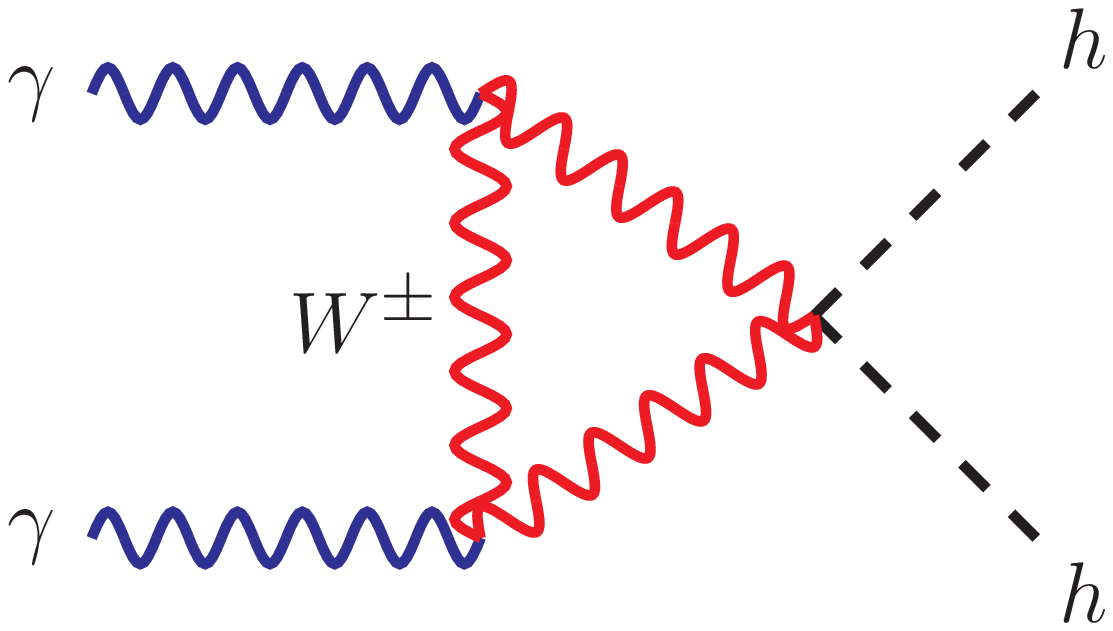,width=0.49\textwidth}  \epsfig{file=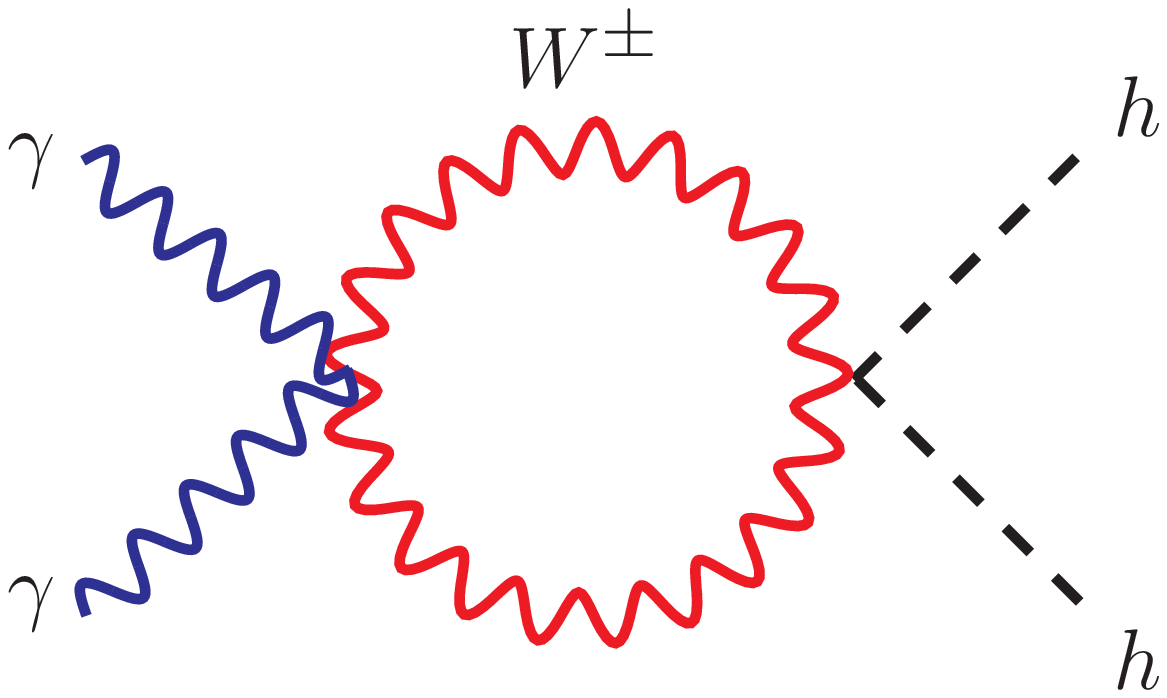,width=0.49\textwidth} \end{center} \end{minipage} & $\mathcal{M}_{c_{2v}}$ \\ \hline
\end{tabular}    
\end{center}
\caption{Diagrams for the di-Higgs production at a photon collider ($\gamma \gamma \to h h$) in the SM, with the corresponding partial amplitudes as defined in the text.} \label{tab:diagSM}
\end{table}

\begin{figure}[htb]
	\begin{center}
    \hspace*{-1cm} 
		\epsfig{file=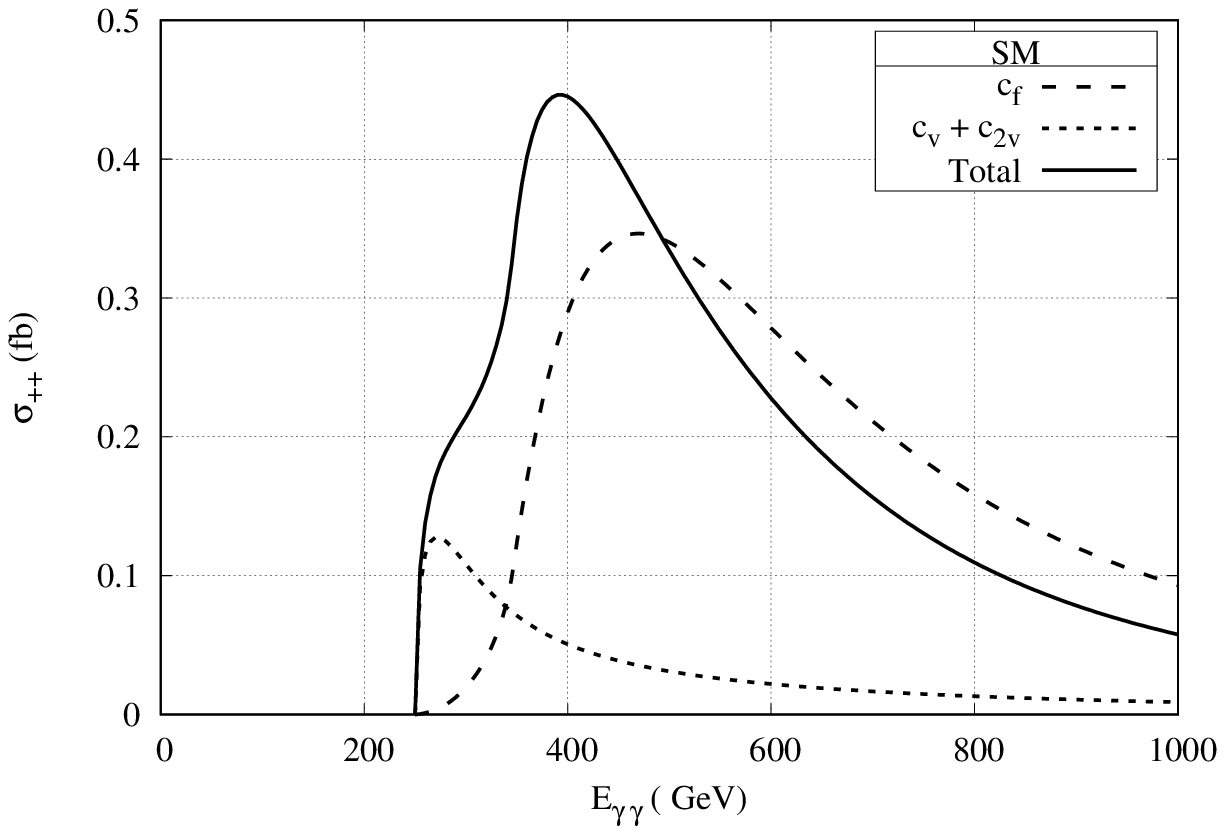,width=0.52\textwidth} 	
		\epsfig{file=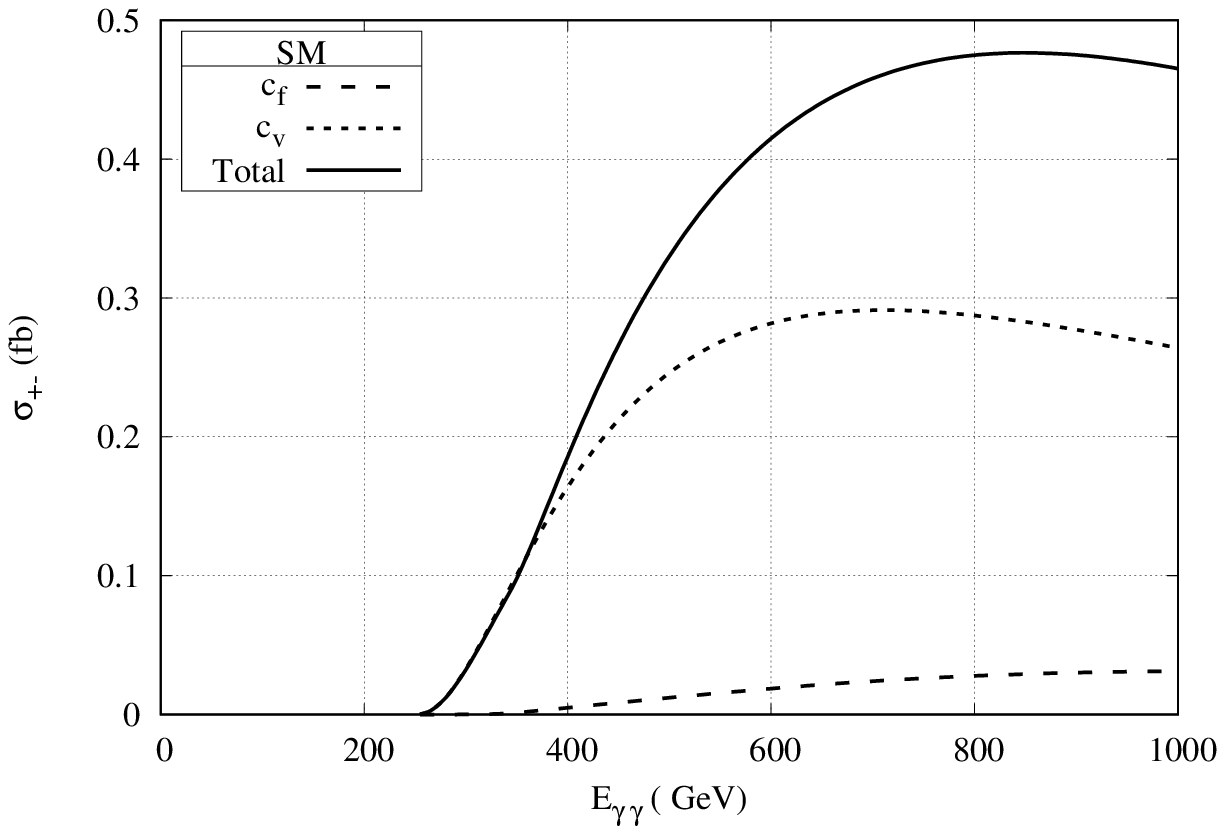,width=0.52\textwidth} 	
	\caption{$\gamma \gamma \to h h$ cross-section for same photon helicities (left panel) and opposite photon helicities (right panel) in the SM. The contributions labeled $c_f$ and $c_v + c_{2v}$ refer to the individual contributions of tops and gauge bosons respectively (C.f. Table \ref{tab:diagSM}). 	\label{fig:sec2-1}} 
	\end{center}
\end{figure}

Di-Higgs production at  a photon collider ($\gamma\gamma \to h h$) occurs only at one-loop level in the SM. The diagrams responsible for the process 
are shown in Table \ref{tab:diagSM}. 
In the process, there are four helicity amplitudes ${\cal M}(\lambda_1,\lambda_2)$, where ($\lambda_1, \lambda_2 = \pm$) are the photon polarisations,  
with only two of them being independent. In fact, parity relates the following pairs:
$$
{\cal M} (+,+) = {\cal M} (-,-),  ~~ 
{\cal M} (+,-) = {\cal M} (-,+)\,.
$$
For each helicity pair, one can define a differential cross section as follows:

\begin{equation}
\frac{d\hat{\sigma}(\lambda_1,\lambda_2)}{d\hat{t}} = \frac{1}{2!}\frac{1}{16\pi\hat{s}^{2}} {\alpha^{2}\alpha_{W}^{2}} \left| {\cal M}(\lambda_1,\lambda_2) \right|^{2} \,,
\label{sec2:eq_1} 
\end{equation} 
for the process $\gamma \gamma \to h h$ at the parton level {\footnote{
Note that a misprint was present in formula (2.5) of \cite{Jikia:1992mt} as already noted in \cite{Asakawa:2008se}.}}.
For further convenience, we can also split each amplitude into three parts, depending on which Higgs couplings enters: $c_f$ for the top Yukawa, $c_v$ for the couplings of a single Higgs to gauge bosons ($W^\pm$), and $c_{2v}$ for the coupling of  two Higgs to $W$'s (see the corresponding diagrams in Table~\ref{tab:diagSM}). The notation is borrowed from the coupling modifiers we will introduce in the next section, and this separation will allow us to study the various BSM contributions in more detail later on. In the SM, the helicity cross-sections are shown in Fig.~\ref{fig:sec2-1}, together with those deriving from the split amplitudes (i.e.~we remove all diagrams except those corresponding to the partial amplitude). This separation is not physical as interference terms are ignored, but it allows us to understand the relative weight of each contribution. 
The individual contributions have cancellations among themselves resulting in a total cross-section that
is much smaller compared to some of the individual contributions. 
In particular, the two amplitudes generated by gauge bosons, $M_{c_v}$ and $M_{c_{2v}}$, feature large destructive interference driven by gauge invariance: for this reason, we will always consider them together in the following and define $M_{c_v + c_{2v}}$ that includes interference.  
It is also to be noted that only box diagrams contribute to the channels having different photon helicities, i.e. $\sigma_{+-}$, whereas all kind of diagrams (box, triangle, self-energy) contribute to channels having same photon helicities,
i.e. $\sigma_{++}$, thereby resulting in some interesting features in $\sigma_{++}$. 

\begin{figure}[tbh]
	\begin{center}
		\epsfig{file=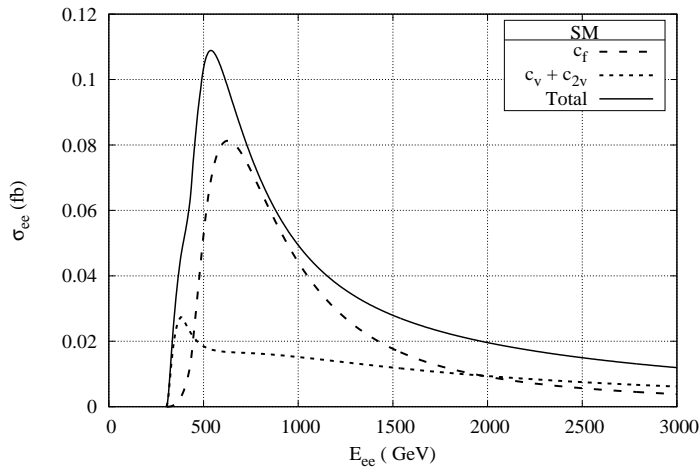,width=0.62\textwidth} 	
	\caption{Production cross-section ($e^+ e^- \to h h$) in the SM. The contributions labeled $c_f$ and $c_v+c_{2v}$ refer to the individual contributions of tops and gauge bosons respectively (C.f. Table~\ref{tab:diagSM}). 	\label{fig:sec2-2} }
	\end{center}
\end{figure}

The photons for the $\gamma \gamma$ collider can be obtained from  laser back-scattering at $e^+ e^-$ colliders. Denoting $\hat{s}$ and $s$ as the center-of-mass energies of the $\gamma\gamma$ and $e^+ e^-$ systems respectively, the total cross section of $e^- e^+ ( \gamma\gamma ) \rightarrow hh$ can be evaluated by convoluting the parton $\gamma \gamma \to h h$ cross-section with the photon luminosity function $f_{\gamma}(x,y)$ \cite{Ginzburg:1982yr}, as follows:
\begin{equation}
\sigma = \int_{4m_{h}^{2}/s}^{y_{m}^{2}} d\tau \frac{dL_{\gamma\gamma}}{d\tau} \left[ \frac{1+\xi_{1}^{\gamma}\xi_{2}^{\gamma}}{2} \hat{\sigma}_{++}(\hat{s}) + \frac{1-\xi_{1}^{\gamma}\xi_{2}^{\gamma}}{2} \hat{\sigma}_{+-}(\hat{s}) \right] \,,
\end{equation}
where $\xi^{\gamma}_{1,2}$ are the mean photon helicities of the two beams (see \cite{Ginzburg:1982yr} for the detailed formulas of the mean photon helicities), and the differential luminosity takes the form 
\begin{eqnarray}
\frac{dL_{\gamma\gamma}}{d\tau} =  \int_{\tau/y_{m}}^{y_{m}} \frac{dy}{y} f_{\gamma}(x,y) f_{\gamma}(x,\tau/y) \,,
\end{eqnarray}
where $\tau=\hat{s}/s$, $y=E_{\gamma}/E_{b}$ with $E_{\gamma}$ and $E_{b}$ being the energy of photon and electron beams respectively, and the maximal energy fraction of photon $y_{m}=x/(1+x)$ with $x=4E_{b}\omega_{0}/m_{e}^{2}$ where $\omega_{0}$ is the laser photon energy and $m_{e}$ is the electron mass. 
The photon luminosity spectrum is given by \cite{Ginzburg:1982yr} 
\begin{eqnarray}
f_{\gamma}(x,y) &=& \frac{1}{D(x)} \left[ \frac{1}{1-y} + 1-y-4r(1-r) -2\lambda_{e}\lambda_{\gamma} r x(2r-1)(2-y) \right] \,,\\
D(x) &=& \left( 1 - \frac{4}{x} - \frac{8}{x^{2}} \right) \ln (1+x) + \frac{1}{2} + \frac{8}{x} - \frac{1}{2(1+x)^{2}} \nonumber\\
&& + 2\lambda_{e}\lambda_{\gamma} \left[ \left( 1+\frac{2}{x} \right) \ln(1+x) -\frac{5}{2} + \frac{1}{1+x} - \frac{1}{2(1+x)^{2}} \right] \,,
\end{eqnarray}
where $r=\frac{y}{x(1-y)}$ and $\lambda_{e}(\lambda_{\gamma})$ is the helicity of the electron (photon). 
In our analysis, we set the dimensionless parameter $x=4.8$ (giving $y_{m}=0.82$) and $\lambda_{e_1}=\lambda_{e_2}=0.45$, $\lambda_{\gamma_{1}} = \lambda_{\gamma_{2}} = -1$ \cite{Jikia:1992mt}. 
In using this prescription as given in \cite{Jikia:1992mt}, we are assuming that a very high degree
of electron beam polarization ($90\%$) and correspondingly very high laser circular polarization ($100\%$) of the photon beam is achievable. 
The mean photon helicities of the two beams are defined by \cite{Ginzburg:1982yr}: 
\begin{equation}
\xi_{i}^\gamma (x,y) = \frac{C_{20}}{C_{00}} \,,
\end{equation}
with
\begin{eqnarray}
C_{00} &=& \frac{1}{1-y} + 1-y -4r(1-r) -2\lambda_{e} \lambda_{\gamma}rx(2r-1)(2-y) \,,\\
C_{20} &=& 2\lambda_{e} rx \left[ 1+(1-y)(2r-1)^{2} \right] - \lambda_{\gamma}(2r-1)\left( \frac{1}{1-y} + 1-y \right) \,.
\end{eqnarray}
The total $e^+ e^- \to  h h$ is shown in Fig.~\ref{fig:sec2-2}. 


\section{Minimal Composite Models \label{section:2} } 	

We will consider first \emph{Minimal Composite Higgs} models, where the vacuum is only misaligned along one direction that breaks the electroweak (EW) symmetry. 
This implies that the composite Higgs does not mix with other pNGBs, and that the misalignment can be described in terms of a single angle $\theta$~\cite{Kaplan:1983fs}, defined as follows:
\beq
v = f \sin(\theta)\,.
\label{eq:1} 
\eeq 
The equation above provides the relation between the EW symmetry breaking scale $v$ and the compositeness scale $f$.
This definition is independent of the coset $\mathcal{G}/\mathcal{H}$ that the model is based upon, and leads to an angle varying from $\theta = 0$ (where the EW symmetry is restored) to $\theta = \pi/2$ (corresponding to a Technicolor-like theory~\cite{Cacciapaglia:2014uja}).

Following Eq.~\eqref{eq:1}, independent of the coset,  the masses of the EW gauge bosons $W^\pm/Z$ are given by:
\beq
m_W^2 (\theta) = \frac{g^2 f^2}{4} \sin^2 \theta \equiv \frac{g^2 v^2}{4}\,, \quad m_Z^2 (\theta) = \frac{1}{c_W^2} m_W^2 (\theta)\,,
\eeq
where the relation between the two is guaranteed by the custodial symmetry embedded in $\mathcal{G}/\mathcal{H}$~\cite{Georgi:1984af}. The coset-independence of the relation between the $W^\pm/Z$ masses and the compositeness scale leads to universal couplings of the Higgs boson to the electroweak gauge bosons, as observed in Refs.~\cite{Liu:2018vel,Liu:2018qtb}.
These couplings can be elegantly expressed in terms of derivatives with respect to the misalignment angle, as follows: 
\bea
g_{WWh} &=& \frac{1}{f} \frac{\partial m_W^2 (\theta)}{\partial \theta} = \frac{2 m_W^2}{v} \cos \theta \,, \\
g_{WWhh} &=& \frac{1}{f^2} \frac{\partial^2 m_W^2 (\theta)}{\partial \theta^2} = \frac{2 m_W^2}{v^2} \cos 2\theta \,, 
\eea
and so on. Similar results also stand for the couplings to the $Z$ boson, which are related via the custodial symmetry.
For convenience, we will use the parameterisation in Refs.~\cite{Grober:2010yv, Contino:2012xk}, which reads
\beq
{\cal L} = m_W^2 W^+_\mu W^{-,\mu} \left( 1 + 2 c_v \frac{h}{v} + c_{2v} \frac{h^2}{v^2} + \dots \right)\,,
\label{eq:gauge1} 
\eeq
with
\beq
c_v = \cos \theta = \sqrt{1-\xi}\,, \qquad c_{2v} = \cos 2\theta = 1-2\xi\,;
\eeq
where $\xi = v^2/f^2 \equiv \sin^2 \theta$.

\begin{table}[tb]
	\begin{center} 	\begin{tabular}{|c|c|c|c|c|c|} \hline 
			Model & $h f \bar{f} (c_f)$ & $h h f \bar{f} (c_{2f})$ & $ h W^+ W^- (c_v)$ & $h h W^+ W^- (c_{2v})$
			          &  $c_{3h}$ \\ \hline 
			MCHM4 \cite{Agashe:2004rs} & $\sqrt{1 - \xi}$  &  $ -\xi$  & $\sqrt{1-\xi}$ & $1-2\xi$ & $\sqrt{1-\xi}$    \\ \hline 
			MCHM5 \cite{Contino:2006qr} & $\frac{1 - 2\xi}{\sqrt{1 - \xi}}$  &  $-4 \xi$  & $\sqrt{1-\xi}$ & $1-2\xi$ 
			         & $\frac{1 - 2 \xi}{\sqrt{1 - \xi}}$ \\ \hline      
			MCHM5-Higgs & $\frac{1 - 2\xi}{\sqrt{1 - \xi}}$  &  $-4 \xi$  & $\sqrt{1-\xi}$ & $1-2\xi$ 
                   & $\lambda_h$ \\ \hline		\end{tabular}
	\caption{The 3 benchmarks for the Higgs couplings as a function of $\xi = v^2/f^2 \equiv \sin^2 \theta$, with $\theta$ the misalignment angle~\cite{Grober:2010yv, Contino:2012xk}.	\label{table:1} } 
	\end{center}
\end{table}

The couplings of the composite Higgs to SM fermions, in particular the tops, are not universal. 
Nevertheless, they can also be expressed in terms of derivatives with respect to the misalignment angle, as follows:
\bea
g_{ffh} = \frac{1}{f} \frac{\partial m_t (\theta)}{\partial \theta}\,, \quad 
g_{ffhh} = \frac{1}{f^2} \frac{\partial^2 m_t (\theta)}{\partial \theta^2}\,, \quad \dots
		\label{eq:2} 
\eea 
The expressions now depend on the details of the model, and in particular on the dependence of the top mass on the misalignment angle $\theta$. 
We will consider here two scenarios: the first is realised in the SO(5)/SO(4) CHM with top partners in the spinorial representation of the global SO(5) (MCHM4). For this model, we have:
\beq
m_t (\theta) = \frac{\lambda f}{\sqrt{2}} \sin \theta \;\; \Rightarrow \;\; \left\{ \begin{array}{l}
g_{ffh} =\displaystyle \frac{m_t}{v} \cos \theta\,, \\
g_{ffhh} =\displaystyle  - \frac{m_t}{v^2} \sin^2 \theta\,.
\end{array} \right.
\eeq
We remind the reader that this is a much more general case, which can also be realised in other models, depending on the representation of the top partners.
The second case is realised in the SO(5)/SO(4) model with the fundamental representation of SO(5) (MCHM5), for which:
\beq
m_t (\theta) = \frac{\lambda f}{\sqrt{2}} \sin 2\theta \;\; \Rightarrow \;\; \left\{ \begin{array}{l}
g_{ffh} =\displaystyle  \frac{m_t}{v} \frac{\cos 2\theta}{ \cos \theta}\,, \\
g_{ffhh} =\displaystyle  - \frac{m_t}{v^2} 4 \sin^2 \theta\,.
\end{array} \right.
\eeq
Again, this case can be realised in other cosets as well.
It is also interesting to note that cases where the top mass receives both contributions can also be realised~\cite{Agugliaro:2018vsu}, for which:
\beq
m_t (\theta) = \frac{f}{\sqrt{2}} (\lambda_1 \sin \theta + \lambda_2 \sin 2 \theta) \;\;
\Rightarrow \;\;
g_{ffhh} = - \frac{m_t}{v^2} \frac{\lambda_1 + 8 \lambda_2 \cos \theta}{\lambda_1 + 2 \lambda_2 \cos \theta} \sin^2 \theta\,.
\eeq
The coefficient of the quartic coupling, which, as we will see, plays the most important role in the di-Higgs production, interpolates between the two cases we consider.
For convenience, we will parameterise the coupling modifier following Ref. \cite{Contino:2010mh} as:
\begin{equation}
{\cal L} =  - m_t \left( \bar{t}_L t_R \right)
\left( 1 + c_f \frac{h}{v} +  \frac{c_{2f}}{2}  \frac{h^2}{v^2} + \dots \right) + h.c. 
\label{eq:fermion1} 
\end{equation}

The last relevant coupling modifier is associated to the Higgs trilinear coupling. Its value in CHMs is highly dependent on the details of the model, as the calculation for the Higgs potential varies greatly depending on the coset and on the choice of top partner representations. We have therefore decided to focus on three benchmarks: first we consider the results obtained in the MCHM4~\cite{Agashe:2004rs} and MCHM5~\cite{Contino:2006qr} specific models, as reported in Table \ref{table:1}, while a third benchmark has the top coupling modifiers of MCHM5 but a generic trilinear coupling modifier. The latter case is chosen to explore the numerical effect of the trilinear coupling. The effect of anomalous Higgs couplings (including trilinear one) within the context of the SM was studied in \cite{Jikia:1992zw}.

For the photon initiated di-Higgs production process, a summary of the benchmark cases considered in the numerical study is reported in Table \ref{table:1}.
Besides the SM diagrams (with modified couplings), the calculation needs to be extended by including the quartic coupling of two Higgs bosons to the top quark, as shown in Table~\ref{tab:diagCH}.

\begin{table}[tb]
\begin{center}
\begin{tabular}{|c|c|}
\hline
Diagrams & Amplitude \\
\hline
\begin{minipage}{0.6\textwidth} 
\begin{center} 
\epsfig{file=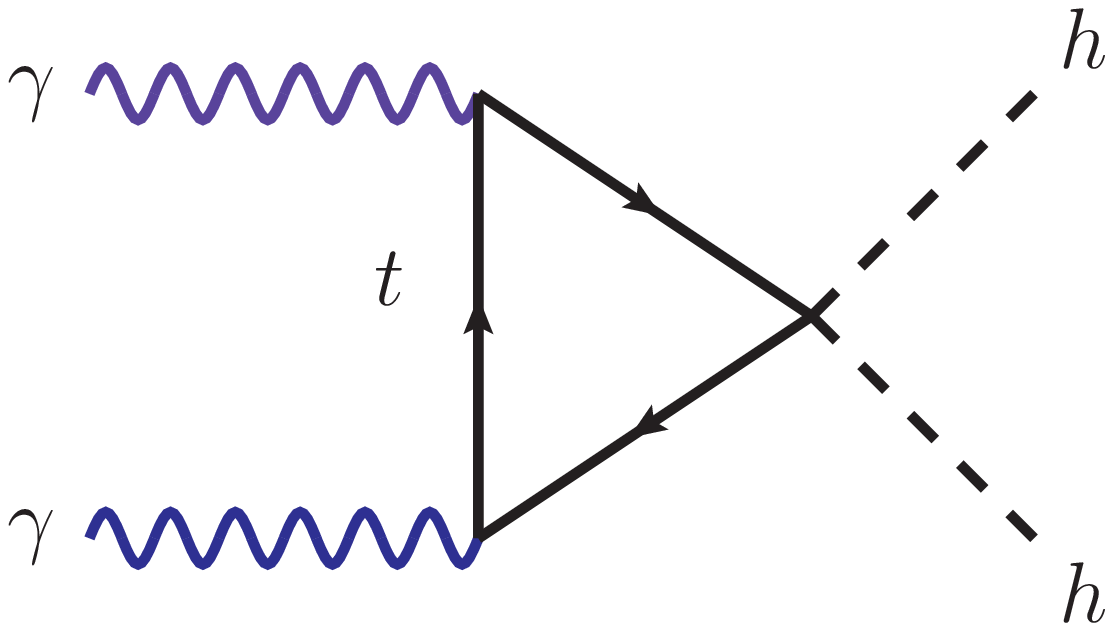,width=0.49\textwidth}  \end{center} \end{minipage} & $\mathcal{M}_{c_{2f}}$ \\ \hline
\end{tabular}    
\caption{Additional diagram in minimal composite Higgs models, with the corresponding partial amplitudes.\label{tab:diagCH}}
\end{center}
\end{table}

\subsection{Numerical results}

\begin{figure}[tb]
	\begin{center}
$\begin{array}{cc}		
{\bf MCHM4} & \phantom{xxx} {\bf MCHM5} \\
\hspace*{-1cm}
\epsfig{file=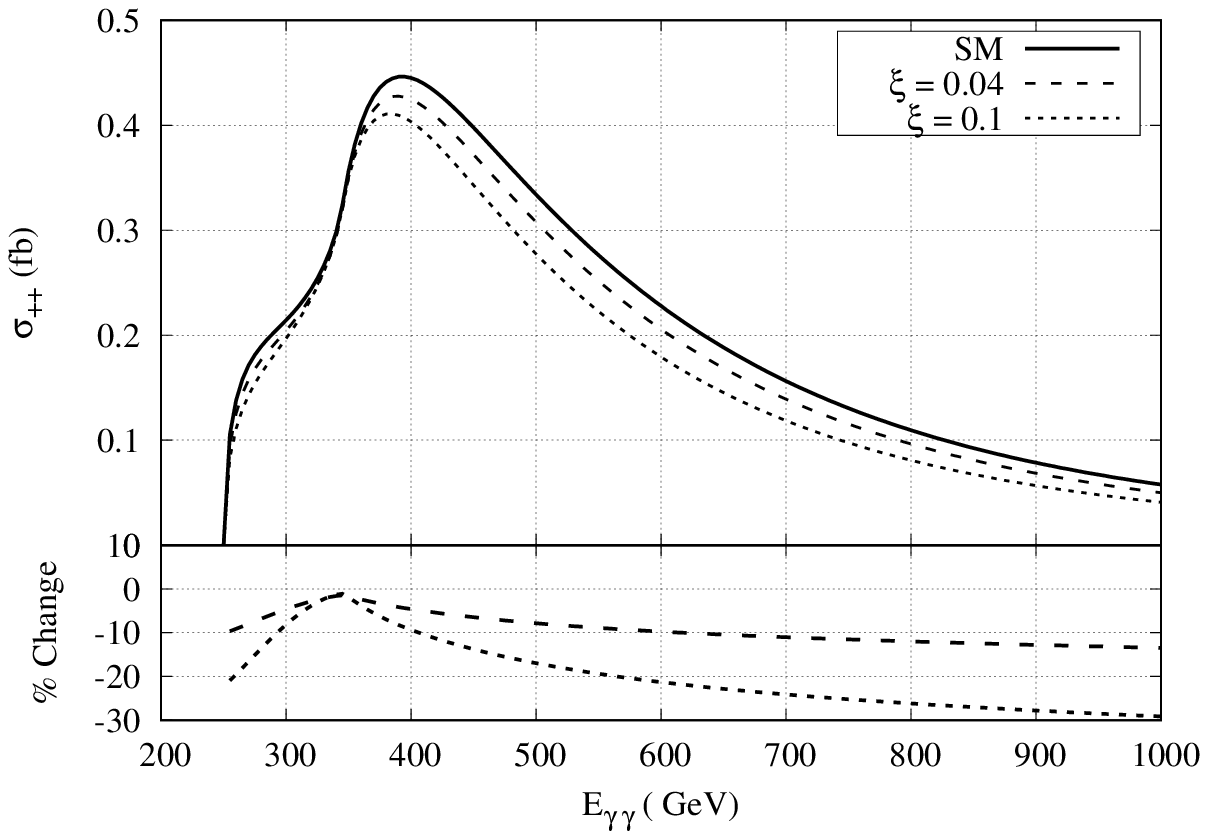,width=0.52\textwidth} 	&   
\epsfig{file=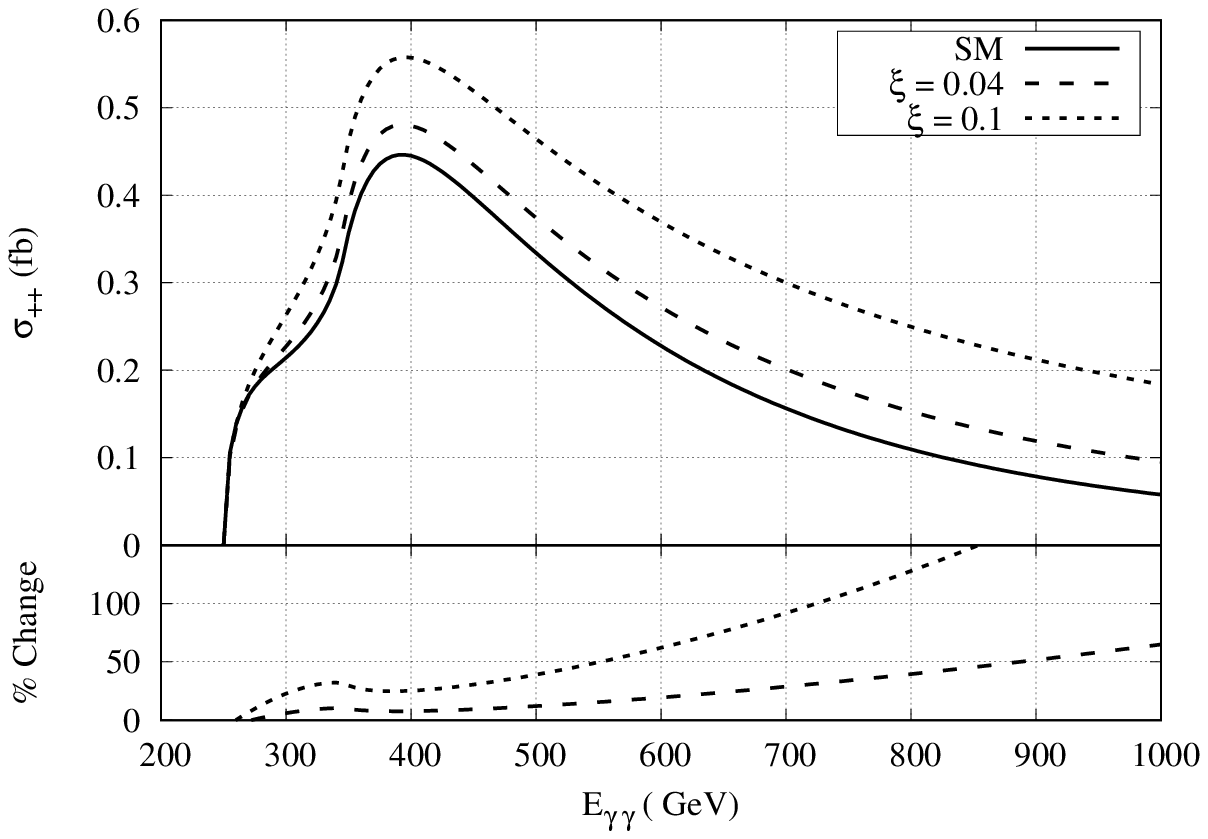,width=0.52\textwidth} \\
\hspace*{-1cm}
\epsfig{file=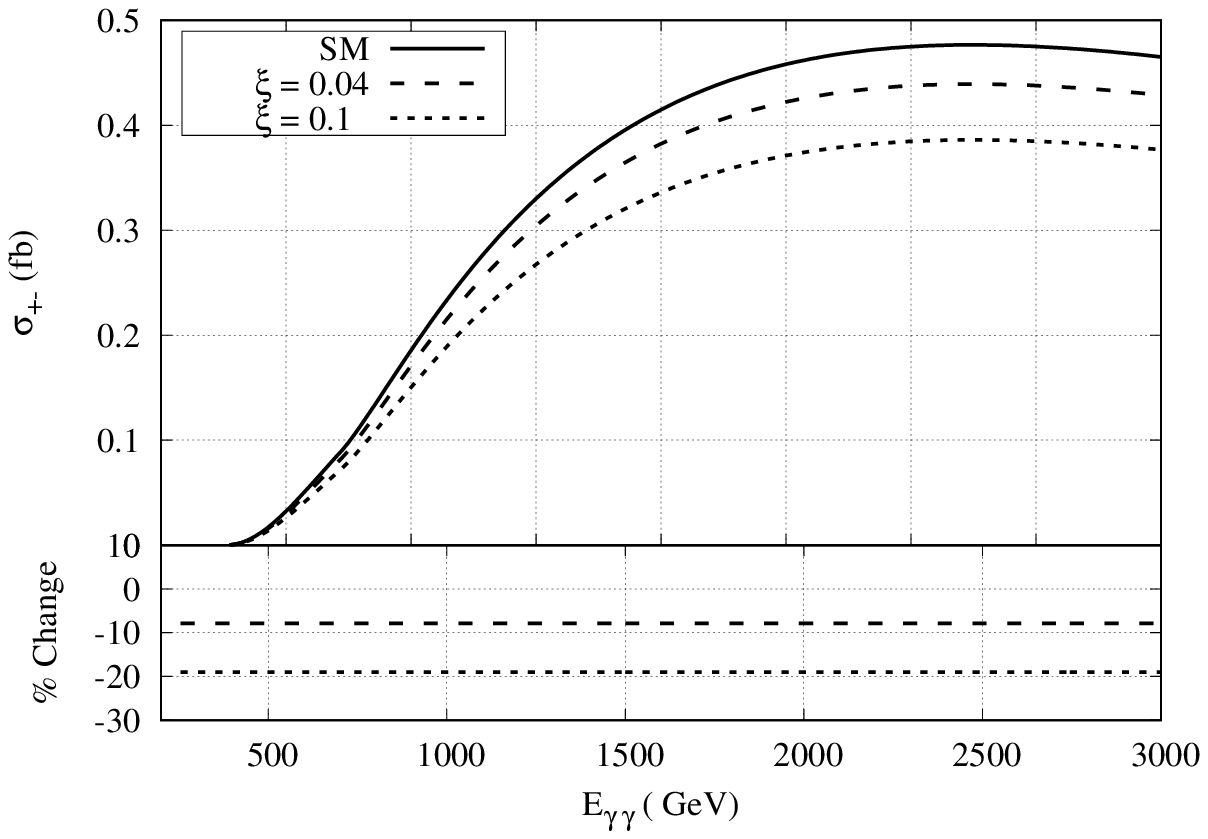,width=0.52\textwidth} & 
\epsfig{file=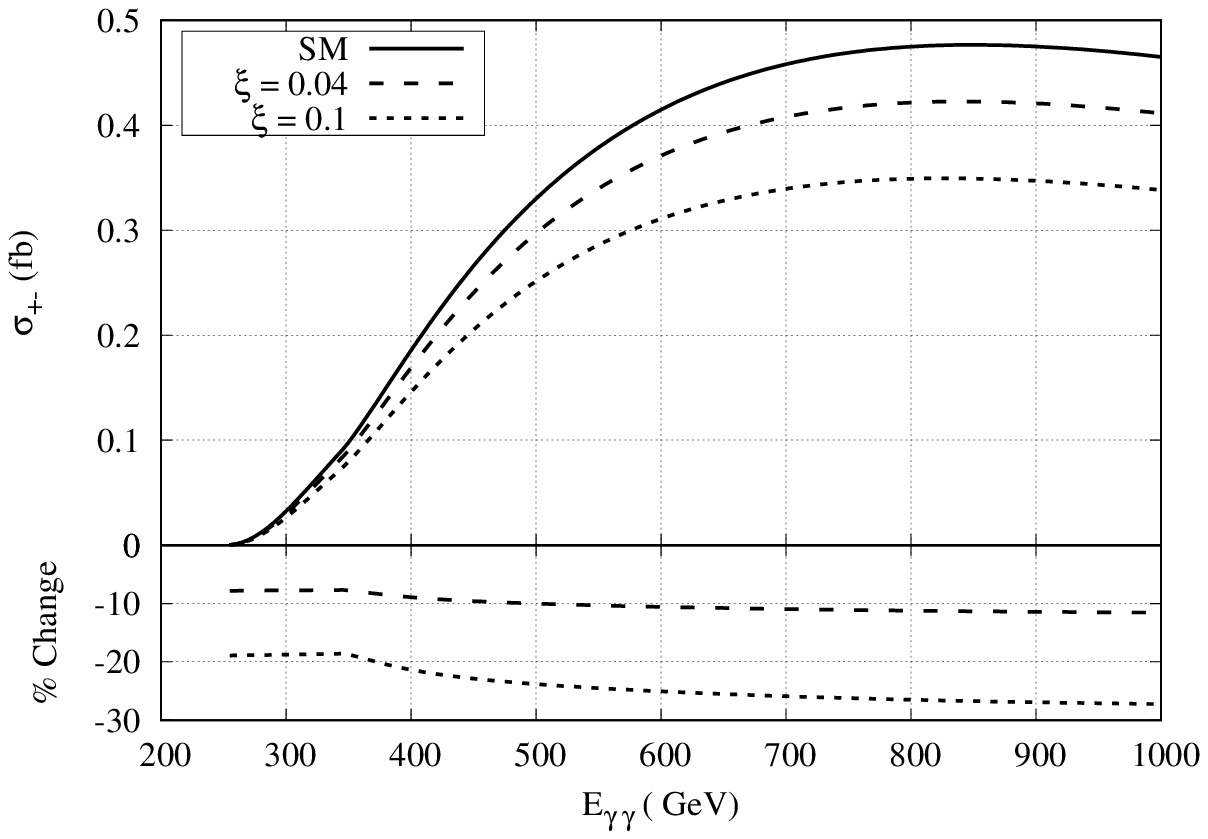,width=0.52\textwidth}
\end{array}	$	
\caption{$\gamma \gamma \to h h$ helicity cross-section for same photon helicities (top panels) and opposite photon helicities (bottom panels) in the MCHM4 (left-column) and MCHM5 (right-column) benchmarks.  
\label{fig:sec3_1} }
\end{center}
\end{figure}

The helicity cross-sections as a function of the di-photon center-of-mass energy are shown in Fig.~\ref{fig:sec3_1}, for two values of $\xi = 0.1,\ 0.04$ and compared to the SM value. The main difference between the MCHM4 and MCHM5 benchmarks is a larger top quartic coupling (c.f. Table~\ref{table:1}) for the latter. This explains the increase in the $\sigma_{++}$ cross-section compared to the SM one, while for the MCHM4 case we observe a systematic decrease.
This result highlights the importance of this new coupling in affecting the photon fusion production of two Higgs bosons. In the Appendix we also report the results for the partial contributions to the cross-section. The increase in
the cross-section for the same helicity photons ($\sigma_{++}$) could be more than $50\%$ in MCHM5 and is the result of 
larger top quartic coupling ($c_{2f}$). 

The modifications of the SM couplings, which always result in reducing their strength, only contribute to a reduction of the cross-section compared to the SM one, as highlighted by the $\sigma_{+-}$ cross-section, which is only sensitive to the box diagrams.

\begin{figure}[tb]
	\begin{center}
    $\begin{array}{cc}
    {\bf MCHM4} & {\bf MCHM5} \\
    \hspace*{-1cm}
    \epsfig{file=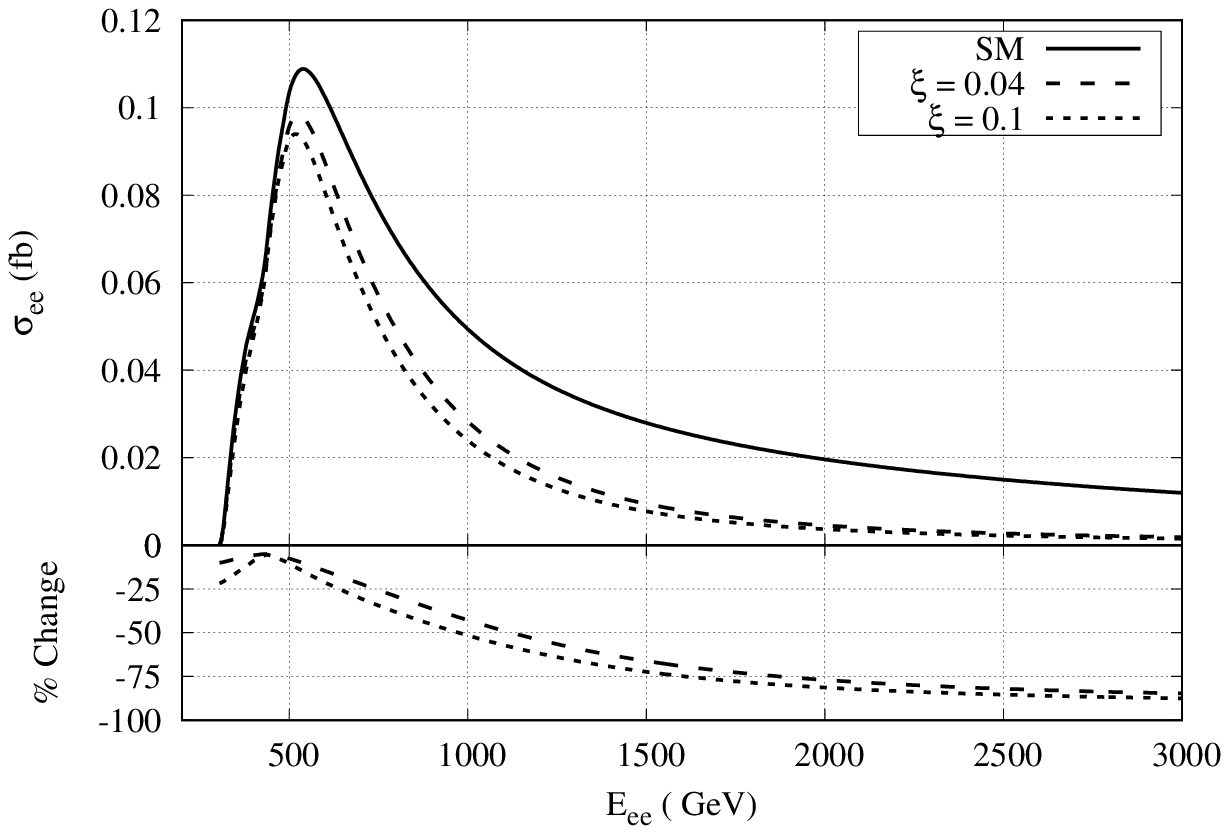,width=0.52\textwidth}  &  
    \epsfig{file=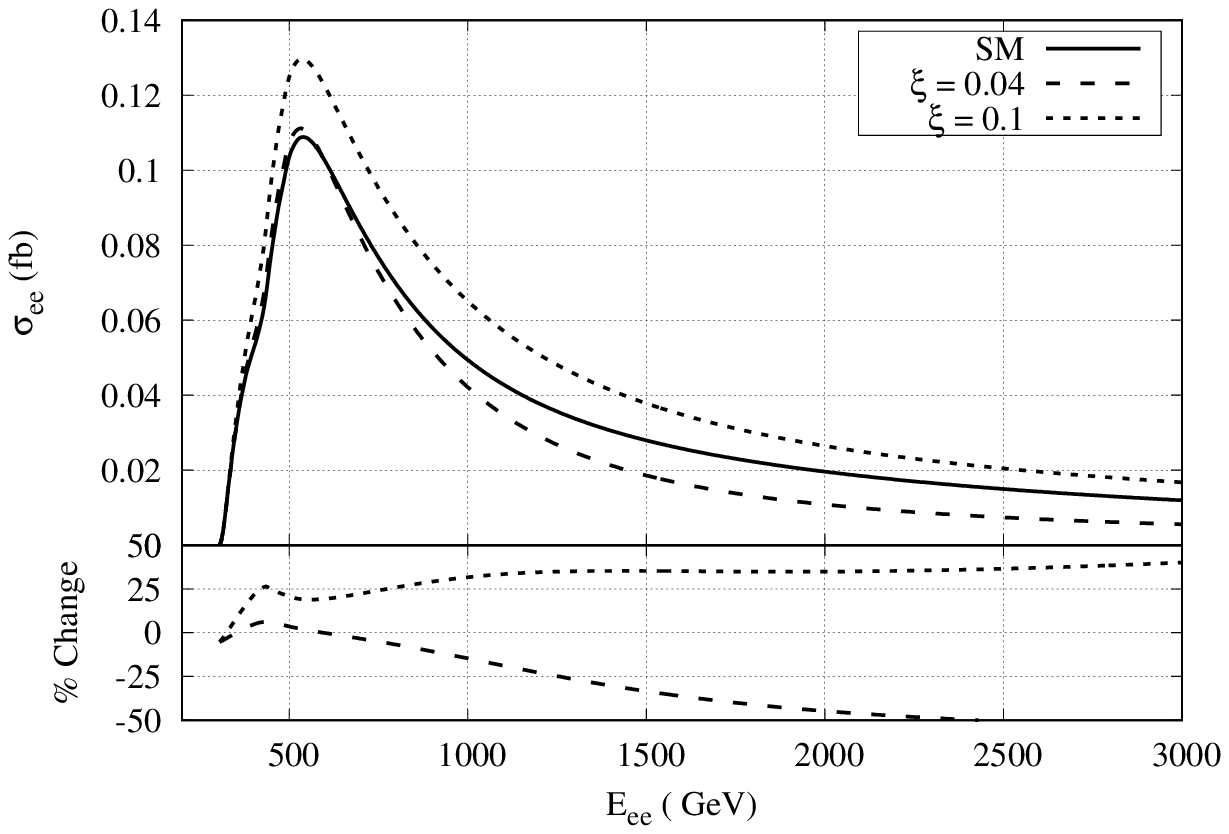,width=0.52\textwidth}	
    \end{array}$
	\caption{$e^+ e^- \to h h$ cross-section as a function of the electron-positron center of mass energy $E_{ee}$ in the MCHM4 (left panel) and MCHM5 (right panel) benchmarks.}  	\label{fig:sec3_2} 
	\end{center}
\end{figure}

In Fig.~\ref{fig:sec3_2} we report the $e^+ e^-$ cross-section as a function of the electron-positron center-of-mass energy for the same benchmarks. In the MCHM4 benchmark, we observe a systematic decrease in the total cross-section, with sizeable effects emerging for center-of-mass energies above $500$~GeV.  Thus, this scenario can only be tested at a high-energy version of the collider. On the other hand, the MCHM5 benchmark can feature an increase in the cross-section compared to the SM one, driven by the $\bar{t} t h h$ coupling $c_{2f}$. This effect can go up to $20\%$ above the $\bar{t} t$ threshold (for $\xi = 0.1$). 

\begin{figure}[htb]
\begin{center} 
    \hspace*{-1cm}
    \epsfig{file=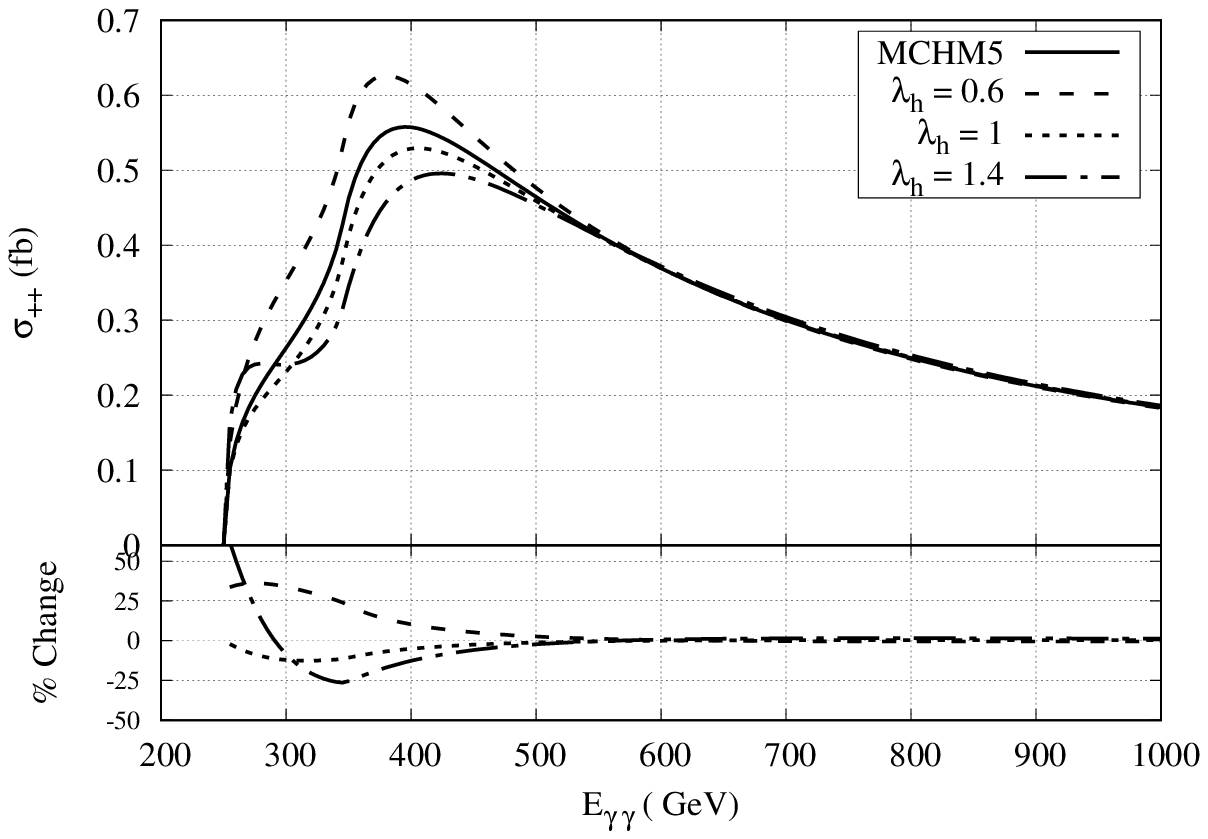,width=0.52\textwidth} 
    \epsfig{file=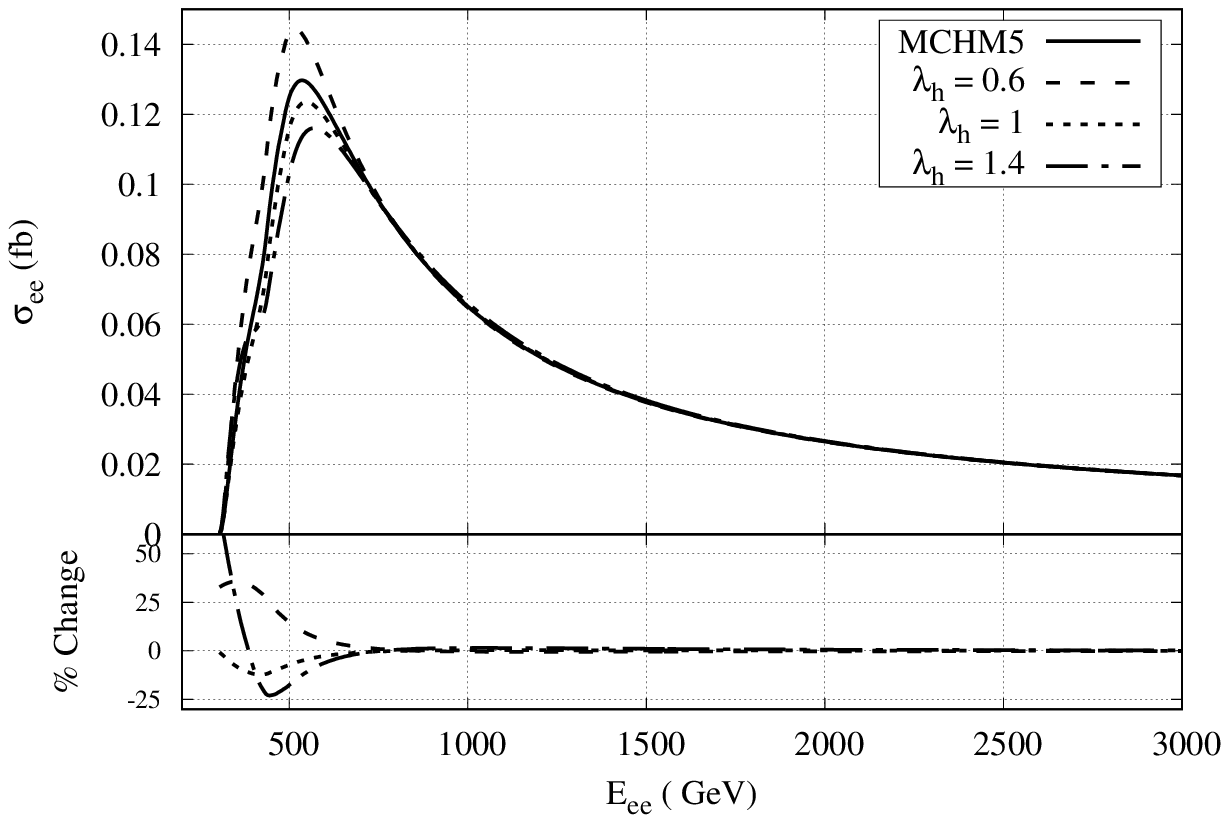,width=0.52\textwidth} 	
    \caption{$\gamma \gamma \to h h$ cross-section for same photon helicity (left panel) and
	$e^+ e^- \to h h$ cross-section (right panel) in MCHM5-Higgs, compared to the MCHM5 benchmark. Here we fix $\xi = 0.1$. The percentage change in above plots represents change when the value of trilinear Higgs coupling is taken 
	as an independent parameter as compared to the MCHM5 values for $\xi=0.1$.}  	
	\label{fig:sec3_3} 
\end{center} 
\end{figure}

\begin{figure}[tb]
	\begin{center}
		\epsfig{file=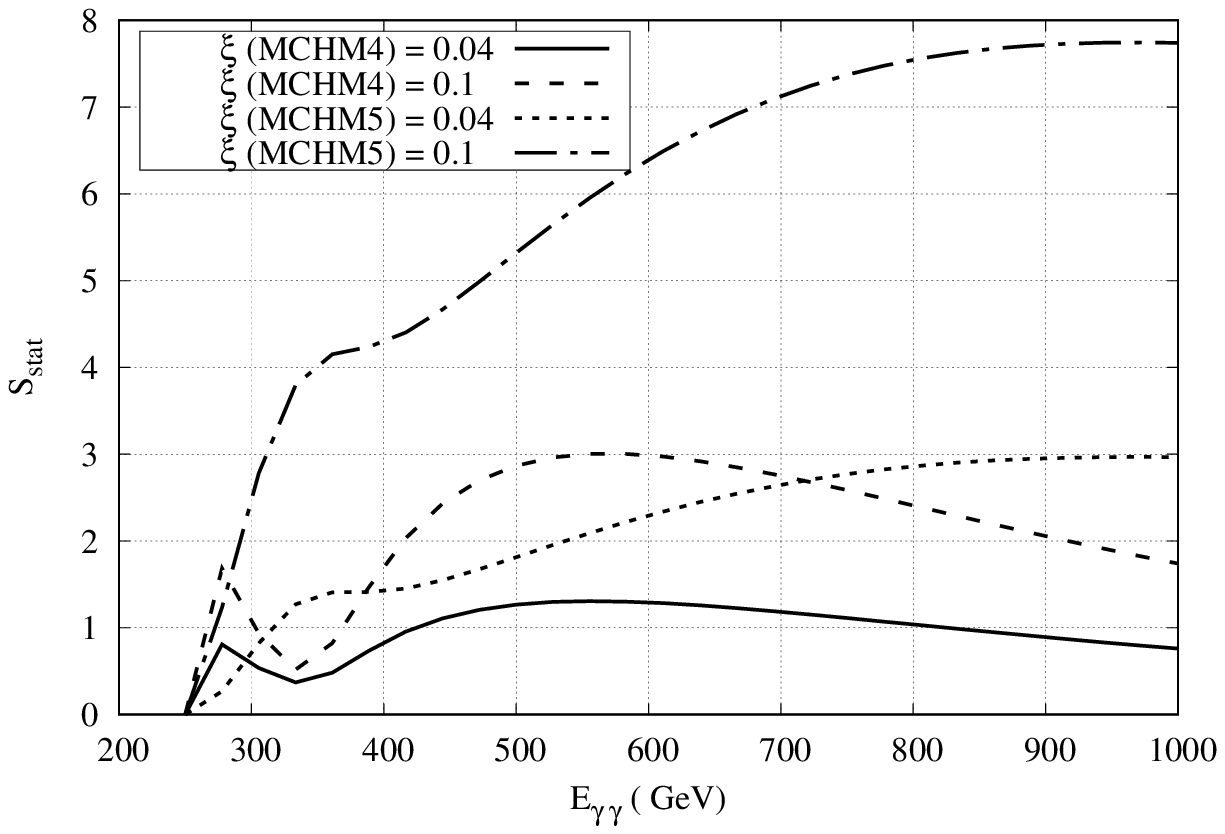,width=0.62\textwidth}  
		\caption{Statistical Sensitivity for $\gamma\gamma \to h h$ in MCHM4 and MCHM5.}  
		\label{fig:sec3_4} 
	\end{center}
\end{figure}

In Fig.~\ref{fig:sec3_3} we have further shown the results in the MCHM5 case, whereby the trilinear Higgs coupling is taken
to be another independent parameter. This scenario is given in the third row of Table \ref{table:1}. 
The variation in the cross-sections could be even more than $25\%$ depending upon the value of trilinear Higgs couplings. 
We have not shown the results for the photon production cross-sections for opposite helicity photons as these contributions 
essentially arise from box diagrams and hence are independent on the trilinear Higgs coupling. 

We have also given an estimate of the statistical sensitivity ($S_{stat}$) as given in \cite{Kawada:2012uy}, 
\begin{equation}
S_{stat} = \frac{| N - N_{SM}|}{\sqrt{N_{obs}}}
= \frac{{\cal L}|\eta\sigma - \eta\sigma_{SM}|}{\sqrt{{\cal L}(\eta\sigma+\eta_{BG}\sigma_{BG})}}
\label{eq:ss1}
\end{equation}
with ${\cal L}$ being the Luminosity. $\sigma$ and $\sigma_{SM}$ are the cross section of the Higgs boson production in the CHMs 
and SM, while $L$, $\eta$, $\eta_{BG}$, and $\sigma_{BG}$ are the integrated luminosity, the detection efficiency for the signal, the 
detection efficiency for backgrounds, and the cross section of background processes, respectively. 
For the statistical sensitivity we have used an integrated luminosity of ${\cal L} = 2 \ ab^{-1}$.  The backgrounds
used in the estimation are as given in \cite{Kawada:2012uy} and used the cut efficiencies as estimated in Table V of 
\cite{Kawada:2012uy}. 
	In estimating the statistical significance we have evaluated the cross-section (for signal and backgrounds) as a function of 
	$E_{\gamma \gamma}$ whereas the cut efficiencies as given in \cite{Kawada:2012uy} are taken to be constant. In addition in the 
	results we have not included another significant background $\gamma \gamma \to t \bar{t}$.  
The statistical sensitivities for MCHM4 and MCHM5 are shown in Fig.  \ref{fig:sec3_4}. 

\subsection{Production in association to leptons in the minimal CHMs}

\begin{figure}[htb]
    \begin{center}
    \epsfig{file=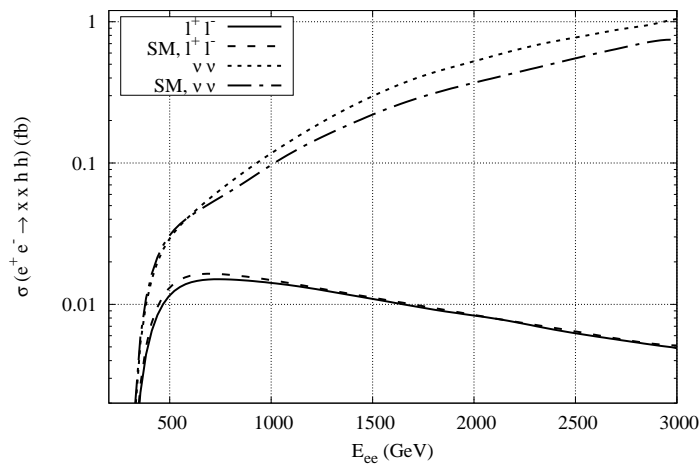,width=0.62\textwidth} 
    \caption{Cross-section for $e^+ e^- \to (\ell^+ \ell^-)/(\nu \bar{\nu}) h h$ in the SM and in the benchmark MCHM5 ($\xi = 0.1$). \label{fig:concl_1}}
    \end{center}
\end{figure} 

The photon initiated di-Higgs production process we consider is a loop mediated process, and one can also have similar processes at tree level at lepton colliders.
In particular, production in association to neutrinos gives typically larger rates in the SM.
It is therefore crucial to study the effects observable from CHMs and compare them to the process we consider, in order to correctly evaluate the relevance of the photon production.
In Fig.~\ref{fig:concl_1} we show the cross-section for the tree level channels 
$\ell^+ \ell^- h h/ \nu \bar{\nu} h h$ as function of the center-of-mass energy. For the evaluation, we used our own implementation of the MCHM5 model using FeynRules \cite{Alloul:2013bka} and MadGraph \cite{Alwall:2011uj}. We should remark that these production channels only depend on the universal modifications of the gauge couplings (thus being the same for all models), and only at a minor level on the model-specific trilinear couplings.
Our results show that the production cross-section for $h h \ell^+ \ell^-$ is much smaller 
compared to the photon one. 
The cross-section for $h h \nu \bar{\nu}$ gives comparable rates to the photon one, however it is much less sensitive to effects from composite models, thus making it less effective in revealing new physics effects.
Furthermore, the presence of a large amount of missing energy (due to neutrinos in the final state) renders it less favorable for a precision study.

These considerations also apply to the cases with resonances, we analyse them in the next two sections.

\section{Introducing a heavy scalar \label{section:4} } 	

\begin{table}[tb]
\begin{center}
\begin{tabular}{|c|c|}
\hline
Diagrams & Amplitude \\
\hline
\begin{minipage}{0.6\textwidth} \begin{center}
 \epsfig{file=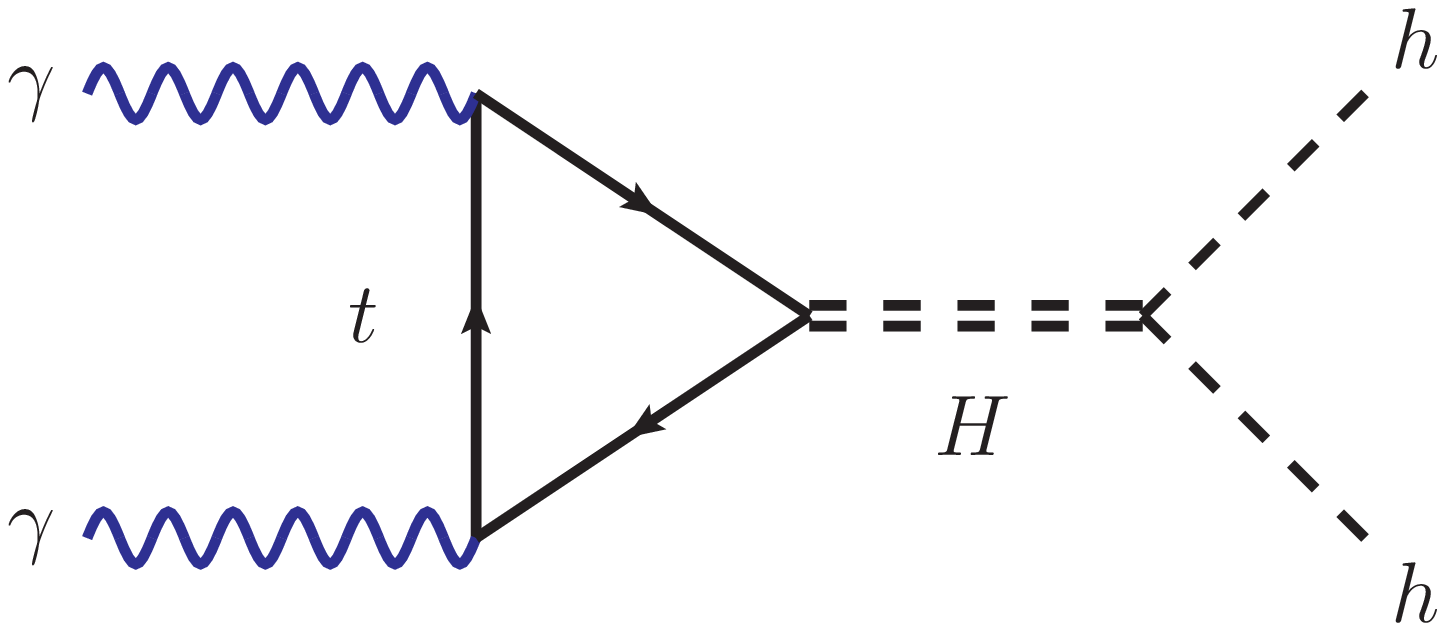,width=0.49\textwidth}  \end{center} \end{minipage}    & $\mathcal{M}_{c_f^H}$ \\ \hline
\begin{minipage}{0.6\textwidth} 
\begin{center} 
\epsfig{file=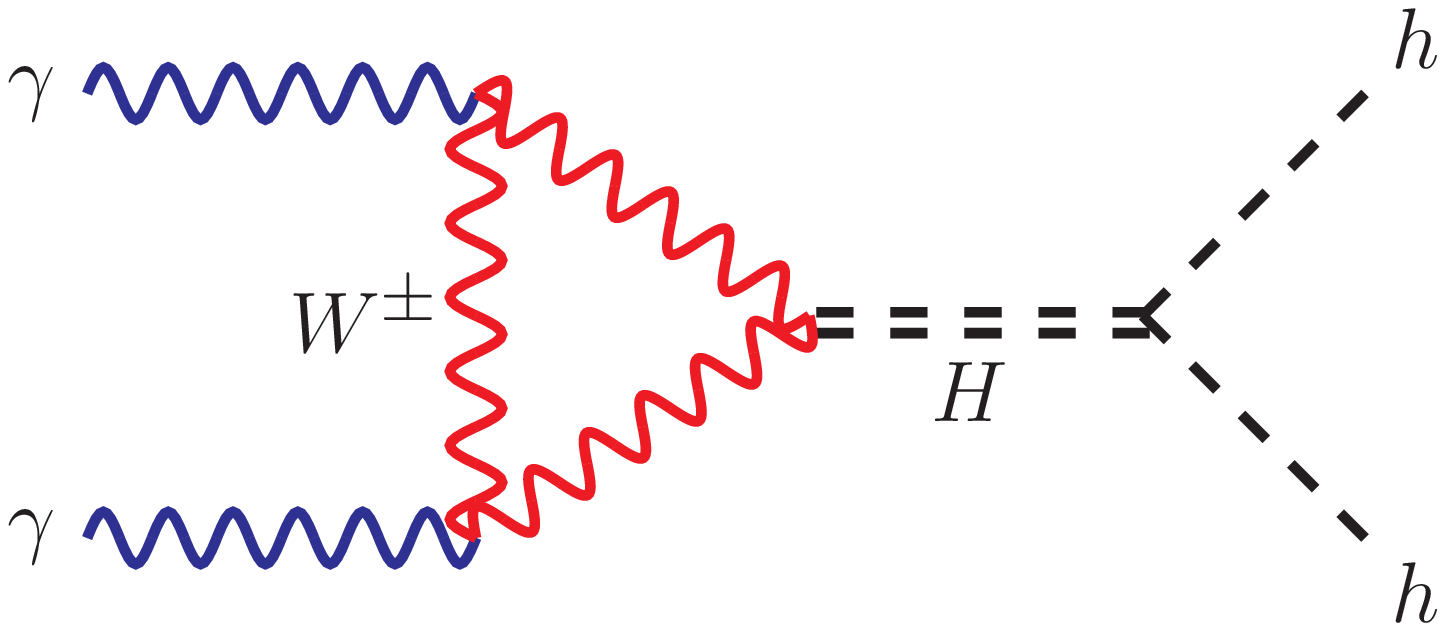,width=0.49\textwidth}  
\epsfig{file=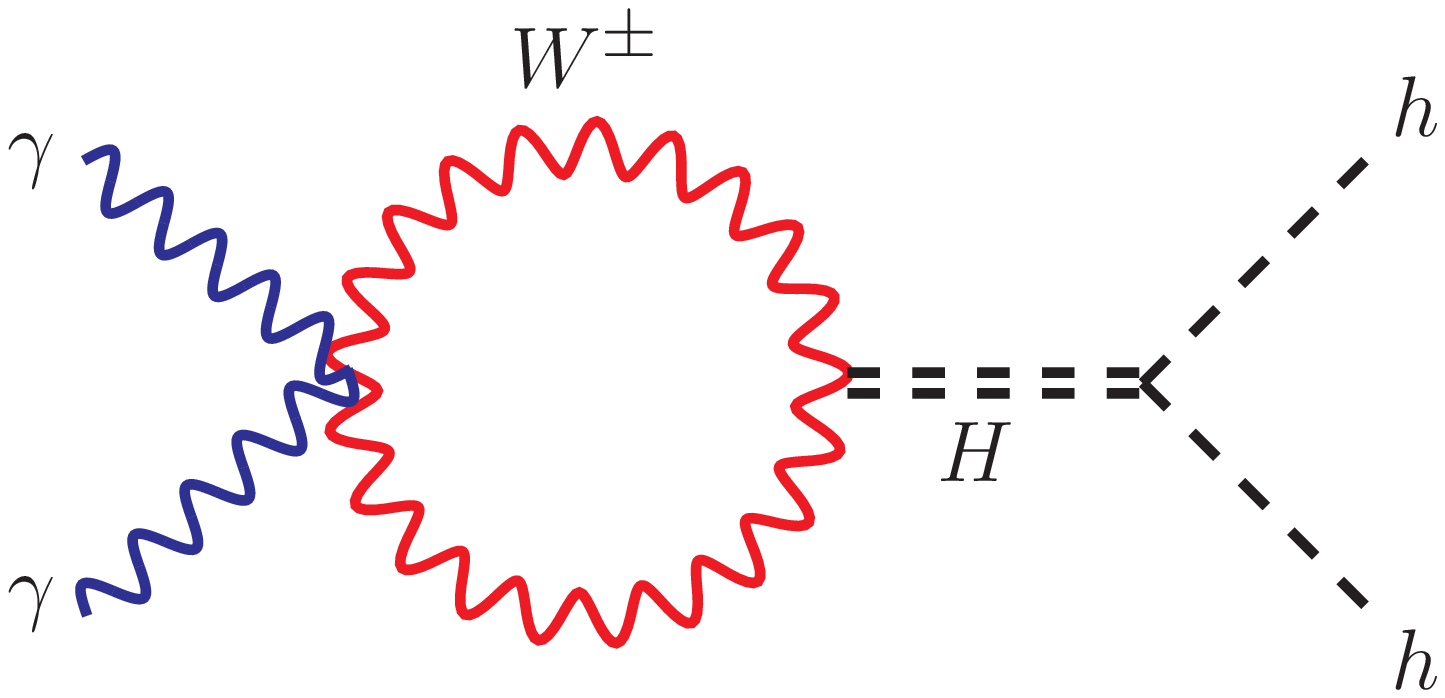,width=0.49\textwidth}  
\end{center} \end{minipage}& $\mathcal{M}_{c_v^H}$ \\  \hline
\end{tabular}    
\caption{Additional diagrams with the heavy scalar $H$, with the corresponding partial amplitudes as defined in the text.} \label{tab:diagH}
\end{center}
\end{table}

The presence of a rather light scalar resonance in the spectrum of CHMs has been shown to help in reducing the constraints on the misalignment angle~\cite{BuarqueFranzosi:2018eaj}.
This situation seems to be rather common and realistic, as there are indications from several sources of its presence if the theory has a near scale-invariant behavior, as needed for modern composite Higgs models. As an example, lattice calculations~\cite{Witzel:2018gxm,Appelquist:2020xua,Witzel:2020hyr} for a QCD-like theory with 4 light flavours (to generate a composite Higgs model) and 6 or 8 heavy ones feature a $0^{++}$ scalar resonance with a mass lower than half of the $\rho$ mass. This result has been found also in gravity dual theories~\cite{Elander:2020fmv,Elander:2017cle} and more recently in holographic realisations~\cite{Pomarol:2019aae,Elander:2020csd,Erdmenger:2020flu}.

For our purposes, the presence of a relatively light scalar can be encoded in the addition of a second heavier ``Higgs'' $H$, with couplings parameterised in analogy to those of the SM Higgs:
\begin{multline}
\mathcal{L} \supset m_W^2 W_\mu^+ W^{-,\mu} \left( 1 + 2 c_v \frac{h}{v} + 2 c_v^H \frac{H}{v} +  c_{2v} \frac{h^2}{v^2} + \dots \right) - \\
m_t \bar{t} t \left( 1 + c_f \frac{h}{v} + c_t^H \frac{H}{v} + \frac{c_{2t}}{2} \frac{h^2}{v^2} +  \dots \right)\,.
\end{multline}

The coefficients $c_x$ and $c_x^H$, for $x = f, v$, can be computed in specific scenarios, and take into account the mixing between the two states.
Furthermore, the model contains a new coupling between the heavy state $H$ and the SM-like Higgs coming from the kinetic term of the pNGB Higgs. This coupling entails derivatives as follows: 
\beq \label{derHhh}
\mathcal{L} \supset c_{Hhh}\ H \partial_\mu h \partial^\mu h \to - \frac{1}{2} c_{Hhh} \frac{\hat{s} - 2 m_h^2}{v}\ H h h\,.
\eeq
The coupling $c_{Hhh}$ enters in the di-Higgs production via s-channel production of $H$, as shown in Table~\ref{tab:diagH}, thus can be effectively replaced by the second term in Eq.~\eqref{derHhh}, where $\hat{s}$ is the invariant mass of the two $h$ system (i.e., the center-of-mass energy of the partonic process).

Following Ref.~\cite{BuarqueFranzosi:2018eaj}, we defined three benchmark points which pass all experimental constraints  summarised in Table~\ref{table:BM12}. As it can be seen, the value of $\xi$ can be allowed to be as large as $0.3$, while larger values for the quartic coupling to fermions are also allowed. Quartic couplings involving $H$ are not reported as they are irrelevant for our purposes.

\begin{table}[htb]
	\begin{center}
		\begin{tabular}{|c|c|c|c|c|c|c|} \hline
			Benchmark 1 & \multicolumn{6}{c|}{$m_H = 610$~GeV, $\xi = 0.306$, $\Gamma_H = 498$~GeV, $k'_G = 1.5$ } \\ \hline
			& $c_f/c_f^{H}$ & $c_{2f}$ & $c_v/c_v^{H}$ & $c_{2v}$
			&  $c_{3h}$ & $c_{Hhh}$\\ \hline
			$h$  & $0.9199$  &  $-0.7814$  & $0.8791$ & $0.5562$ & $\lambda_h$  & $-$  \\ \hline
			$H$  & $3.507$  &  $\dots$  & $0.3054$ & $\dots$
			& $-$ & $0.4149$ \\ \hline       \hline
			Benchmark 2 & \multicolumn{6}{c|}{$m_H = 800$~GeV, $\xi = 0.197$, $\Gamma_H = 350$~GeV, $k'_G = 1.8$ } \\ \hline
			& $c_f/c_f^{H}$ & $c_{2f}$ & $c_v/c_v^{H}$ & $c_{2v}$
			&  $c_{3h}$ & $c_{Hhh}$\\ \hline
			$h$  & $0.9102$  &  $-0.4627$  & $0.9305$ & $0.7381$ & $\lambda_h$ & $-$    \\ \hline
			$H$  & $2.368$  &  $\dots$  & $0.3109$ & $\dots$
			& $-$ & $0.4001$\\ \hline   \hline
			Benchmark 3 & \multicolumn{6}{c|}{$m_H = 1000$~GeV, $\xi = 0.0646$, $\Gamma_H = 47.6$~GeV, $k'_G = 1.$ } \\ \hline
			& $c_f/c_f^{H}$ & $c_{2f}$ & $c_v/c_v^{H}$ & $c_{2v}$
			&  $c_{3h}$ & $c_{Hhh}$\\ \hline
			$h$  & $0.9572$  &  $-0.1498$  & $0.9741$ & $0.9038$ & $\lambda_h$ & $-$    \\ \hline
			$H$  & $0.6896$  &  $\dots$  & $0.0511$ & $\dots$
			& $-$ & $0.1270$\\ \hline
		\end{tabular}
	\caption{Couplings of the Higgs $h$ and of the heavier state $H$, for 3 benchmark points. The parameter $k'_G$ characterises the coupling of the heavy resonance to the gauge bosons (see Ref.~\cite{BuarqueFranzosi:2018eaj} for more details).\label{table:BM12}}
	\end{center}
\end{table}

\subsection{Numerical results}

\begin{figure}[htb]
	\begin{center}
	\hspace*{-1cm}
		\epsfig{file=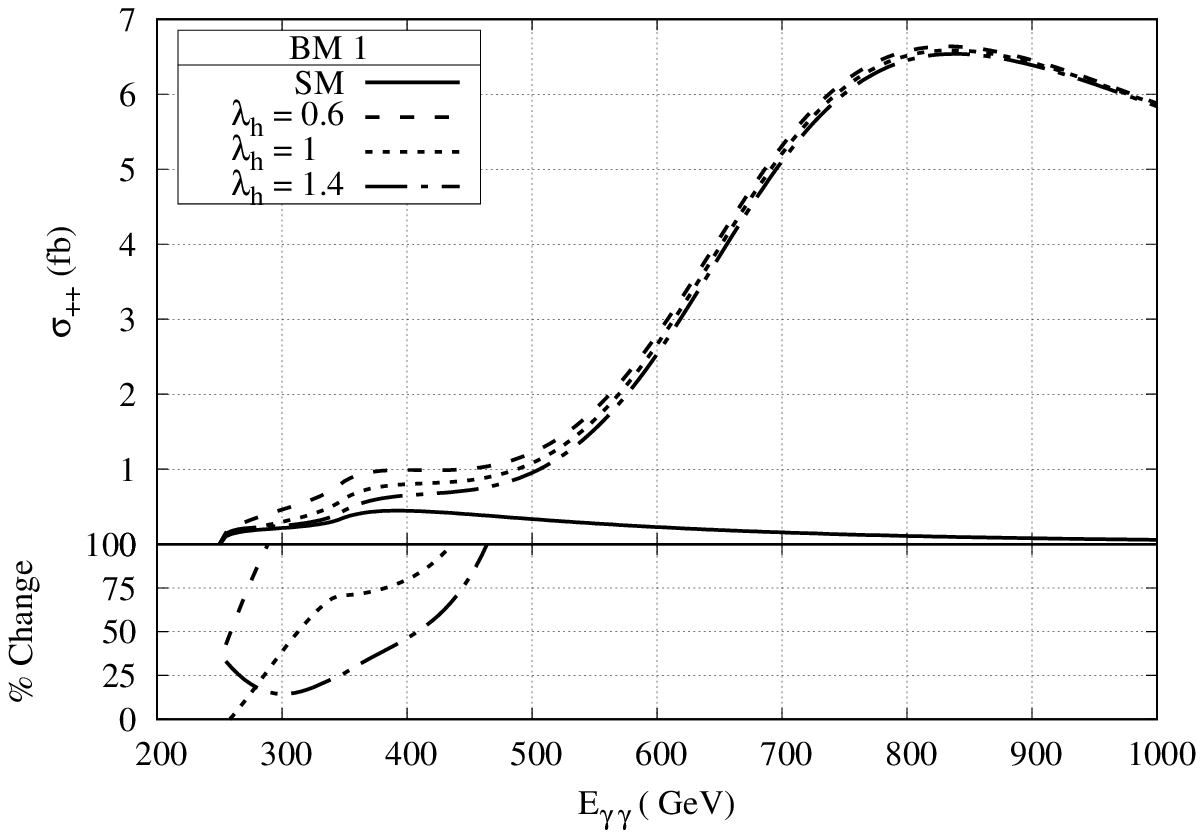,width=0.52\textwidth} 	
		\epsfig{file=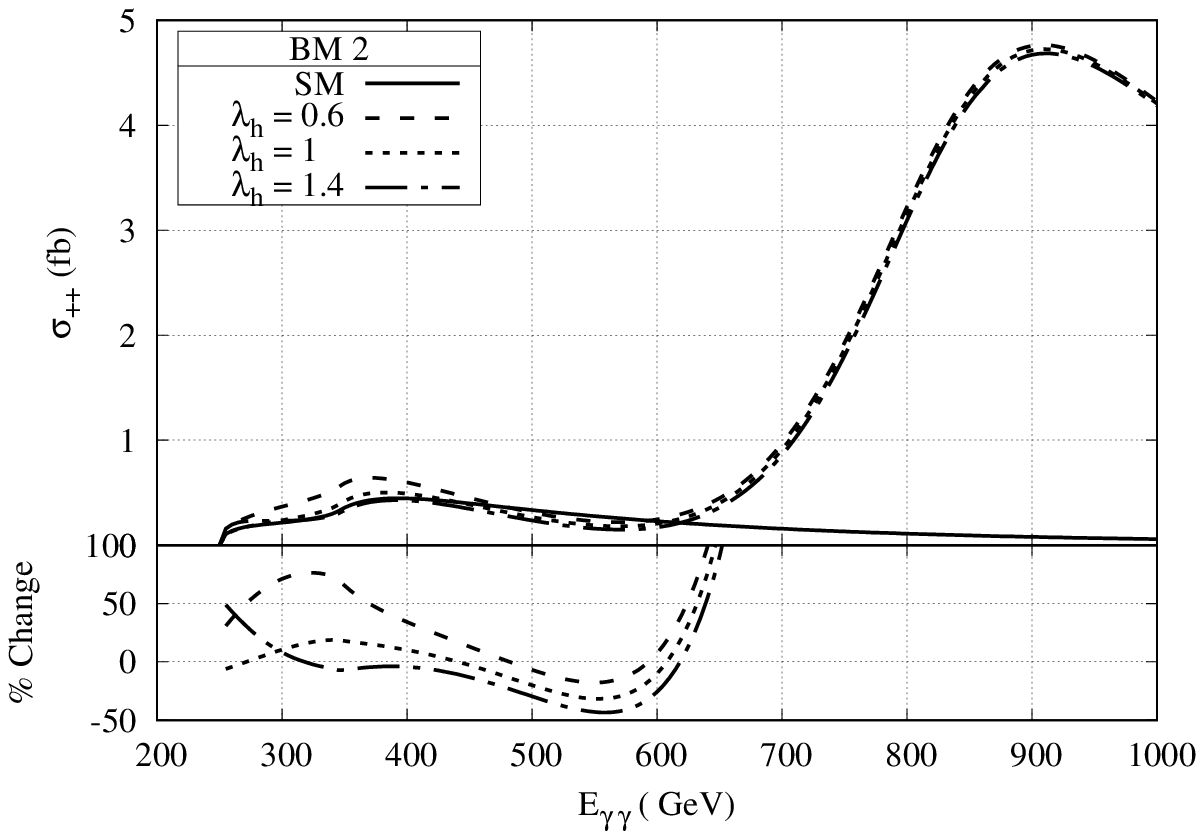,width=0.52\textwidth} \\
	\hspace*{-1cm}
		\epsfig{file=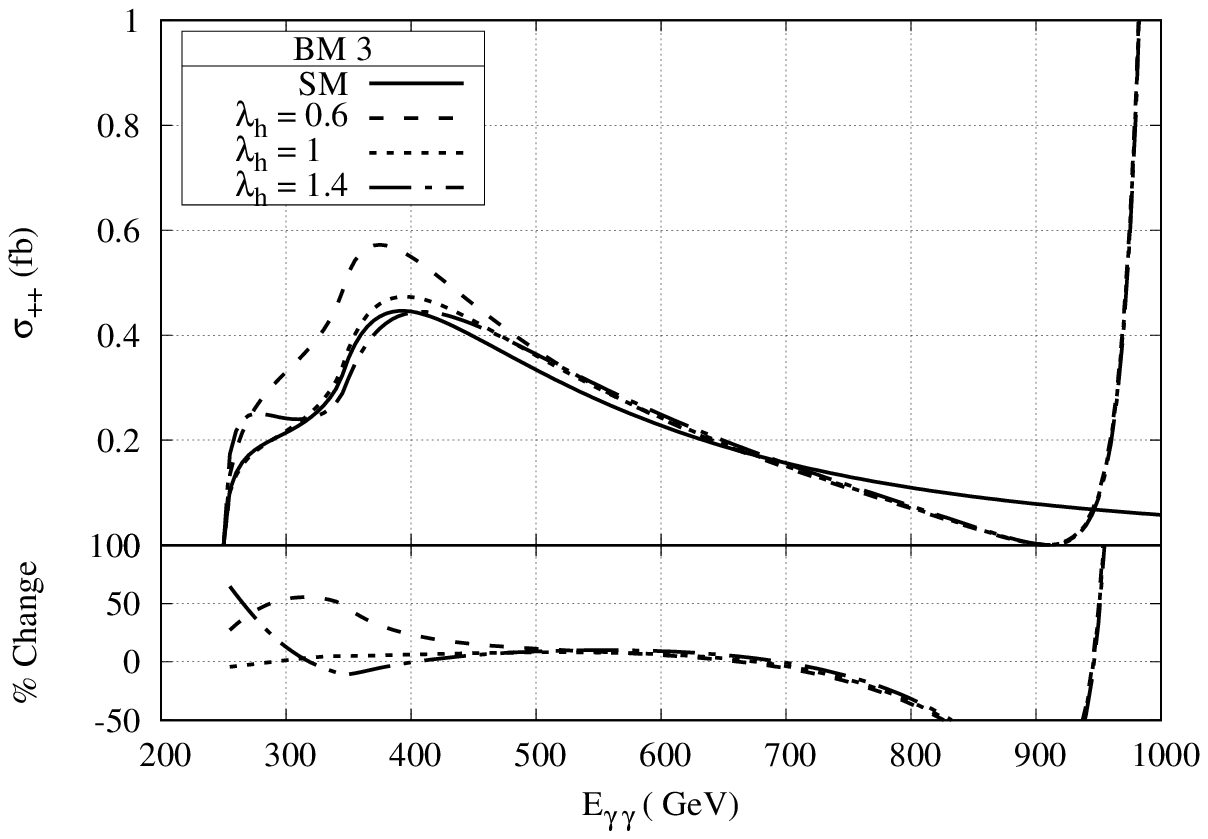,width=0.52\textwidth} 	
		\epsfig{file=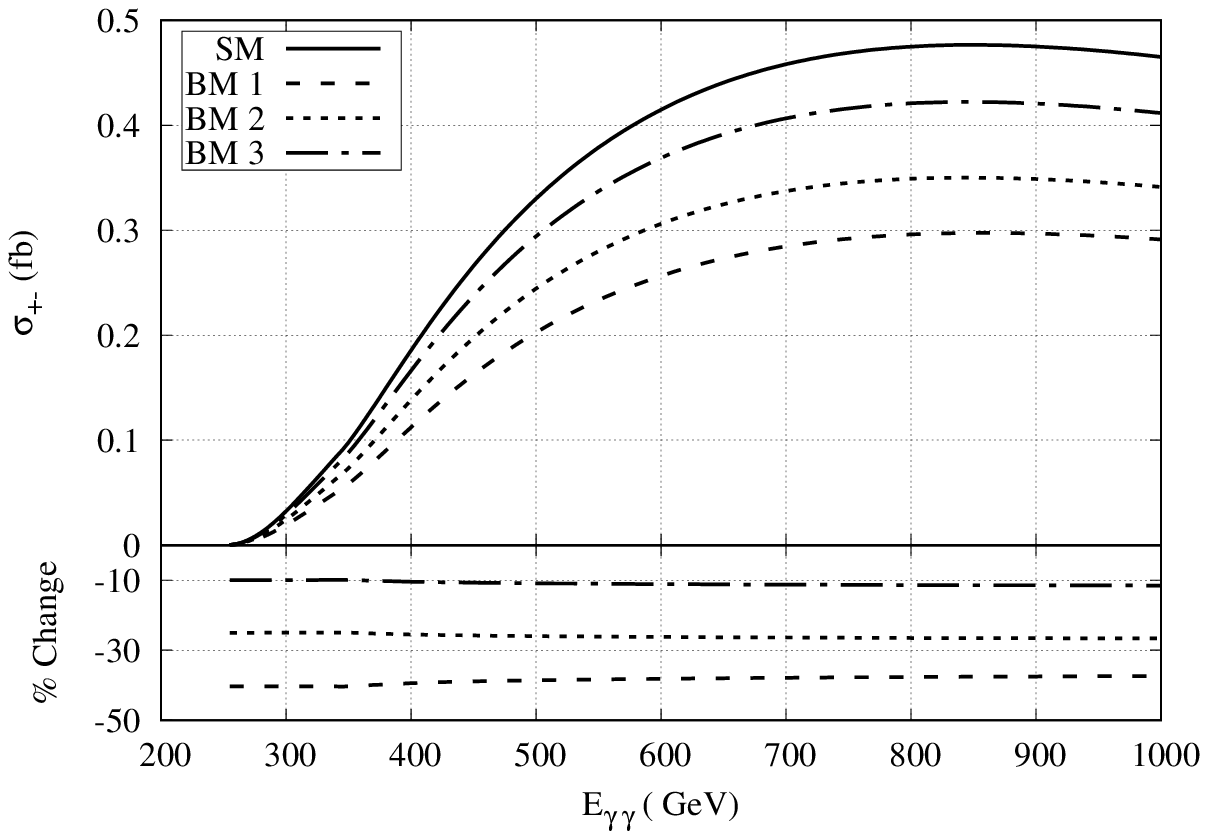,width=0.52\textwidth} 
	\caption{$\gamma \gamma \to h h$ helicity cross-section in the model with an additional heavy scalar $H$. The benchmark 
	points (BM) for the above results are given in Table \ref{table:BM12}.}  	\label{fig:sec4_1} 
	\end{center}
\end{figure}

In Fig.~\ref{fig:sec4_1} we show the helicity cross-sections for the three benchmark points shown in Table~\ref{table:BM12}. 
The first three plots show these benchmark points separately, for three values of the SM-Higgs quartic coupling, $\lambda_h = 0.6,\ 1,\ 1.4$. 
The plots clearly show a substantial enhancement around the mass of $H$, which also depends on the width of this state. 
Thus, in BM1 (where the width of a singlet scalar Higgs is very large), the enhancement is spread over wide values of the 
di-photon energy, while it becomes narrower for BM3 as the width of the heavy Higgs is relatively smaller. 
This effect derives from the new s-channel diagrams mediated by $H$ as shown in Table~\ref{tab:diagH}. These diagrams
		 contribute only to  $\sigma_{++}$. 
		The $\lambda_h$ dependence is visible for low center-of-mass energies. In this region one can observe effects of the trilinear 
		SM Higgs coupling ($c_{3h}$) diagrams with the SM Higgs as mediator in s-channel. These diagrams are given in Table \ref{tab:diagSM}. 
    In the bottom-right panel, we show the $\sigma_{+-}$ cross-section: as it involves only box diagrams, 
	it is not affected by the Higgs trilinear coupling nor by the $H$. The reduction is therefore only due to the reduction in the SM Higgs couplings, 
	which are more marked for BM1 compared to BM3. 

\begin{figure}[htb]
	\begin{center}
	\hspace*{-1cm}
		\epsfig{file=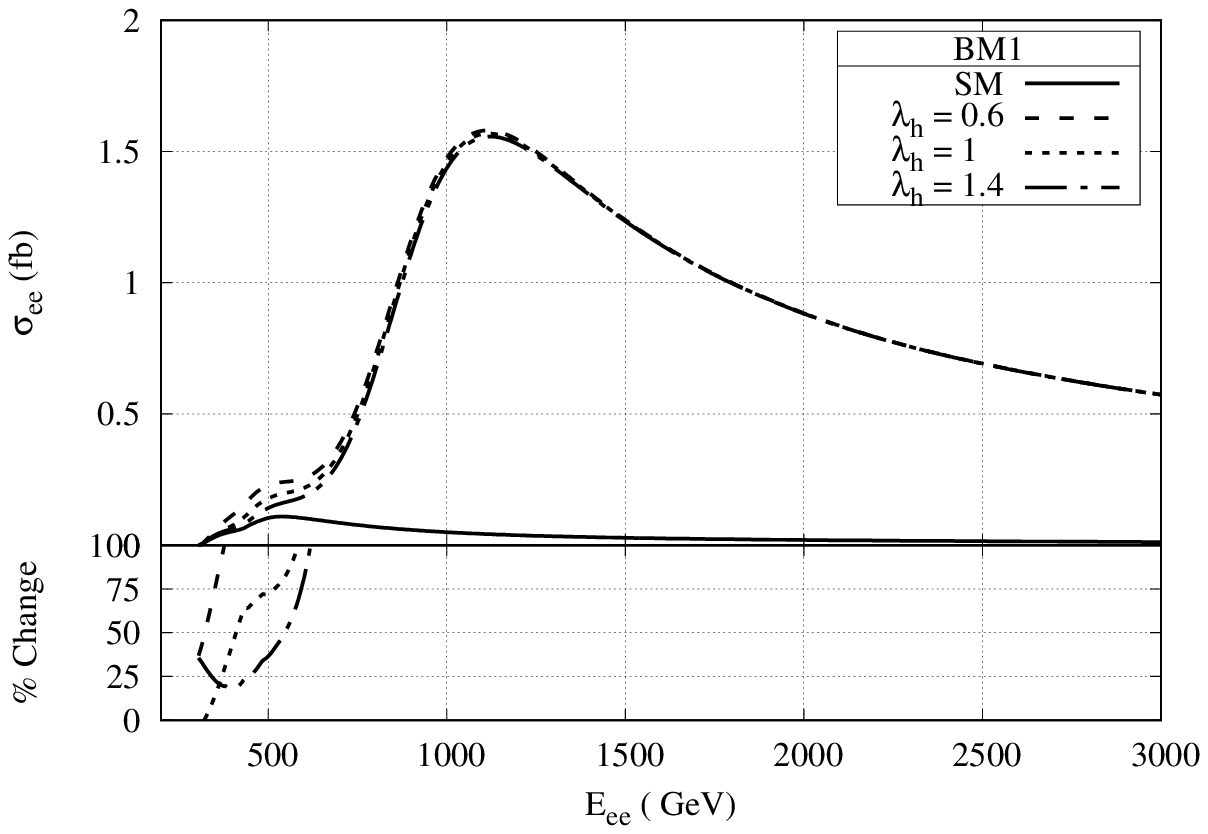,width=0.52\textwidth} 	
		\epsfig{file=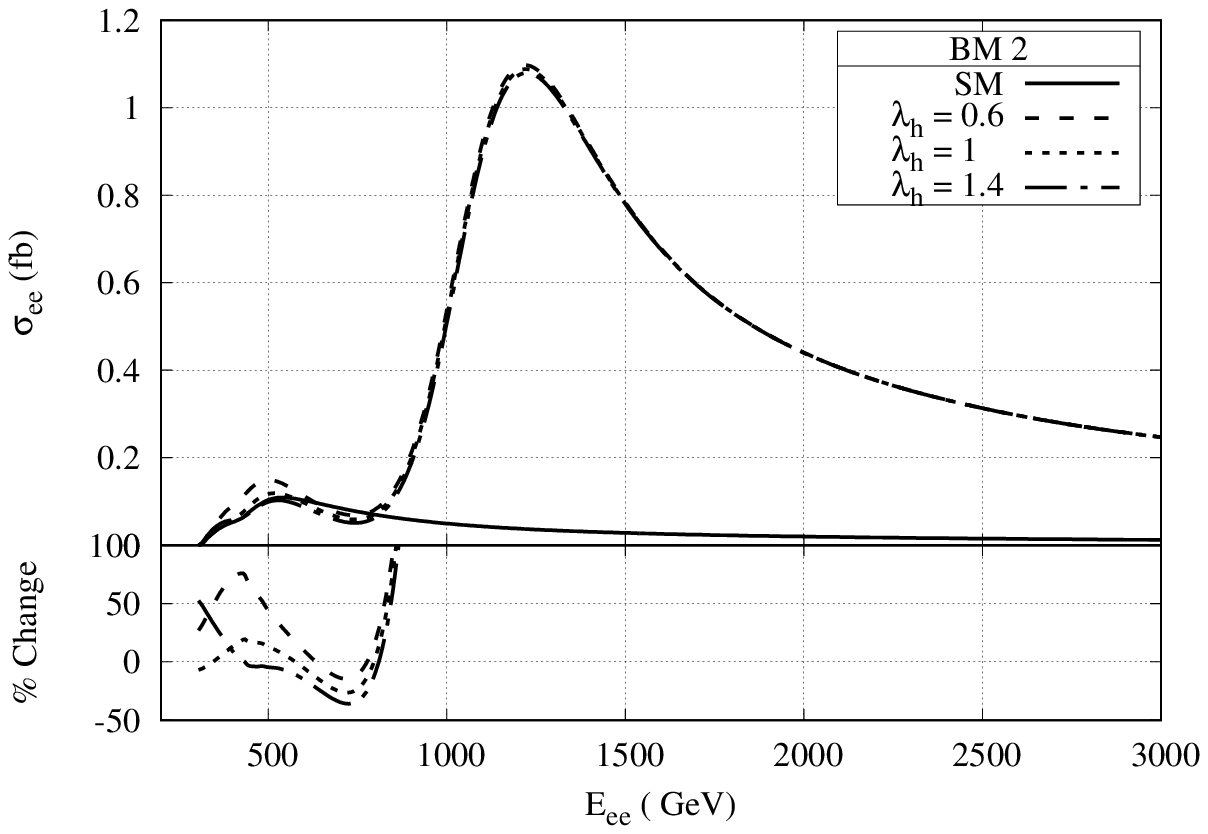,width=0.52\textwidth} 	\\ 
		\epsfig{file=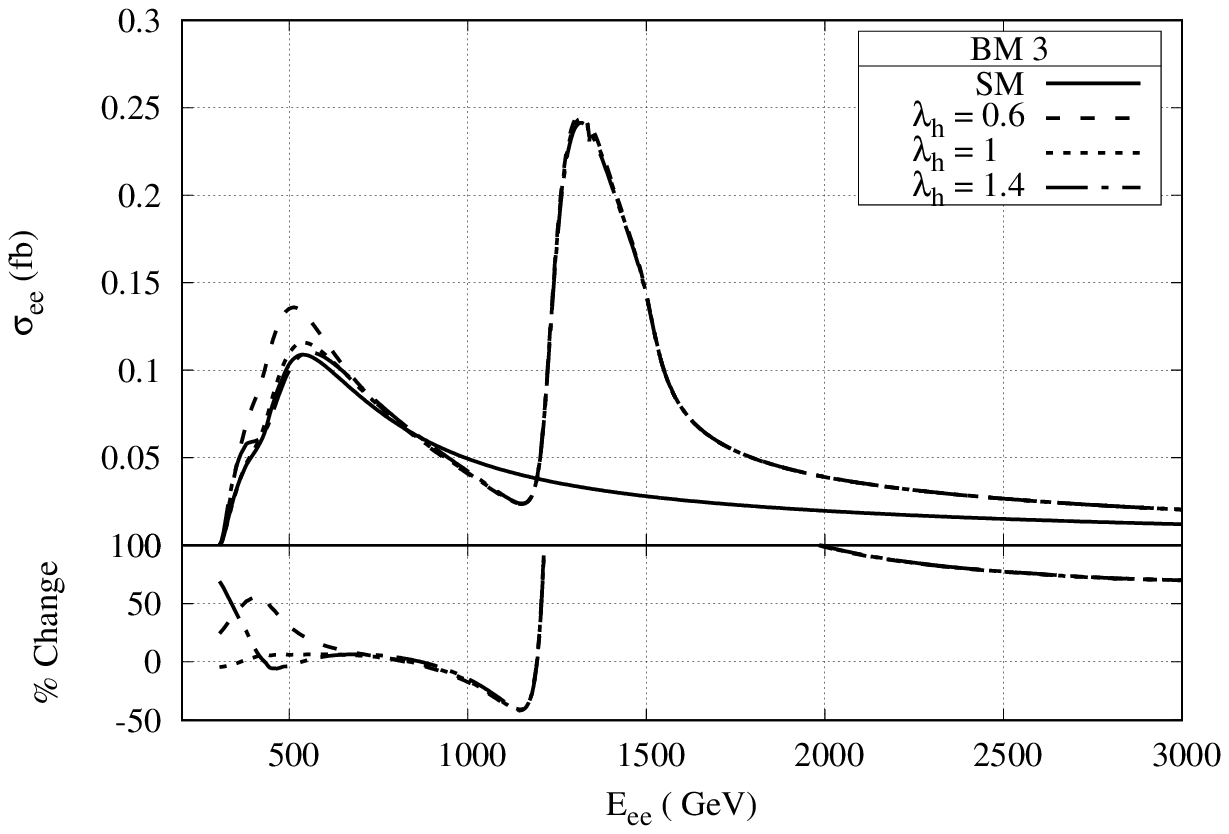,width=0.52\textwidth} 	
	\caption{$e^+ e^- \to h h$ cross-section in the model with an additional heavy scalar $H$. The benchmark 
	points (BM) for the above results are given in Table \ref{table:BM12}.}   	\label{fig:sec4_2} 
	\end{center}
\end{figure}

\begin{figure}[htb]
	\begin{center}
		\epsfig{file=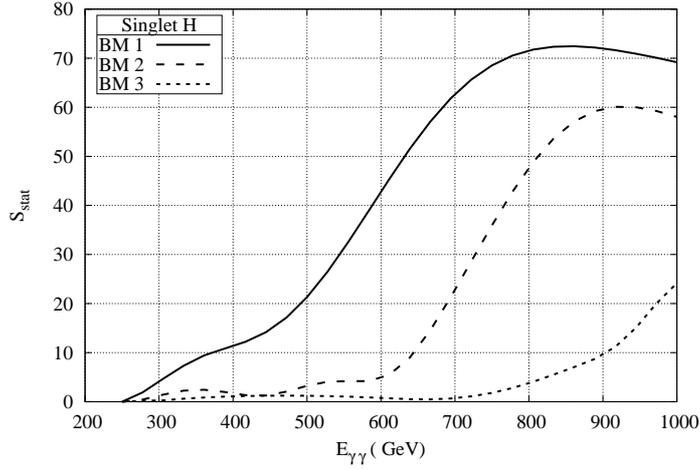,width=0.62\textwidth} 	
		\caption{Statistical Sensitivity as a function of $\gamma\gamma$ collider energy for the process $\gamma\gamma \to h h $
			for BM1, BM2 and BM3 (as given in Table \ref{table:BM12})  in the model with an additional heavy singlet Scalar $H$.}   	
		\label{fig:sec4_3} 
	\end{center}
\end{figure}

In Fig.~\ref{fig:sec4_2} we show the $e^+ e^-$ cross-sections for the three benchmark points, again for three values of $\lambda_h$. 
	As expected, for BM1 and BM2, which feature a relatively lighter $H$, sizeable enhancements are present at 
	low energies, and the cross-section shows major deviations near the mass of  the heavy Higgs.
As the $H$ is rather heavy for BM3,
its effect at lower energies (say up to 1 TeV) is suppressed, whereas the
impact can be significant around its mass.
     The statistical 
	sensitivity for the three benchmark points (as given in Table \ref{table:BM12}) is shown in Fig. \ref{fig:sec4_3} and shows highest
	significance among all the models we have considered/discussed. 

\section{Introducing top partners \label{section:5} } 	

\begin{table}[tb]
\begin{center}
\begin{tabular}{|c|c|}
\hline
Diagrams & Amplitude \\
\hline
\begin{minipage}{0.6\textwidth} \begin{center}
 \epsfig{file=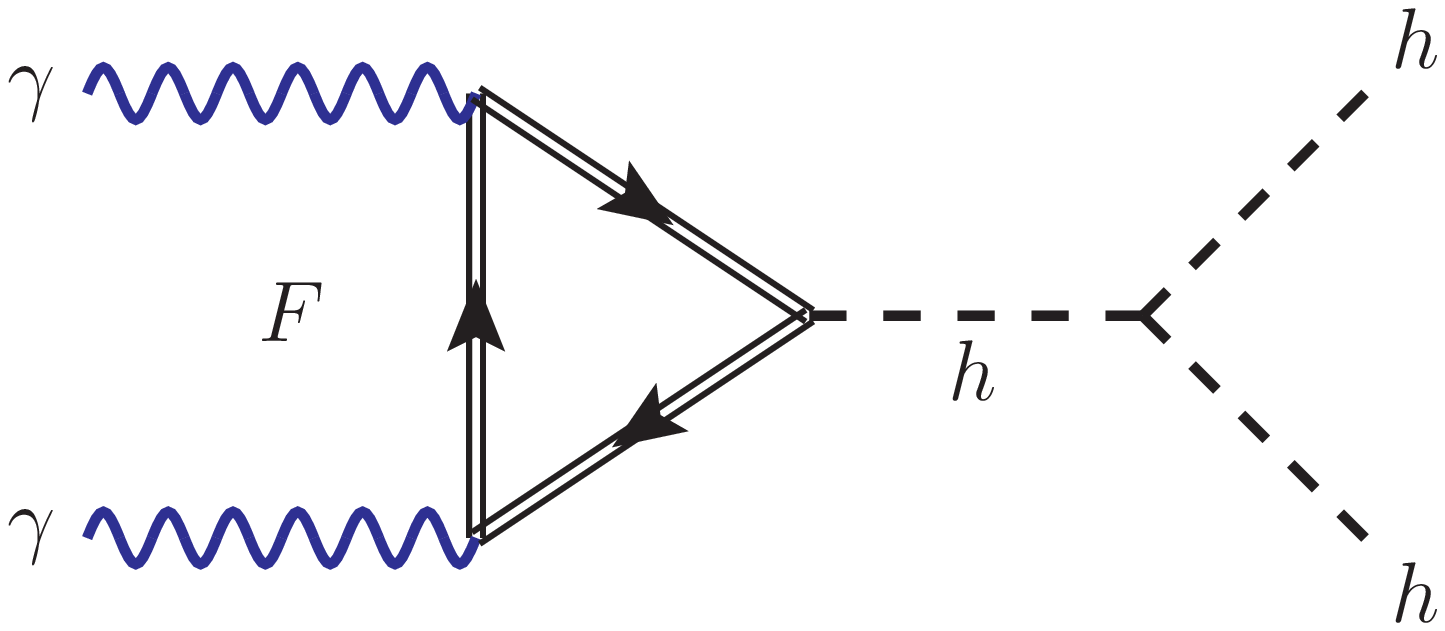,width=0.49\textwidth}  
 \epsfig{file=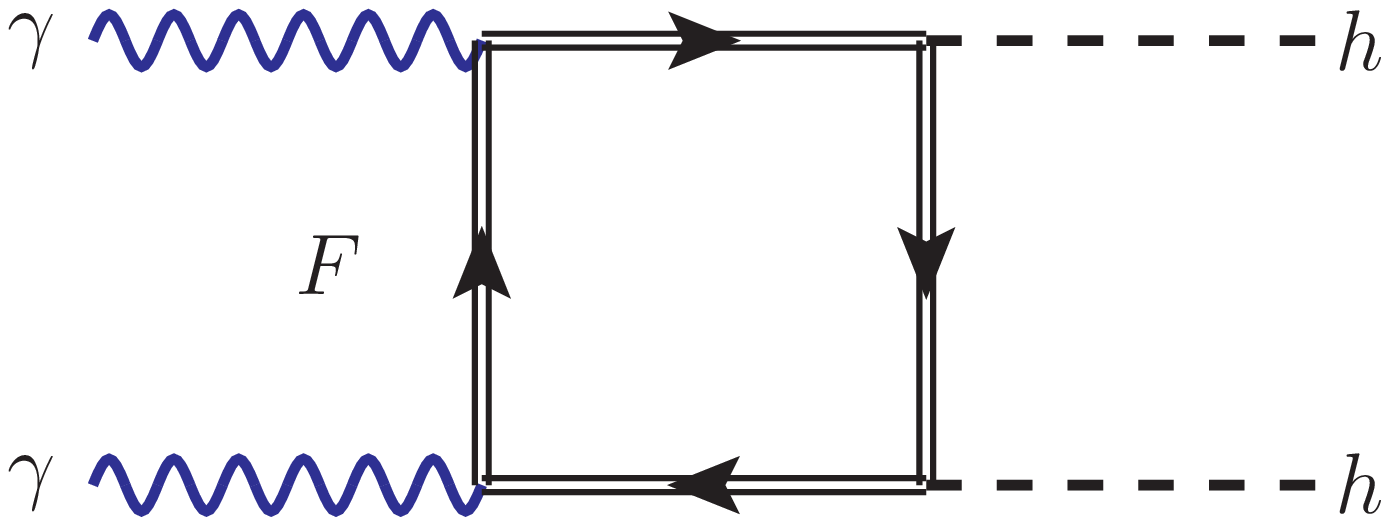,width=0.49\textwidth} \end{center} \end{minipage}    & $\mathcal{M}_{c_F}$ \\ \hline
\begin{minipage}{0.6\textwidth} \begin{center} 
\epsfig{file=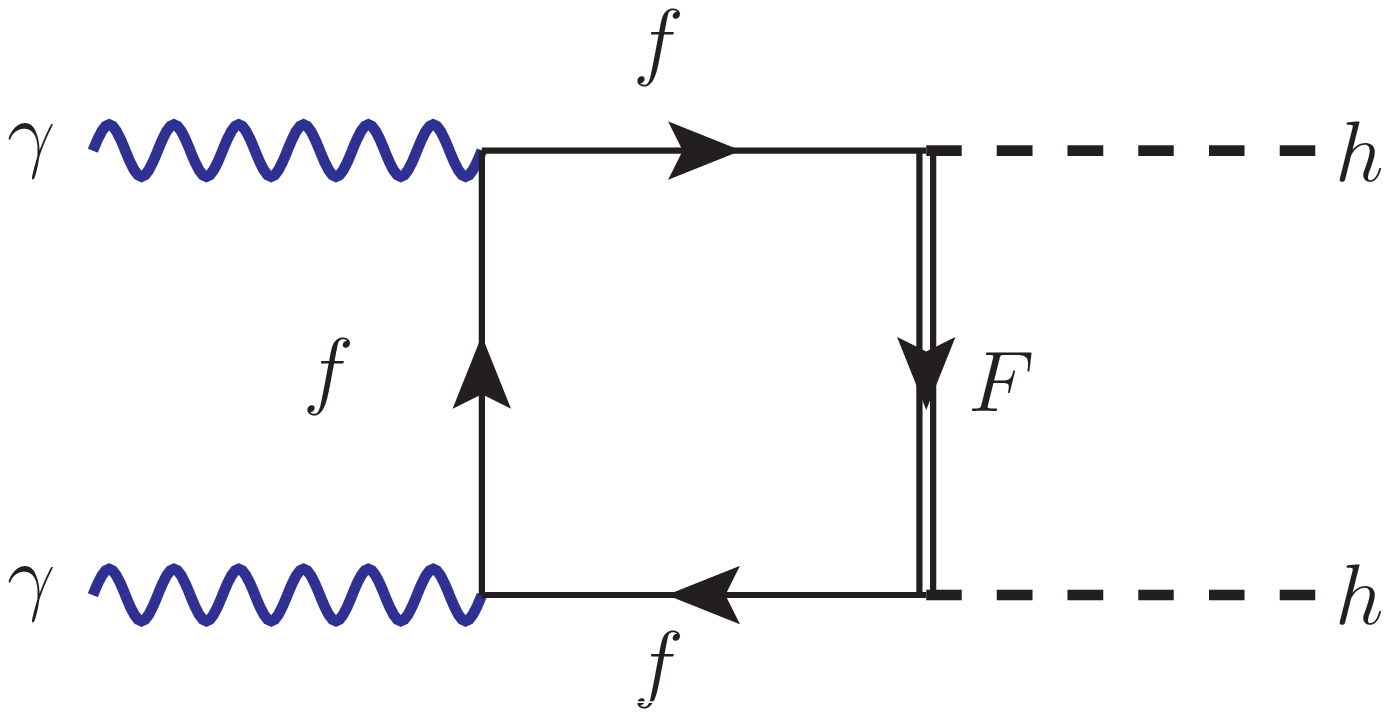,width=0.49\textwidth} 
\epsfig{file=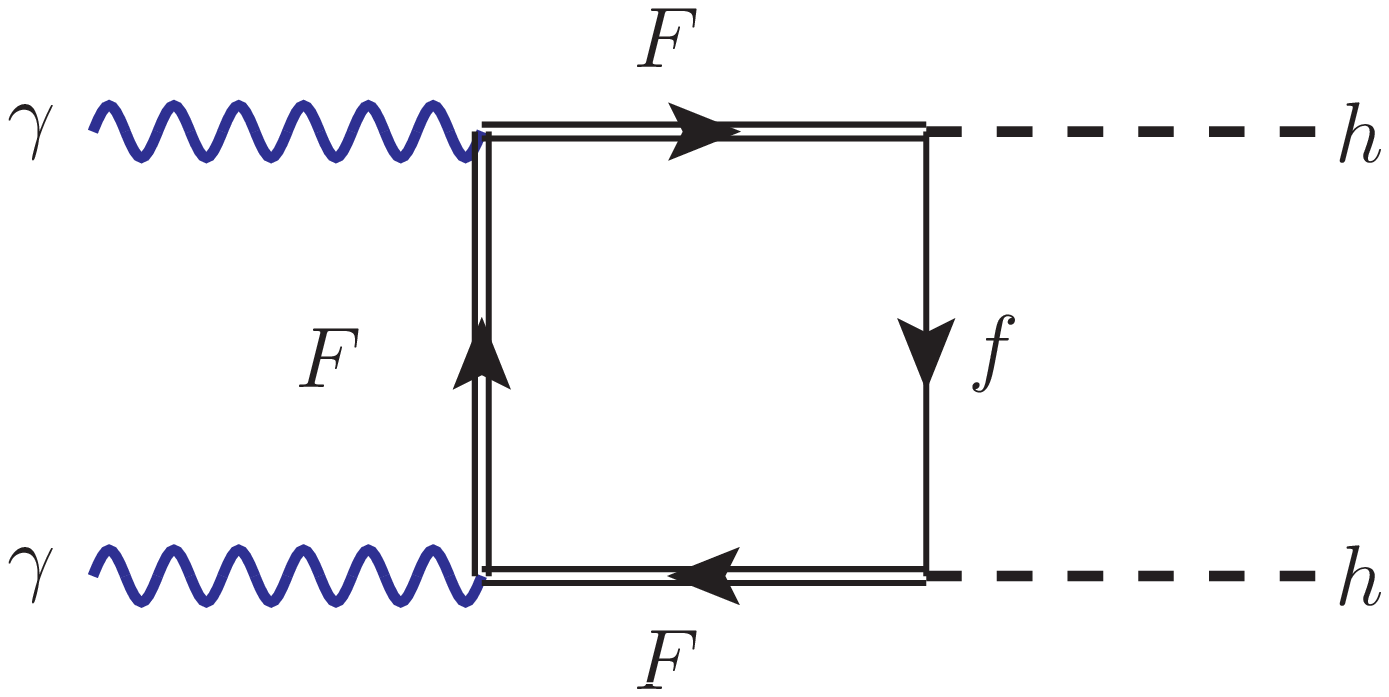,width=0.49\textwidth}\end{center} \end{minipage}& $\mathcal{M}_{c_{fF}}$ \\ \hline
\begin{minipage}{0.6\textwidth} \begin{center} 
\epsfig{file=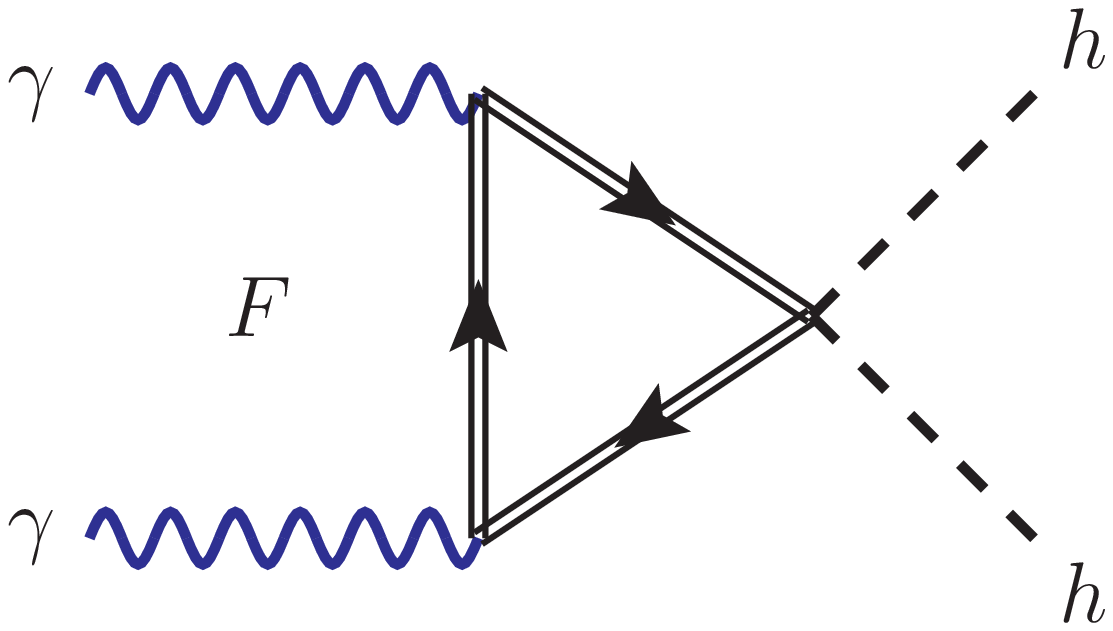,width=0.49\textwidth}   \end{center} \end{minipage} & $\mathcal{M}_{c_{2F}}$ \\ \hline
\end{tabular}    
\caption{Additional diagrams with top partners, with the corresponding partial amplitudes as defined in the text.} \label{tab:diagTP}
\end{center}
\end{table}

Top partners are a necessary ingredient in any CHM that implements partial compositeness~\cite{Kaplan:1991dc} for the top~\cite{Panico:2015jxa}. Typically, they are required to be light if responsible for generating the Higgs (misalignment) potential at loop level~\cite{Matsedonskyi:2012ym}. Furthermore, loops involving top partners may also help softening the constraints from electroweak precision on the misalignment angle~\cite{Grojean:2013qca,Ghosh:2015wiz}.

For the di-Higgs production we are focusing on, top partners may also be relevant as they enter the production mode at one loop level, as shown in Table~\ref{tab:diagTP}.
To study their effect, while avoiding the complications of plunging in specific models, here we construct an effective Lagrangian that only includes the most relevant components. In practice, we will introduce a singlet $S$ and a doublet $Q = ( U, D)^T$ vector-like composite fields, which then mix with the elementary top fields via the misalignment angle. The most general Lagrangian reads:
\begin{multline}
- \mathcal{L}_{TP} = M_Q \left( \bar{U}_L U_R + \bar{U}_R U_L + \bar{D}_L D_R + \bar{D}_R D_L \right) + M_S \left( \bar{S}_L S_R + \bar{S}_R S_L \right) +  \\
 y_L f \cos \theta\, \left( \bar{t}_L U_R + \bar{b}_L D_R\right) + y_R f \cos \theta\, \bar{S}_L t_R + \mbox{h.c.} \\
 - y'_L f \sin \theta\, \bar{t}_L S_R - y'_R f \sin \theta\, \bar{U}_L t_R + \mbox{h.c.}
 \label{eq:LTP}
\end{multline}
As the pNGB Higgs $h$ is associated to the misalignment angle, its couplings can be extracted by taking derivatives with respect to $\theta$, as follows:
\begin{equation}
\mathcal{L}_{h + h^2} = \frac{\partial \mathcal{L}_{TP}}{\partial \theta} \frac{h}{f} + \frac{1}{2} \frac{\partial^2 \mathcal{L}_{TP}}{\partial \theta^2} \frac{h^2}{f^2}\,,
\end{equation}
and subsequently rotating the fields to the mass eigenstate basis.
To simplify the analysis, in the following we will consider two separate cases, which include only the singlet or only the doublet respectively.
The effective Lagrangian for top-like state $T$ and a bottom-like one $B$ thus reads:
\begin{eqnarray}
\mathcal{L}_{\rm eff} &=& - m_t \left\{ \left( 1 + c_f  \frac{h}{v} + \frac{c_{2f}}{2}  \frac{h^2}{v^2} + \dots \right) \bar{t}_L t_R + \left( c_T \frac{h}{v} + \frac{ c_{2T}}{2} \frac{h^2}{v^2} + \dots \right) \bar{T}_L T_R + \right. \nonumber \\
& & \left.  \left( c_{tT} \frac{h}{v} + \dots \right) \bar{t}_L T_R + \left( c_{Tt} \frac{h}{v} + \dots \right) \bar{T}_L t_R \right\} + \mbox{h.c.} \nonumber \\
&& - m_t \left\{ \left( c_B \frac{h}{v} + \frac{c_{2B}}{2}  \frac{h^2}{v^2} + \dots \right) \bar{B}_L B_R + \right. \nonumber \\
& & \left.  \left( c_{bB} \frac{h}{v} + \dots \right) \bar{b}_L B_R + \left( c_{Bb} \frac{h}{v} + \dots \right) \bar{B}_L b_R \right\} + \mbox{h.c.} \label{eq:LeffTP}
\end{eqnarray}
where the second term, containing the bottom quark and a bottom partner, only appears for the doublet case, and we display the relevant couplings (the dots indicating higher orders in the Higgs field).

The top partners also affect the couplings of a single Higgs to gluons and photons, which have been measured accurately at the LHC. As it was already observed in the literature~\cite{Gillioz:2012se,Bizot:2015zaa}, however, there is a cancellation at work such that the reduction of the top couplings can be compensated by the coupling of top partners. In fact, the effect of the heavy top partners can be encoded in an effective top Yukawa modifier as follows:
\begin{eqnarray}
c_f^{\text{eff},gg} &=& c_f + c_T \frac{m_t}{m_T} + c_B \frac{m_t}{m_B}\,, \\
c_f^{\text{eff},\gamma\gamma} &=& c_f + c_T \frac{m_t}{m_T} + \frac{1}{4} c_B \frac{m_t}{m_B}\,.  
\end{eqnarray}
Thus, it suffices that the above combinations are close to unity to avoid strong constraints. On the other hand, a looser direct bound on $c_f$ is imposed by the measurement of the $t\bar{t} h$ cross section~\cite{Sirunyan:2020icl,Aaboud:2017rss}.

Top partner masses are also constrained by direct searches at the LHC, which look for direct decays into SM final states: $T \to W b, Z t, h t$ and $B \to W t, Z b, h b$. Current bounds lie in the $1.3 \div 1.5$~TeV range, depending on the relative branching ratios~\footnote{C.f. \href{https://twiki.cern.ch/twiki/bin/view/CMSPublic/PhysicsResultsB2G}{CMS B2G Working Group} and the \href{https://atlas.web.cern.ch/Atlas/GROUPS/PHYSICS/CombinedSummaryPlots/EXOTICS/}{ATLAS Exotics} results.}. Note that composite Higgs models typically contain additional pNGBs, possibly lighter than the top partners: they may provide additional decay modes which could, in some cases, reduce the constraints~\cite{Bizot:2018tds,Xie:2019gya,Benbrik:2019zdp,Cacciapaglia:2019zmj,Brooijmans:2020yij}. Nevertheless, it seems unlikely that top partners below $1$~TeV are allowed.
In the following, we will consider benchmark models with values of the top partner masses around $1.5$~TeV, and with a misalignment angle $\theta = 0.2$ (corresponding to $\xi \approx 0.04$). This choice guarantees that the benchmarks are not yet excluded and will be difficult to test even at the HL-LHC run~\cite{CidVidal:2018eel}.

\subsection{Singlet case \label{subsection:5_1}}

The Lagrangian in Eq.~\eqref{eq:LTP} can be reduced to
\begin{equation}
- \mathcal{L}_S = M_S\, \bar{S}_L S_R  + y_R\ f\ \cos \theta\, \bar{S}_L t_R - y'_L\ f\ \sin \theta\, \bar{t}_L S_R + \mbox{h.c.}
\end{equation}
so that only two couplings are relevant, $y_R$ and $y'_L$. 
It is convenient to first define an angle, $\alpha_R$, characterising the degree of compositeness of the right-handed top, as follows:
\begin{equation}
\sin \alpha_R = \frac{y_R f}{M}\,, \quad M = \sqrt{M_S^2 + y_R^2 f^2}\,.
\end{equation}
The remaining free parameter is $y'_L$, which we trade for the top mass:
\beq
m_t = \frac{y'_L f \ \sin \alpha_R\ \sin 2 \theta}{2 \sqrt{1- \sin^2 \alpha_R\ \sin^2 \theta}} + \dots
\label{eq:mtopSing}
\eeq
at leading order in $y'_L$.
This expression is close to the one obtained in the MCHM5 scenario in the previous section, so we will use the latter as a control to evaluate the net effect of the top partners.

In the mass eigenstate basis, the relevant couplings match the first term in Eq.~\eqref{eq:LeffTP}, with $T$ being the heavier vector-like state. In Table~\ref{table:TopPartners} we report the numerical values for 4 choices of $\alpha_R$. 
In the last column, we also report the values of the effective couplings $c_f^{\rm eff}$ entering the single Higgs couplings to gluons and photons: they are nearly independent of $\alpha_R$ and match the value in MCHM5. Thus, even if $c_f$ can be much smaller than 1, as for $\alpha_R = \pi/10$, the measurements of the Higgs couplings do not exclude this benchmark. Note also that a similar sum rule is absent for the quartic coupling, which is always substantially smaller than for MCHM5 case.

\begin{table}[tb]
	\begin{center}
{\footnotesize \begin{tabular}{|c|c|c|c|c|c|c|c|c|c|} \hline
			Benchmark 1 & \multicolumn{7}{c|}{$M = 1500$~GeV, $\theta = 0.2$, $m_{\rm top} = 173$~GeV } & \multicolumn{2}{c|}{$c_f^{\rm eff}$}\\ \hline
			& $M_T$ & $c_f$ & $c_{2f}$ & $c_{T}$ & $c_{2T}$ & $c_{tT}$ & $c_{Tt}$ & $c_f^{\text{eff},gg}$ & $c_f^{\text{eff},\gamma\gamma}$   \\ \hline
$\alpha_R = \pi/3$  & $1480$ & $0.965$ & $-0.0497$ & $-0.212$ & $-0.250$ & $-0.603$ & $-0.220$ & \multicolumn{2}{c|}{$0.940$} \\ \hline 
$\alpha_R = \pi/4$ & $1496$ & $0.945$ & $-0.0596$ & $-0.0427$ & $-0.167$ & $-1.034$ & $-0.291$ & \multicolumn{2}{c|}{$0.940$} \\ \hline  
$\alpha_R = \pi/6$ & $1525$ & $0.906$ & $-0.0685$ & $0.293$ & $-0.0923$ & $-1.77$ & $-0.347$ & \multicolumn{2}{c|}{$0.939$}\\ \hline  
$\alpha_R = \pi/10$ & $1608$ & $0.810$ & $-0.0709$ & $1.204$ & $-0.0749$ & $-3.123$ & $-0.430$ & \multicolumn{2}{c|}{$0.939$}\\ \hline
\multicolumn{10}{|c|}{} \\ \hline
			Benchmark 2 & \multicolumn{7}{c|}{$M = 1500$~GeV, $\theta = 0.2$, $m_{\rm top} = 173$~GeV } & \multicolumn{2}{c|}{$c_f^{\rm eff}$} \\ \hline
			& $M_F$ & $c_f$ & $c_{2f}$ & $c_{F}$ & $c_{2F}$ & $c_{fF}$ &  $c_{Ff}$  & $c_f^{\text{eff},gg}$ & $c_f^{\text{eff},\gamma\gamma}$    \\ \hline
$\alpha_L = \pi/3$  & T: $1481$ & $0.965$ & $-0.0497$ & $-0.212$ & $-0.250$ & $-0.220$ & $-0.603$ & $0.910$ & $0.933$\\ 
			      & B: $1478$ & $0$ & $0$ & $-0.255$ & $-0.250$ & $-0.150$ & $0$  &  &\\ \hline 
$\alpha_L = \pi/4$ & T: $1496$ & $0.945$ & $-0.0596$ & $-0.0427$ & $-0.167$ & $-0.291$ &  $-1.034$ & $0.920$ & $0.935$\\ 
			     & B: $1485$ & $0$ & $0$ & $-0.169$ & $-0.166$ & $-0.173$ & $0$ & & \\ \hline  
$\alpha_L = \pi/6$ & T: $1525$ & $0.906$ & $-0.0685$ & $0.293$ & $-0.0923$ & $-0.347$ & $-1.77$ & $0.929$ & $0.937$\\ 
			     & B: $1493$ & $0$ & $0$ & $-0.0843$ & $-0.0826$ & $-0.149$ & $0$  & & \\ \hline  
$\alpha_L = \pi/10$ & T: $1608$ & $0.810$ & $-0.0709$ & $1.204$ & $-0.0749$ & $-0.430$ & $-3.123$ & $0.936$ & $0.939$\\ 
			       & B: $1497$ & $0$ & $0$ & $-0.0321$ & $-0.0314$ & $-0.101$ & $0$ & & \\ \hline
\multicolumn{10}{|c|}{} \\ \hline
Control & \multicolumn{7}{c|}{MCHM5 with $\theta = 0.2$} & \multicolumn{2}{c|}{$c_f^{\rm eff}$} \\ \hline
 & - & $0.940$ & $-0.158$ & - & - & - & - & \multicolumn{2}{c|}{$0.940$}\\ \hline
 		\end{tabular}}
	\caption{Couplings of the Higgs $h$ to top (bottom) quarks and top partners relevant for our calculation. Benchmark 1 corresponds to the singlet, while benchmark 2 to the doublet (here, $F = T,B$ and $f = t,b$). The ``control'' corresponds to the model MCHM5 without light top partners.}
		\label{table:TopPartners}
	\end{center}
\end{table}

\begin{figure}[htb]
	\begin{center}
	\hspace*{-1cm}
		\epsfig{file=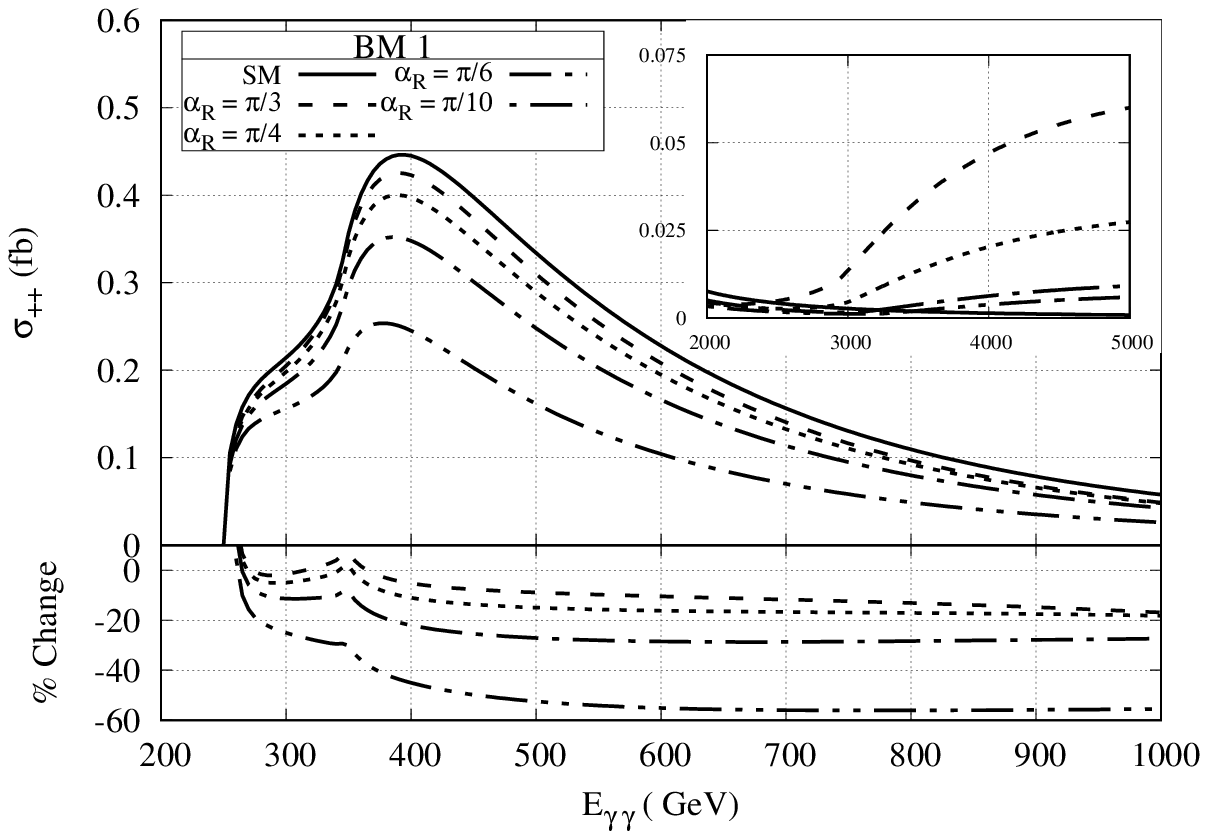,width=0.52\textwidth} 	
		\epsfig{file=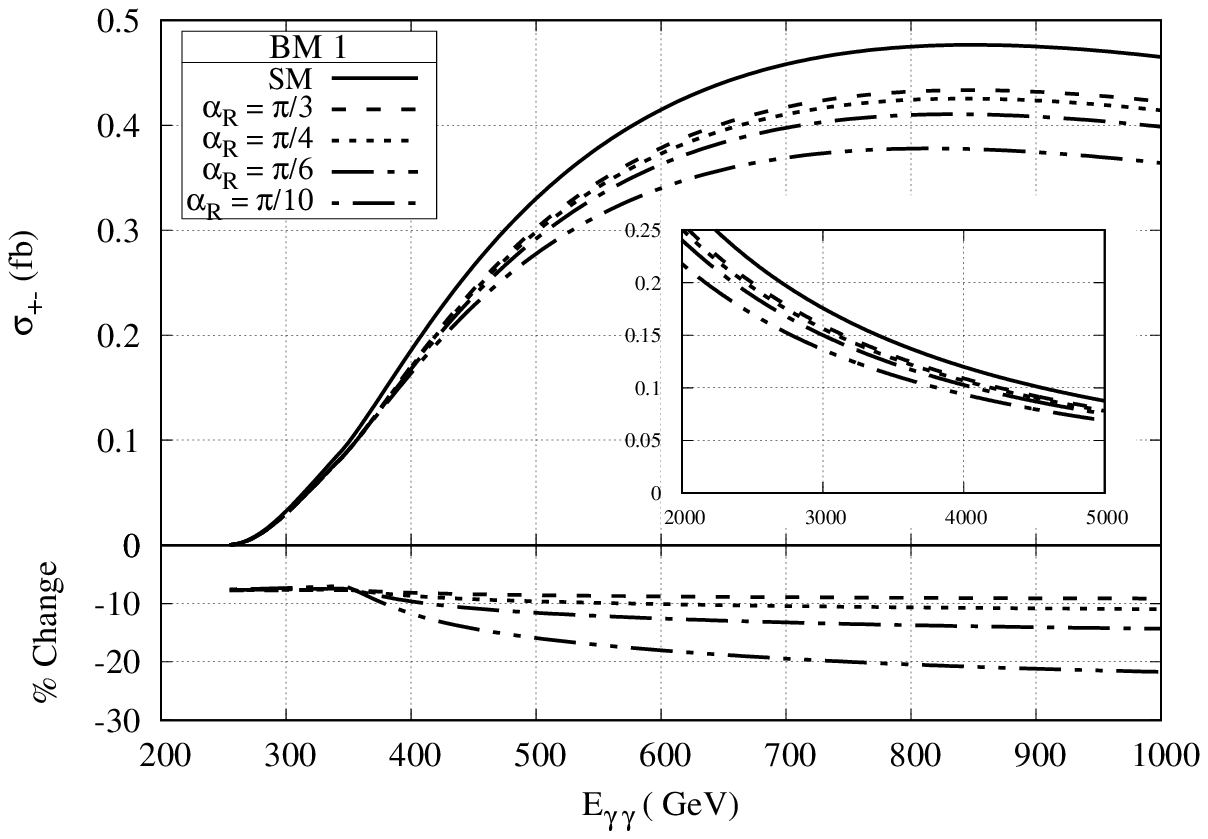,width=0.52\textwidth} \\
	\hspace*{-1cm}
		\epsfig{file=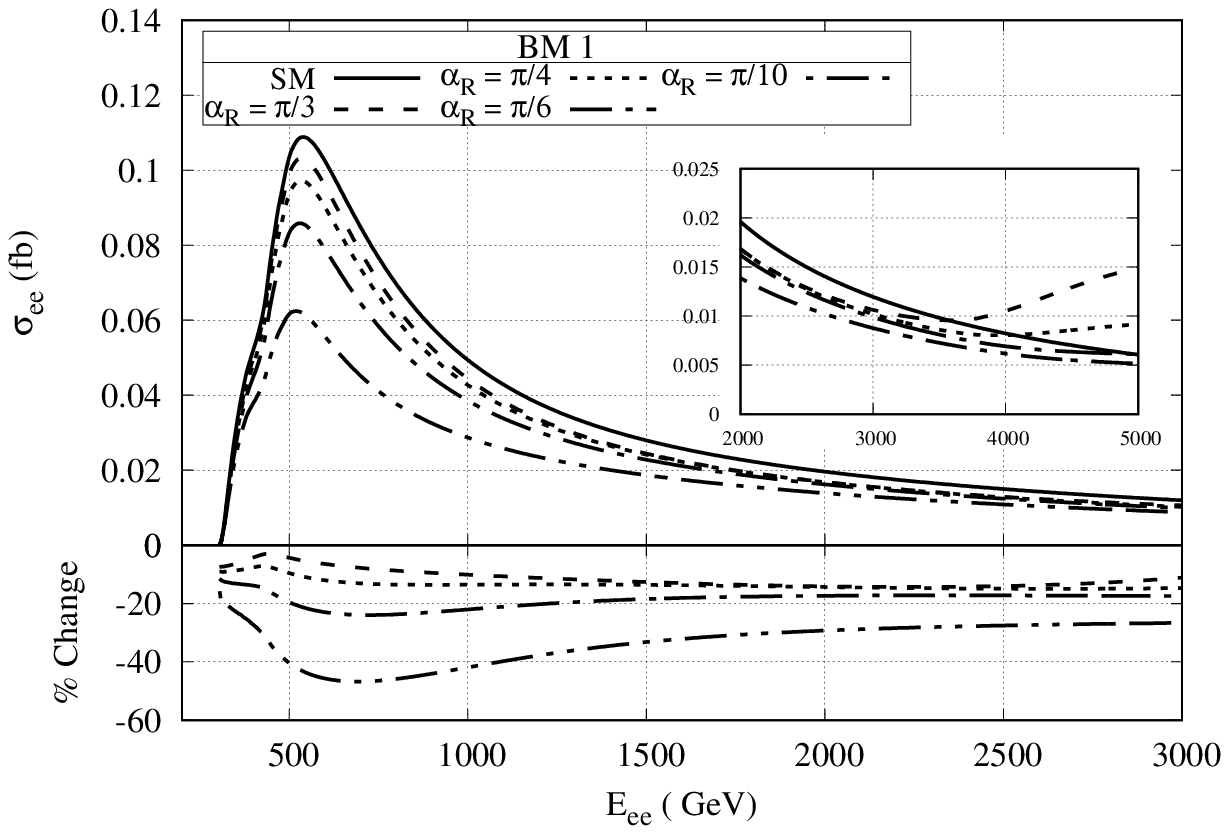,width=0.52\textwidth} 	
    \caption{$\gamma \gamma \to h h$ cross-section for same photon helicities (top left panel), opposite photon helicities (top right panel) and $e^+ e^- \to h h$ (bottom panel) in the model having an additional singlet top partner. The BM1 point parameters are given as Benchmark 1
    	 in Table \ref{table:TopPartners}.} 
    	\label{fig:sec5_1} 
	\end{center}
\end{figure}

\begin{figure}[htb]
	\begin{center}
		\epsfig{file=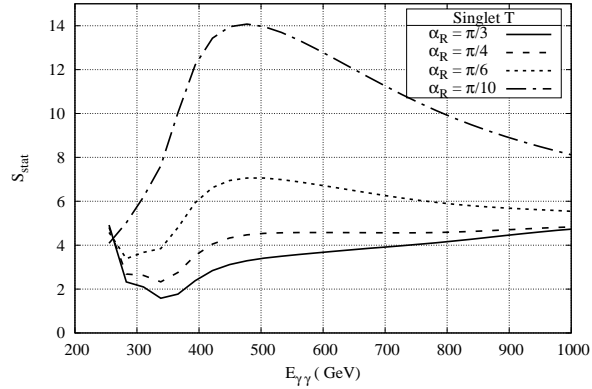,width=0.52\textwidth} 	
		\caption{Statistical Sensitivity as a function of $\gamma\gamma$ collider energy for the process $\gamma\gamma \to h h $ in the model with an additional singlet Top partner. The benchmark point parameters are given as Benchmark 1 in Table \ref{table:TopPartners}. }   	
		\label{fig:sec5_1a} 
	\end{center}
\end{figure}

\subsection{Doublet case \label{subsection:5_2}}

In this case, the Lagrangian in Eq.~\eqref{eq:LTP} is reduced to:
\begin{eqnarray}
- \mathcal{L}_{Q} = M_Q \left( \bar{U}_L U_R + \bar{D}_L D_R \right) + y_L f \cos \theta\, \left( \bar{t}_L U_R + \bar{b}_L D_R\right) - y'_R f \sin \theta\, \bar{U}_L t_R + \mbox{h.c.}
\end{eqnarray}
As for the singlet case, we can define the degree of compositeness of the left-handed top (and bottom) as
\begin{equation}
\sin \alpha_L = \frac{y_L f}{M}\,, \quad M = \sqrt{M_Q^2 + y_L^2 f^2}\,.
\end{equation}
The remaining free parameter is $y'_R$, which we trade for the top mass, given by the analog of Eq.~\eqref{eq:mtopSing} with $L \leftrightarrow R$.
Moreover, the spectrum now also contains heavy bottom partners, which couple to the Higgs via the couplings in the second term of Eq.~\eqref{eq:LeffTP}.

In Table~\ref{table:TopPartners} we report the numerical values for four benchmark values of $\alpha_L$: the couplings of the top partner $T$ are the same as in the singlet case (up to a reverse of the chirality for the mixed couplings $c_{tT}$ and $c_{Tt}$). The new ingredient is, therefore, the presence of the bottom partner. In the last column, we report the effective couplings $c_f^{\rm eff}$. The difference between gluon and photon couplings, as well as the discrepancy from the MCHM5 case, are due to loops of $B$.

\begin{figure}[htb]
	\begin{center}
	\hspace*{-1cm}
		\epsfig{file=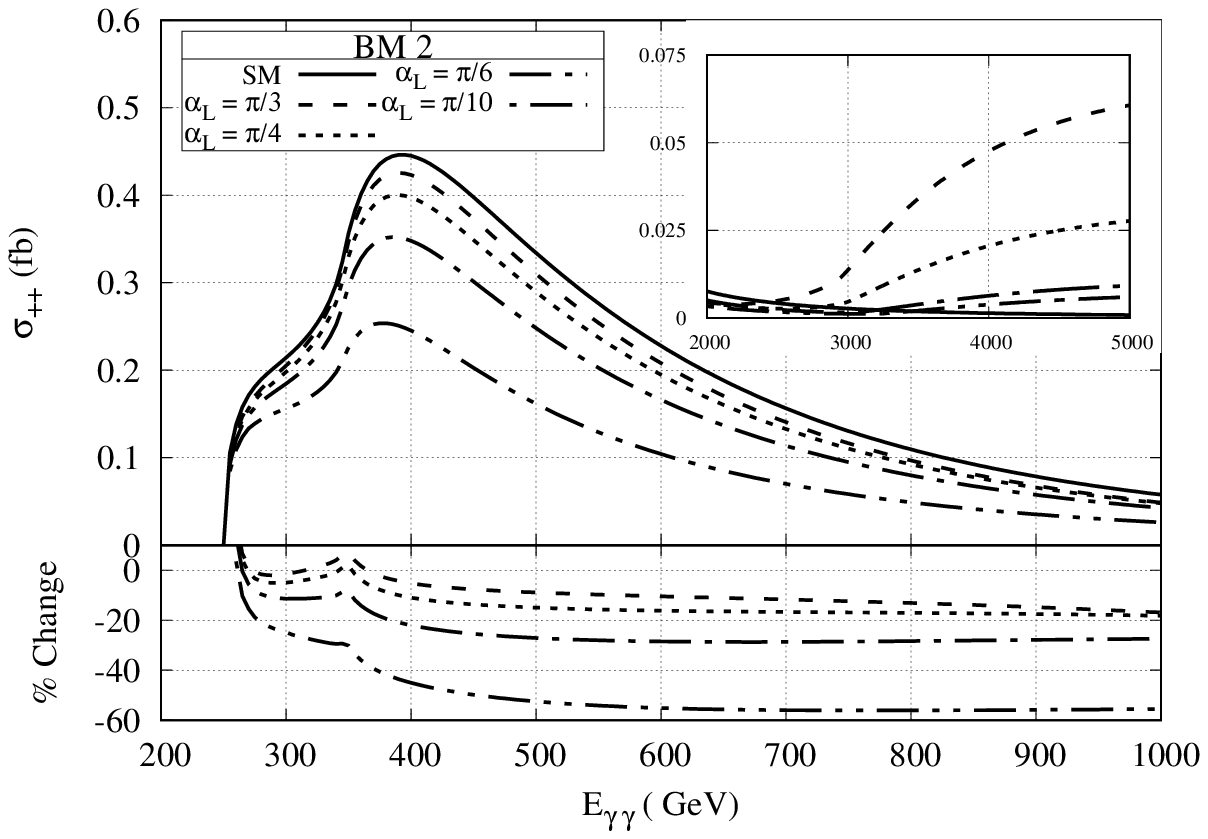,width=0.52\textwidth} 	
		\epsfig{file=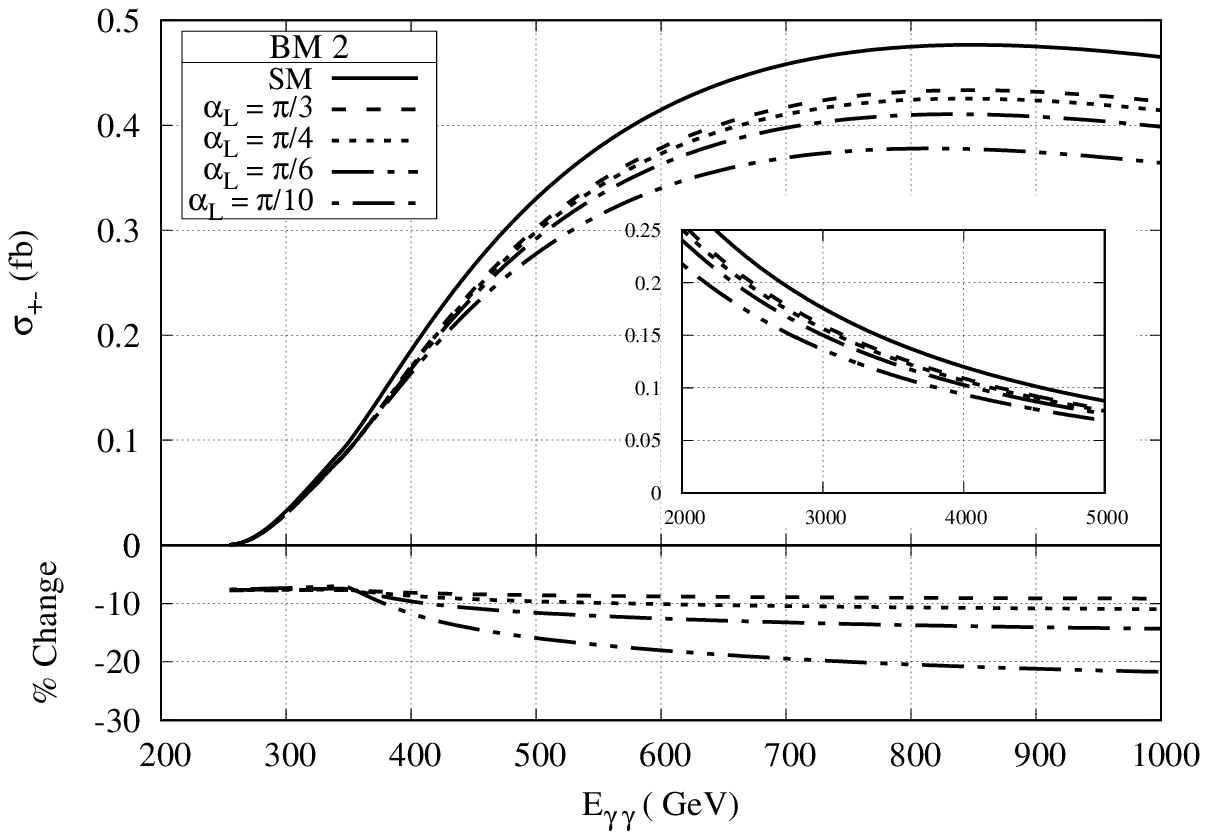,width=0.52\textwidth}  \\
	\hspace*{-1cm}
		\epsfig{file=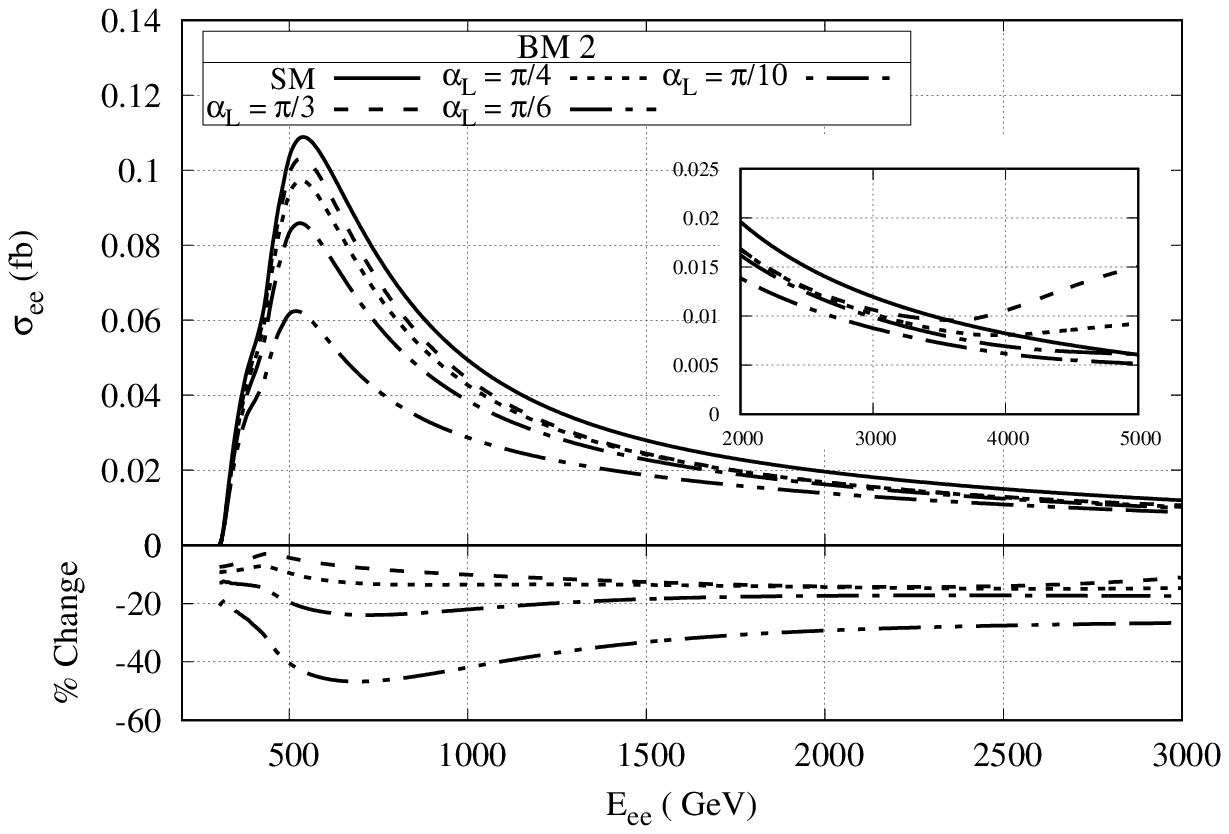,width=0.52\textwidth} 	
	\caption{$\gamma \gamma \to h h$ cross-section for same photon helicities (top left panel), opposite photon helicities (to right
	panel) and $e^+ e^- \to h h$ (bottom panel) model having additional doublet of top and bottom partners. The BM2 point parameters are given in Table \ref{table:TopPartners}.} \label{fig:sec5_2}   
	\end{center}
\end{figure}

\subsection{Numerical results}

In Fig. \ref{fig:sec5_1} (two top panels) we show the results of the helicity cross-sections for BM1 as given in Table \ref{table:TopPartners} for the model where a heavy singlet vector-like top partner is introduced (c.f., section \ref{subsection:5_1}). 
The results are shown for four indicative values of $\alpha_R$ (measure of degree of the compositeness of the right-handed top). 
In all these results there is a universal pattern of reduction in the cross-section as compared to SM values. The reduction 
of the cross-section for 
opposite photon helicities is a universal feature observed in all the earlier models that we have discussed. As explained earlier, 
the reason for this is that this process goes through box diagrams only, which do not receive contributions from additional new 
couplings like the quartic couplings of the Higgs with tops. The other 
couplings tend to be reduced in magnitude as compared to the corresponding SM values, thus resulting in smaller value of the cross-sections. In the case 
of same-helicity photons, as can be seen from BM1 in Table \ref{table:TopPartners}, 
the value of the quartic coupling of tops with the Higgs ($c_{2f}$) is very small and the other couplings ($c_f, c_v, c_{2v}$) 
tend to be reduced in magnitude as compared to the SM values, resulting again in the lower value of cross-section at small center-of-mass
energy. In the inset plots of Fig. \ref{fig:sec5_1}, we show the cross-sections for larger values of the energy (above top partner mass, $M_T$). The substantial increase in the cross-section is the due to the large value of the quartic coupling $c_{2T}$. 
This behavior can also be seen from the Appendix plot Fig. \ref{fig:appendix_3}, where we display individual contributions as given in Tables \ref{tab:diagSM}, \ref{tab:diagCH} and \ref{tab:diagTP}.  

In Fig.\ref{fig:sec5_2} we present the results of the cross-sections for a doublet of top and bottom partners. This model was discussed in Section \ref{subsection:5_2} and the benchmark point BM2 considered 
in our numerical analysis is given in Table \ref{table:TopPartners}. The individual contributions (as given in Tables \ref{tab:diagSM}, 
\ref{tab:diagCH} and 
\ref{tab:diagTP}) are shown in the Appendix, Fig. \ref{fig:appendix_4}. In this case, there are two vector-like quarks, namely a vector-like top ($T$) 
and a vector-like bottom ($B$), in the particle spectrum. 
The numerical results are similar to those of a singlet, proving that the effect of the bottom partner is negligible.

The results of the statistical significance in case where an additional singlet Vector Like Top quark is present are shown in Fig. \ref{fig:sec5_1a}.
The results are similar in the case of the presence of an additional vector like quark doublet.



\section{Conclusion and Outlook \label{section:conclusion} } 	

In this paper we have discussed the Higgs pair production in photon collisions in composite Higgs models, considering first minimal scenarios and then the effect of an extended Higgs sector and vector-like fermion multiplets, typical of these models. Such realizations can be 
studied at electron-positron colliders, namely FCC-ee, CEPC and ILC, which can probe regions up to the TeV center-of-mass energy. We focused on composite models as they can provide novel Higgs pair production mechanisms and interference effects. In particular, photon collisions are sensitive to all modified Higgs couplings and effects stemming from the new quartic Higgs-fermion vertices. 
The coupling responsible for these vertices, absent in the SM and arising from the non-linear nature of the composite Higgs, is the primary cause of enhancements in the photon Higgs pair production.
On the other hand, modifications to the SM-like couplings are universal, i.e.~they do not depend much on the specific model, and only result in mild reductions in the cross-sections. This effect is clearly demonstrated by comparing the results for the case of same-helicity and opposite-helicity photons in the $\gamma\gamma\to hh$ cross section. For the former case, when considering the MCHM5 benchmark with a large value of the Higgs-top quartic coupling, large enhancements of more than 100\% are seen, whereas for the latter this quartic Higgs-fermion coupling does not play a role and the cross section is diminished. Note that this translates to a 20-30\% enhancement of the overall $e^+e^-\to hh$ cross section for the MCHM5 benchmark point.

The presence of an additional scalar resonance $H$ opens up a new s-channel diagram, again only affecting the same-sign helicity photon cross sections. 
This results in large enhancements at the relatively lower diphoton energy and for lower value of $m_H$ and depending on the width $\Gamma_H$.
	This enhancement can be substantially high near the resonant (mass of heavy Higgs) value of diphoton energy. Similar enhancements 
	can be observed in the $e^+e^-\to hh$. Amongst all the models considered in our study this model is the most promising as far as the enhancement 
	from the SM results are concerned.
On adding fermion partners, the results for the cross sections are in general below the SM result for all benchmarks considered. The only exception, where a substantial enhancement can be observed, is when a very large center-of-mass energy is considered, beyond the mass of the top partners. 

The photon-initiated Higgs pair production process, therefore, is a key element for the discovery of deviations from the SM. This would be a strong indication of the composite structure underlying the Higgs sector. 
Our work shows that composite models can be revealed at future lepton colliders, even if the energy is not sufficient to produce new resonances. 
An enhancement or reduction in the total cross section can also be accompanied by different kinematics of the final Higgs pairs, which we leave for a future detailed study.
The detection of Higgs pair production via photon collisions can give precious preliminary indications on the presence of composite resonances, which could be produced and discovered at high-energy hadron colliders, which are planned to follow the electron-positron ones.

\acknowledgments
AB, GC, AD, NG, FM and KS would like to thank CEFIPRA for the financial support on the project 
entitled  "Composite Models at the Interface of Theory and Phenomenology" (Project No. 5904-C). 
The work of NG was also supported in part by CNRS LIA THEP and INFRE-HEPNET of CEFIPRA/IFCPAR. 
GC, AD, NG, DH and KS would like to express a special thanks to the Mainz 
Institute for Theoretical Physics (MITP) of the Cluster of Excellence PRISMA+ 
(Project ID 39083149) for its hospitality and support during part of this work.
This work is also supported in part by the TYL-FJPPL program. The work of
DH is partly supported by KEK and KAKENHI grant number JP15K05066. 

\appendix
\section{Appendix \label{section:appendix} } 	

For completeness we list in the following the relevant helicity amplitudes of the top quark loop with the Higgs and the box contributions. The helicity amplitudes of top quark triangle loop with the Higgs self-coupling is:
\begin{equation}
{\cal M}_{c_{f}} = c_{f} \ c_{3h} \ \left(\frac{y_{t}v}{\sqrt{2}}\right) 
    \ \frac{4m_{h}^{2}m_{t}}{m_{W}^{2}(s-m_{h}^{2})} \ f_{t}(s) \,,
\end{equation}
where the Higgs self-coupling in composite models is given by $\lambda_{hhh} = c_{3h} \ 
\lambda_{hhh}^{SM} = c_{3h} \ 3m_{h}^{2}/v$.
The top quark triangle diagram with $t\bar{t}hh$ coupling is given by
\begin{eqnarray}
{\cal M}_{c_{2f}} = c_{2f} \ \left(\frac{y_{t}v}{\sqrt{2}}\right) \ \frac{4m_{t}}{3m_{W}^{2}} \ f_{t}(s) \,,
\end{eqnarray}
where the triangle loop function of top quark $f_{t}(s)$ is written as
\begin{equation}
f_{t}(s) = \left[ 2 - (s-4m_{t}^{2}) \ C_{0}(s;m_{t}) \right] \,.
\end{equation}
The Passarino-Veltman three-point loop function $C_{0}$ \cite{Passarino:1978jh} is abbreviated as
\begin{equation}
C_{0}(s;m) = C_{0}(p_{1},p_{2};m,m,m) \,,
\end{equation}
where $m$ is the mass of the internal loop particle.
\begin{figure}[htb]
    \begin{center}
    \hspace*{-1cm}
        \epsfig{file=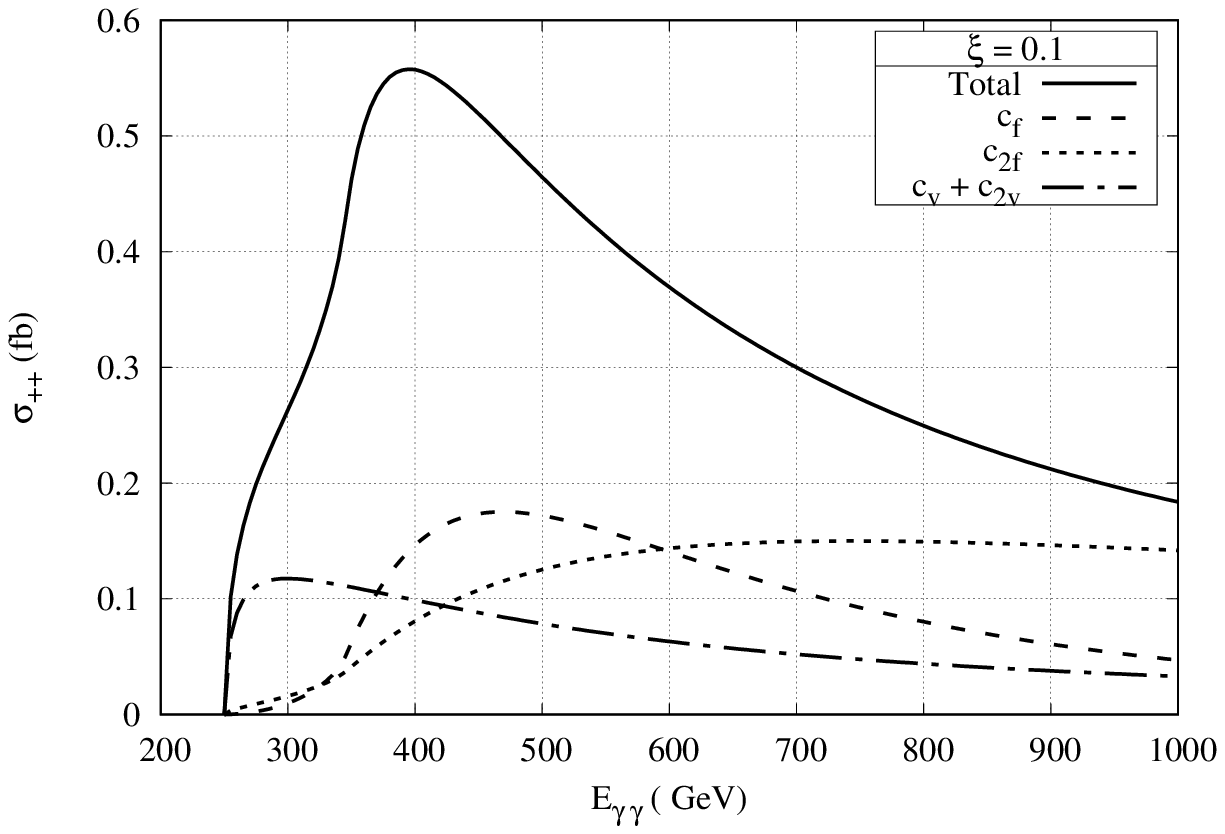,width=0.52\textwidth} 
        \epsfig{file=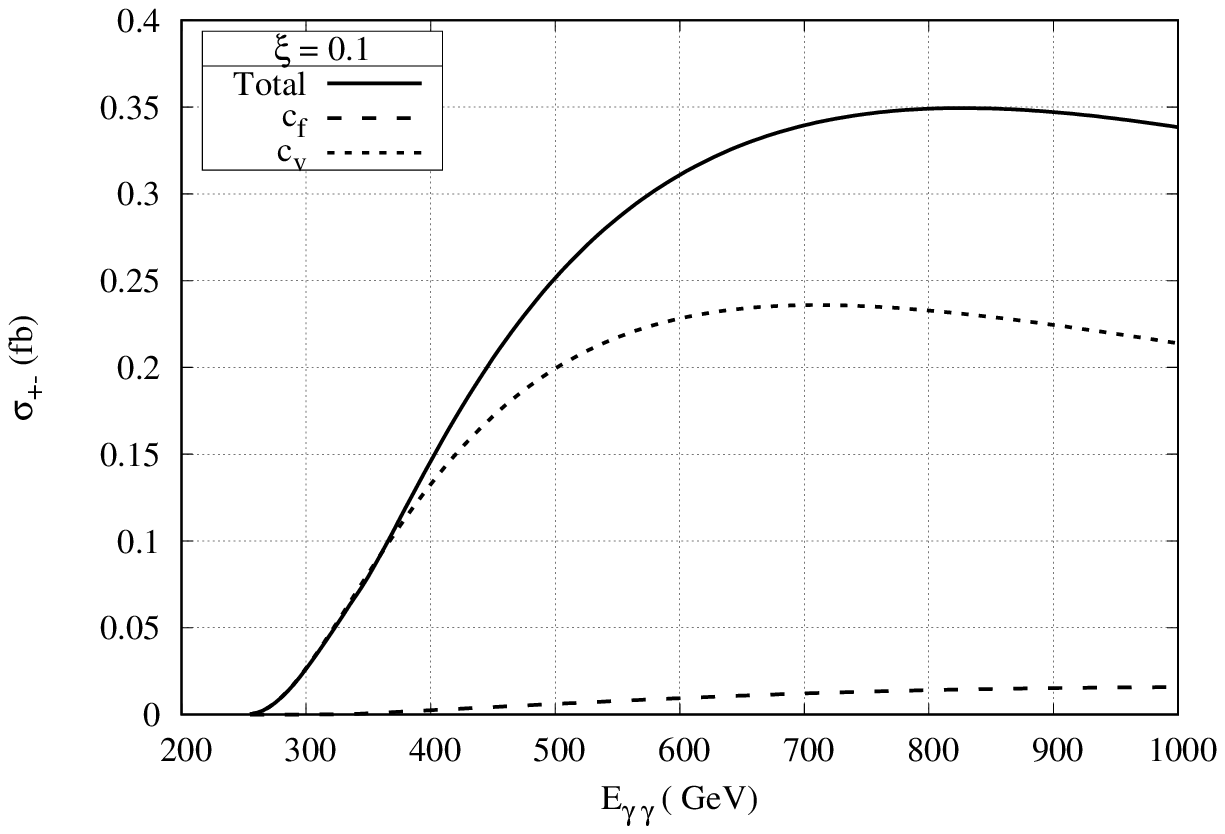,width=0.52\textwidth}  \\
        \epsfig{file=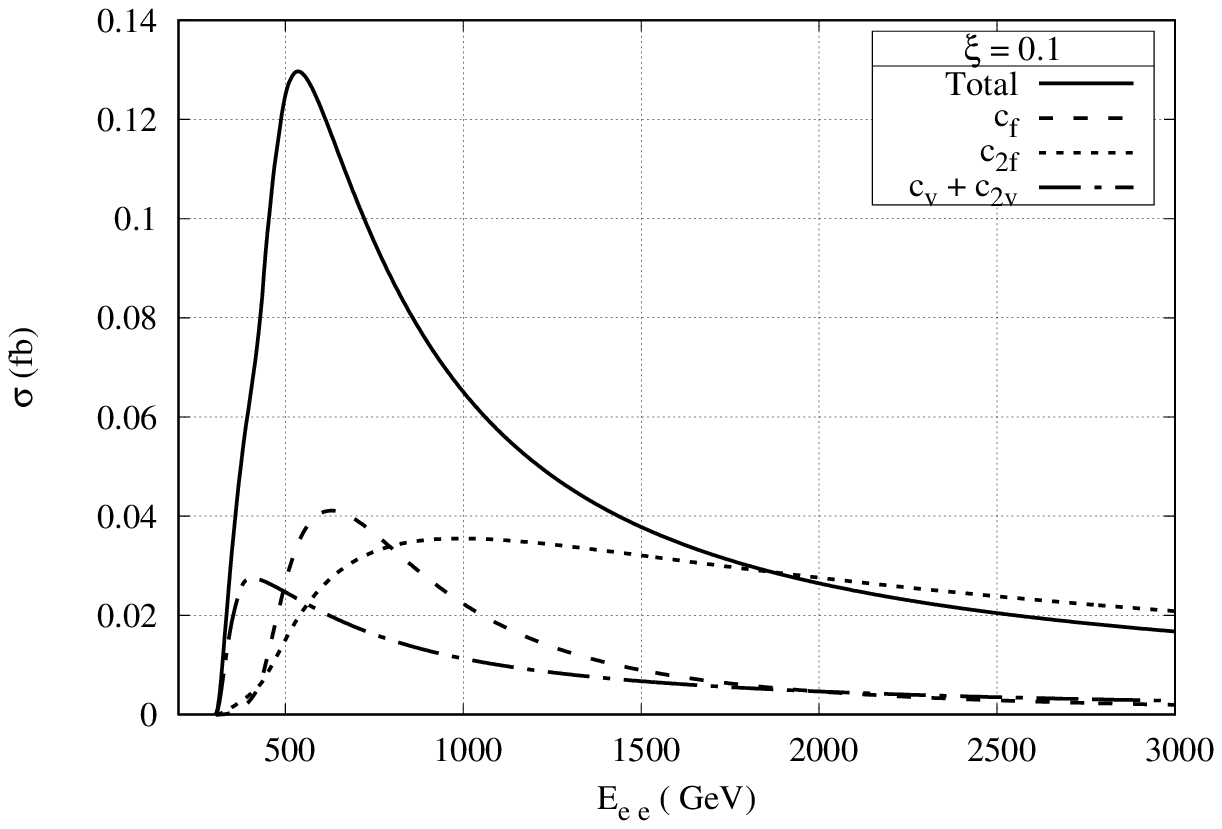,width=0.52\textwidth}
    \caption{MCHM5: Cross-sections for the individual contributions as given in Tables \ref{tab:diagSM} and \ref{tab:diagCH} 
    for $\gamma \gamma \to h h$ same photon helicities (top left panel), opposite photon helicities (top right panel) and
    $e^+ e^- \to hh$ (bottom panel).} 
        \label{fig:appendix_1} 
    \end{center}
\end{figure}
The box diagrams of top quark are written as
\begin{equation}
{\cal M}_{box}^{f}(+,+) = c_{f}^{2} \ {\cal M}_{box}^{SM,f}(+,+) \,, \qquad
{\cal M}_{box}^{f}(+,-) = c_{f}^{2} \ {\cal M}_{box}^{SM,f}(+,-) \,,
\end{equation}
where ${\cal M}_{box}^{SM,f}(\lambda_1,\lambda_2)$ are the top quark box diagrams in SM.
The $W$ boson triangle loop diagrams with $WWh$ and $WWhh$ coupling are given by
\begin{eqnarray}
{\cal M}_{c_v}&=& c_v \ c_{3h} \ \frac{3m_{h}^{2}}{2m_{W}^{2}(s-m_{h}^{2})} \ f_{W}(s) \,,\\
{\cal M}_{c_{2v}} &=& c_{2v} \ \frac{1}{2m_{W}^{2}} \ f_{W}(s) \,,
\end{eqnarray}
with the $W$ boson loop function
\begin{equation}
f_{W}(s) = \left[ 8m_{W}^{2} s \ C_{0}(s;m_{W}) - (6m_{W}^{2} + m_{h}^{2}) \ 
             \left(1+2m_{W}^{2} \ C_{0}(s;m_W) \right) \right]  \,.
\end{equation}

The $W$ boson box diagrams are:
\begin{equation}
{\cal M}^{W}_{box}(+,+) = c_{v}^{2} \ {\cal M}^{SM,W}_{box}(+,+) \,, \qquad
{\cal M}^{W}_{box}(+,-) = c_{v}^{2} \ {\cal M}^{SM,W}_{box}(+,-) \,,
\end{equation}
where ${\cal M}^{SM,W}_{box}(\lambda_1,\lambda_2)$ are the box diagrams of $W$ boson loop in the SM.
The helicity amplitudes of the top quark and $W$ boson box diagrams in the SM are given in Ref.~\cite{Jikia:1992mt}.

The results of individual and total contribution of the helicity cross-sections (top row) and electron-positron 
cross-sections (bottom row) for a particular value of MCHM5 parameter ($\xi = 0.1$) are given in 
Fig.\ref{fig:appendix_1}. As can be seen the new $c_{2f}$ coupling can give substantial contribution to the cross-sections.

\vspace*{0.3cm}
\noindent\underline{\bf Heavy Scalar $H$ : } 
\vspace*{0.3cm}

\noindent The triangle top quark loop with the heavy scalar $H$ can be written as
\begin{eqnarray}
{\cal M}_{c_f^H} = -c_{t}^{H} \ c_{Hhh} \ \left(\frac{y_{t}v}{\sqrt{2}}\right) \        
         \frac{4m_{t}(s-2m_{h}^{2})}{3m_{W}^{2}(s-m_{H}^{2}+im_{H}\Gamma_{H})} \ f_{t}(s) \,,
\end{eqnarray}
where $m_H$ and $\Gamma_H$ are the mass and width of Heavy scalar. The $W$ boson triangle diagrams with $H$ are 
given by
\begin{eqnarray}
{\cal M}_{c_V^H} &=& -c_{V}^{H} \ c_{Hhh} \ \frac{s-2m_{h}^{2}}{2m_{W}^{2}(s-m_{H}^{2}+im_{H}\Gamma_{H})} \ f_{W}(s) \,
\end{eqnarray}

\begin{figure}[htb]
    \begin{center}
    \hspace*{-1cm}
        \epsfig{file=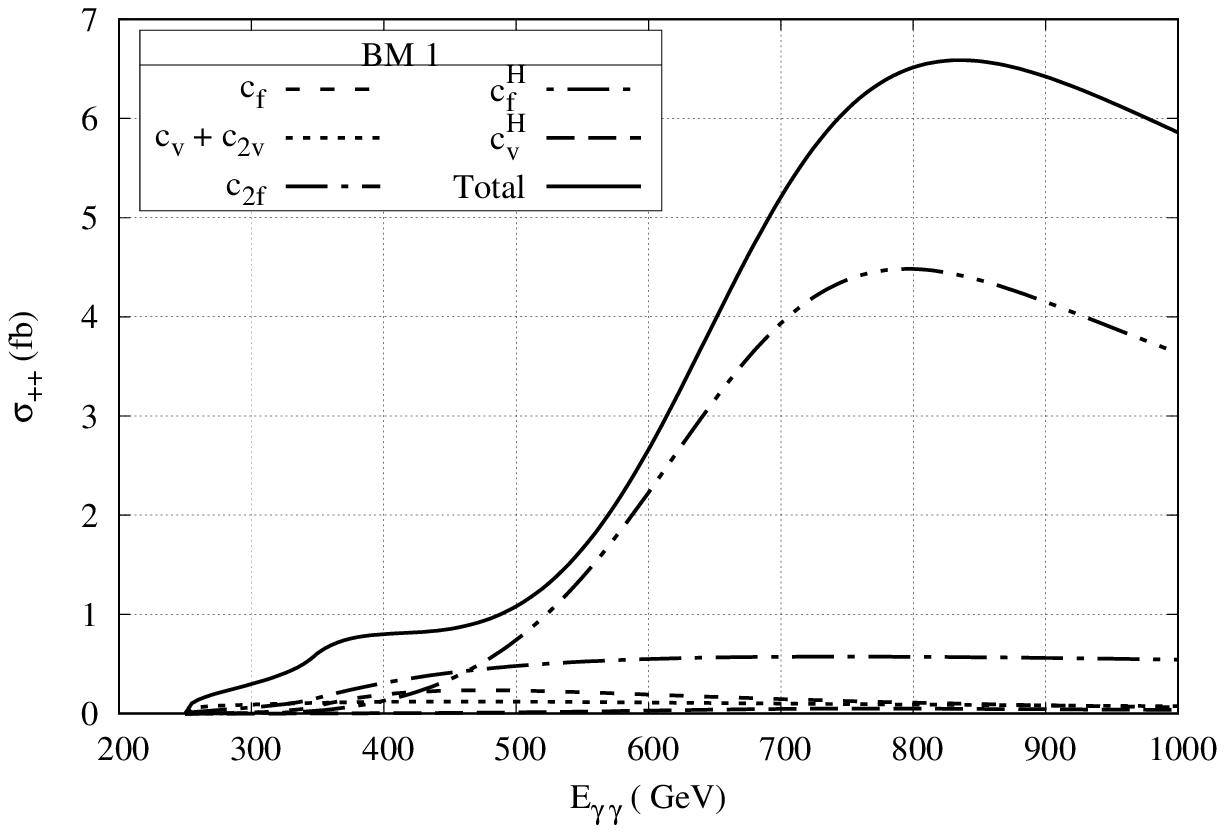,width=0.52\textwidth} 
        \epsfig{file=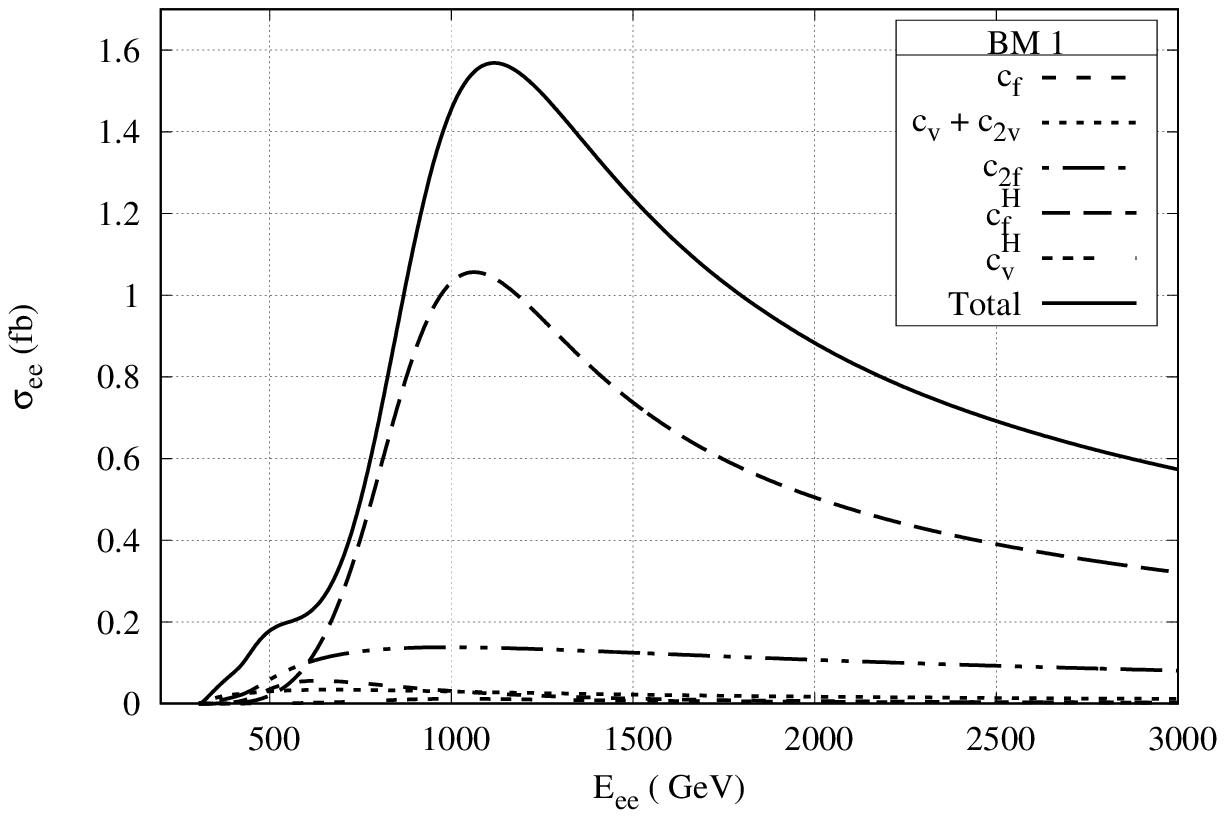,width=0.52\textwidth}
    \caption{Cross-sections for the individual contributions as given in Tables \ref{tab:diagSM} and \ref{tab:diagCH} 
    for $\gamma \gamma \to h h$ same photon helicities (left panel) and
    $e^+ e^- \to hh$ (right panel) in the presence of heavy scalar (as given in Section \ref{section:4}). }  
    \label{fig:appendix_2} 
    \end{center}
\end{figure}

The results of individual contributions in the model where heavy scalar ($H$) is introduced in the spectrum 
(Section \ref{section:4}) are shown in Fig.\ref{fig:appendix_2}. The plots shown are for the benchmark point 1 
of Table \ref{table:BM12}. It is to be noted that the new quartic Higgs-fermionic coupling is substantial in 
this model resulting in a larger value of $c_{2f}$ contribution thereby resulting in substantial deviations from SM 
predictions. 

\begin{figure}[htb]
    \begin{center}
    \hspace*{-1cm}
        \epsfig{file=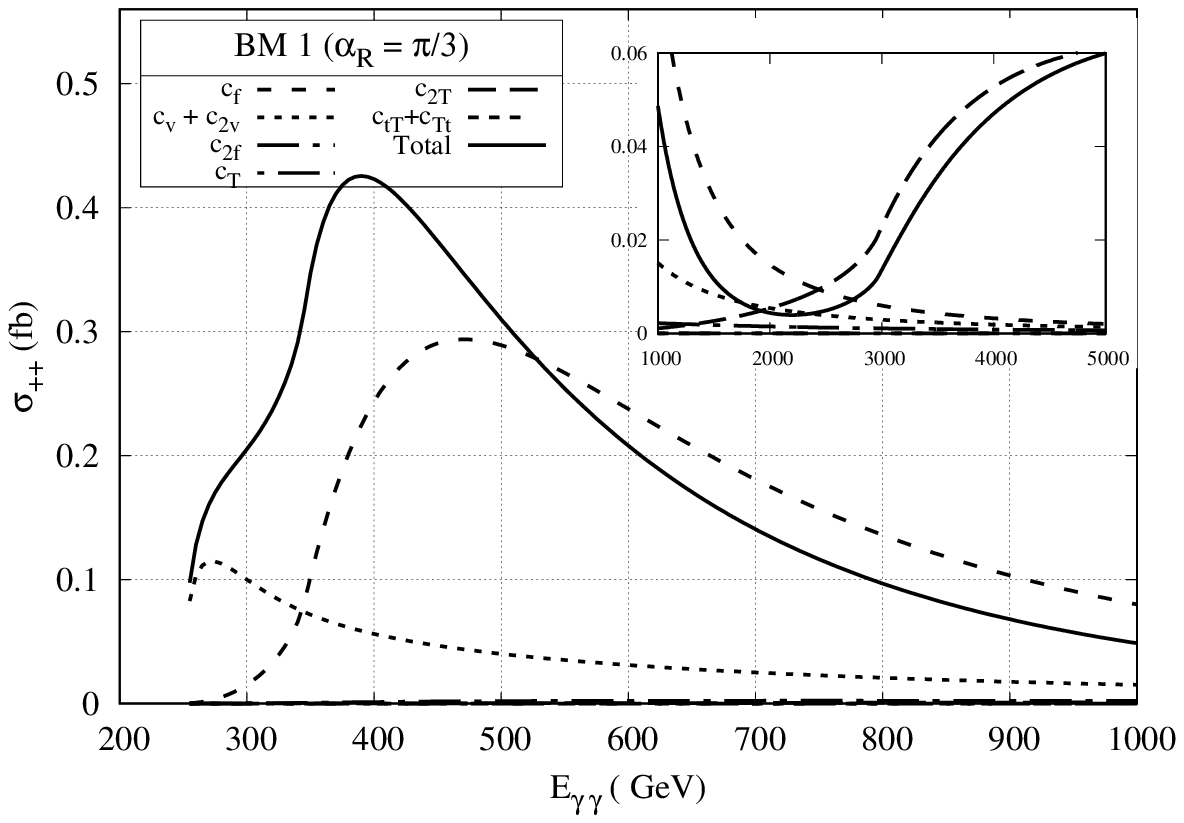,width=0.52\textwidth} 
        \epsfig{file=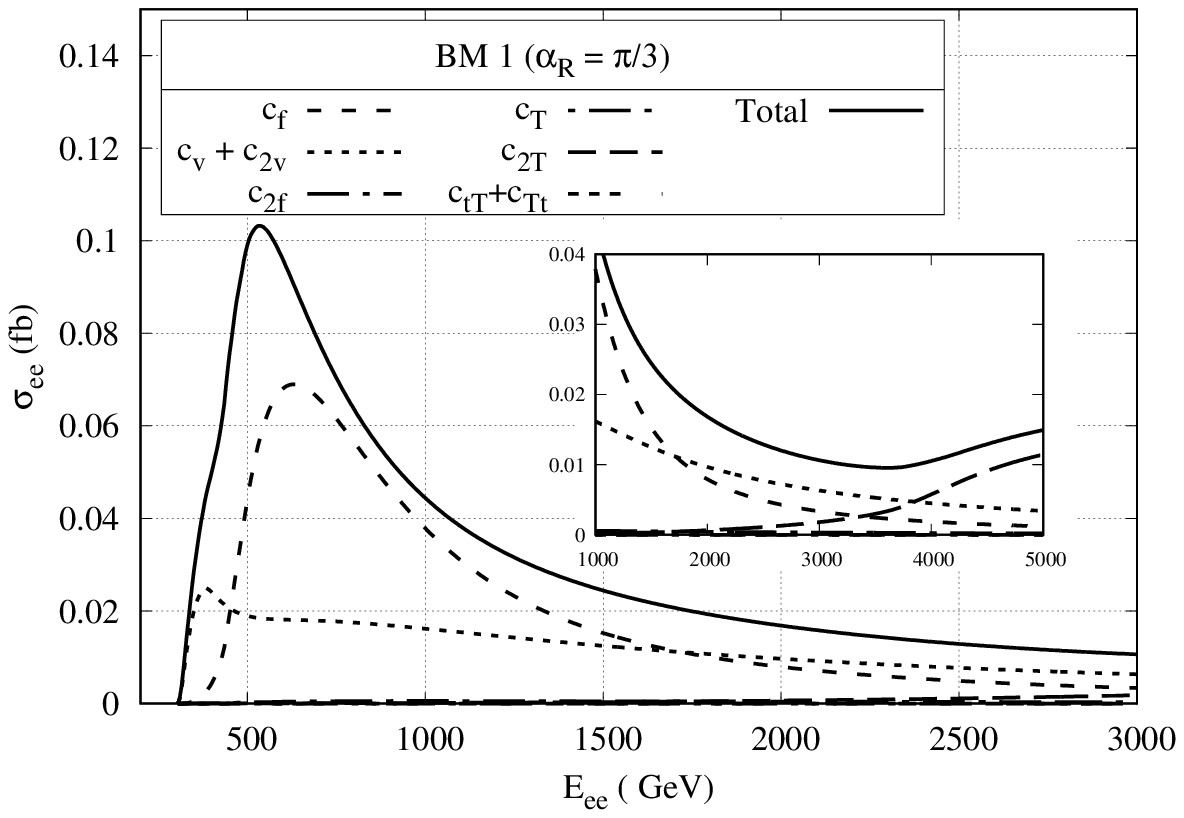,width=0.52\textwidth}
    \caption{Cross-sections for the individual contributions as given in Tables \ref{tab:diagSM} and \ref{tab:diagTP} 
    for $\gamma \gamma \to h h$ same photon helicities (left panel) and
    $e^+ e^- \to hh$ (right panel) in the presence of a Singlet Vector Like Top-quark (as given in Section \ref{subsection:5_1}). }  
    \label{fig:appendix_3} 
    \end{center}
\end{figure}

\vspace*{0.3cm}
\noindent\underline{\bf Vector Like Quark :}
\vspace*{0.3cm}

In the notation below we have labelled $f = t,b$ SM top and bottom quarks and $F = T, B$ Vector Like Top and Bottom quarks. 
The helicity amplitudes given below have same expressions for $(t,T)$ and $(b,B)$ the difference between them arise due to the 
Electromagnetic Charge. In below expressions we have used a parameter $q$ which has values $q = 1$ for $(t,T)$ and $q = 1/4$ 
for $(b,B)$. 

The $F$ quark loop can be written as: 
\begin{equation} 
{\cal M}_{c_{T}} = q^2 \ c_{T} \ c_{3h} \ \left(\frac{y_{t}v}{\sqrt{2}}\right) \ 
   \frac{4m_{h}^{2}m_{T}}{m_{W}^{2}(s-m_{h}^{2})} f_{T}(s) \,,
\end{equation} 
The $T$ quark triangle diagram with $T\bar{T}hh$ coupling:
\begin{equation}
{\cal M}_{c_{2T}} = q^2 \ c_{2T} \ \left(\frac{y_{t}v}{\sqrt{2}}\right) \ \frac{4m_{T}}{3m_{W}^{2}} \ f_{T}(s) \,,
\end{equation}
with
\begin{equation}
f_{T}(s) = \left[ 2 - (s-4m_{T}^{2}) \ C_{0}(s;m_{T}) \right] \,.
\end{equation}
The box diagrams of $T$ quark are:
\begin{eqnarray}
{\cal M}_{box}^{T}(+,+) &=& q^2 \ c_{T}^{2} \ \left(\frac{m_{t}}{m_{T}}\right)^{2} \ {\cal M}_{box}^{SM,T}(+,+), \\ 
{\cal M}_{box}^{T}(+,-) &=& q^2 \ c_{T}^{2} \ \left(\frac{m_{t}}{m_{T}}\right)^{2} \ {\cal M}_{box}^{SM,T}(+,-) \,,
\end{eqnarray}

The helicity amplitudes of $t-T$ box contributions are:
\begin{eqnarray}
{\cal M}_{box}^{t-T}(+,+) &=&
-\frac{8}{3m_{W}^{2}} \ q^2 \ (c_{tT}^{2} + c_{Tt}^{2}) \ 
\Biggl\{ -4B_{0}(s;m_{t},m_{t})+16C_{24[t,t,T]}(s)-2sC_{0[t,t,T]}(s) \nonumber\\
&& + s \left[(u-m_{t}^{2}-m_{T}^{2}) D_{0[t,t,t,T]}(s,t) + (t-m_{t}^{2}-m_{T}^{2}) D_{0[t,t,t,T]}(s,u) \right] \nonumber\\
&& -\left[(t-m_{h}^{2})(u-m_{h}^{2}) + s(m_{t}^{2}+m_{T}^{2}-m_{h}^{2}) \right]D_{0[t,t,t,T]}(t,u) \nonumber\\
&& +4(s-2m_{h}^{2}+2m_{t}^{2}+2m_{T}^{2}) \nonumber\\
&& \times\left( D_{27[t,t,t,T]}(s,t) + D_{27[t,t,t,T]}(s,u) + D_{27[t,t,t,T]}(t,u) -\frac{1}{2} C_{0[t,t,T]}(s) \right) \Biggr\} \nonumber\\
&& + q^2 \ c_{tT} \ c_{Tt} \ m_{t} \ m_{T} \ \Biggl[s(D_{0[t,t,t,T]}(s,t) + D_{0[t,t,t,T]}(s,u) + D_{0[t,t,t,T]}(t,u)) \nonumber\\
&& -4\left(D_{27[t,t,t,T]}(s,t) + D_{27[t,t,t,T]}(s,u) + D_{27[t,t,t,T]}(t,u) - C_{0[t,t,T]}(s) \right) \Biggr] \nonumber\\
&& -\frac{t \ u - m_{h}^{4}}{2m_{W}^{2}s} \ \Biggl\{ (c_{tT}^{2} + c_{Tt}^{2}) \Biggl[ s(D_{13[t,t,t,T]}(s,t) + D_{13[t,t,t,T]}(s,u))  \nonumber\\
&& +(s - 2m_{h}^{2} + 2m_{t}^{2} + 2m_{T}^{2}) \nonumber\\
&& \times\left(D_{23[t,t,t,T]}(s,t) + D_{23[t,t,t,T]}(s,u) + D_{12[t,t,t,T]}(t,u) + D_{22[t,t,t,T]}(t,u) \right) \nonumber\\
&& - \left(C_{0[t,t,T]}(t) + C_{0[t,t,T]}(u) + C_{11[t,t,T]}(t) + C_{11[t,t,T]}(u) + C_{12[t,t,T]}(t) + C_{12[t,t,T]}(u) \right) \Biggr] \nonumber\\
&& + 4 \ q^2 \ c_{tT} \ c_{Tt} \ m_{t} \ m_{T} \nonumber\\
&& \times 
\left( D_{23[t,t,t,T]}(s,t) + D_{23[t,t,t,T]}(s,u) + D_{12[t,t,t,T]}(t,u)  + D_{22[t,t,t,T]}(t,u) \right) \Biggr\}  \nonumber\\
&& + (m_{t} \leftrightarrow m_{T})
\end{eqnarray}

\begin{eqnarray}
{\cal M}_{box}^{t-T}(+,-) &=& \frac{t \ u -  m_h^4}{2s} \ q^2 \ \Biggl\{
   (c_{tT}^{2} + c_{Tt}^{2}) \Bigl[ 
    (s-2m_{h}^{2}+2m_{t}^{2}+2m_{T}^{2}) \nonumber\\
&& \times(D_{23[t,t,t,T]}(s,t)+D_{23[t,t,t,T]}(s,u)+D_{12[t,t,t,T]}(t,u)+D_{22[t,t,t,T]}(t,u)) \nonumber\\
&& + s(D_{13[t,t,t,T]}(s,t) + D_{13[t,t,t,T]}(s,u)) \nonumber\\
&& - (C_{0[t,t,T]}(s)+C_{0[t,t,T]}(t)+C_{11[t,t,T]}(s)+C_{11[t,t,T]}(t)+C_{12[t,t,T]}(s)+C_{12[t,t,T]}(t)) \nonumber\\
&& +c_{tT}c_{Tt} m_{t}m_{T}     
\left[ D_{23[t,t,t,T]}(s,t) + D_{23[t,t,t,T]}(s,u) + D_{12[t,t,t,T]}(t,u) + D_{22[t,t,t,T]}(t,u) \right] \Biggr\} \nonumber\\
&& + (m_{t} \leftrightarrow m_{T}) \,,
\end{eqnarray}

\noindent where the Passarino-Veltman loop functions \cite{Passarino:1978jh} are abbreviated as
\begin{eqnarray}
C_{x[i,j,k]}(s) &=& C_{x}(p_{1},p_{2};m_{i},m_{j},m_{k})  \,,\\
C_{x[i,j,k]}(t) &=& C_{x}(p_{2},p_{3};m_{i},m_{j},m_{k}) \,,\\
C_{x[i,j,k]}(u) &=& C_{x}(p_{1},p_{3};m_{i},m_{j},m_{k}) \,,\\
\tilde{C}_{x[i,j,k]}(s) &=& C_{x}(p_{3},p_{4};m_{i},m_{j},m_{k}) \,, \\ 
D_{x[i,j,k,l]}(s,t) &=& D_{x}(p_{1},p_{2},p_{3};m_{i},m_{j},m_{k},m_{l}) \,,\\
D_{x[i,j,k,l]}(s,u) &=& D_{x}(p_{2},p_{1},p_{3};m_{i},m_{j},m_{k},m_{l}) \,,\\
D_{x[i,j,k,l]}(t,u) &=& D_{x}(p_{1},p_{3},p_{2};m_{i},m_{j},m_{k},m_{l}) \,.
\end{eqnarray}
In Fig. \ref{fig:appendix_3} the results of individual contributions on introduction of a singlet Vector Like Top quark 
(as discussed in Sub-Section \ref{subsection:5_1}) are given. The results are shown for the benchmark point 1 as given in
Table \ref{table:TopPartners}. As can be seen from the table the contribution coming from the quartic Higgs-fermion (SM) vertices 
($c_{2f}$) is relatively small thereby resulting in decrease in cross-sections as compared to SM value. However, 
the quartic Higgs-Vector like fermion vertex ($c_{2T}$) could be substantial and its effects can be seen after the threshold 
of the Vector Like Top quark mass (inset of Fig. \ref{fig:appendix_3}). 

\begin{figure}[htb]
    \begin{center}
    \hspace*{-1cm}
        \epsfig{file=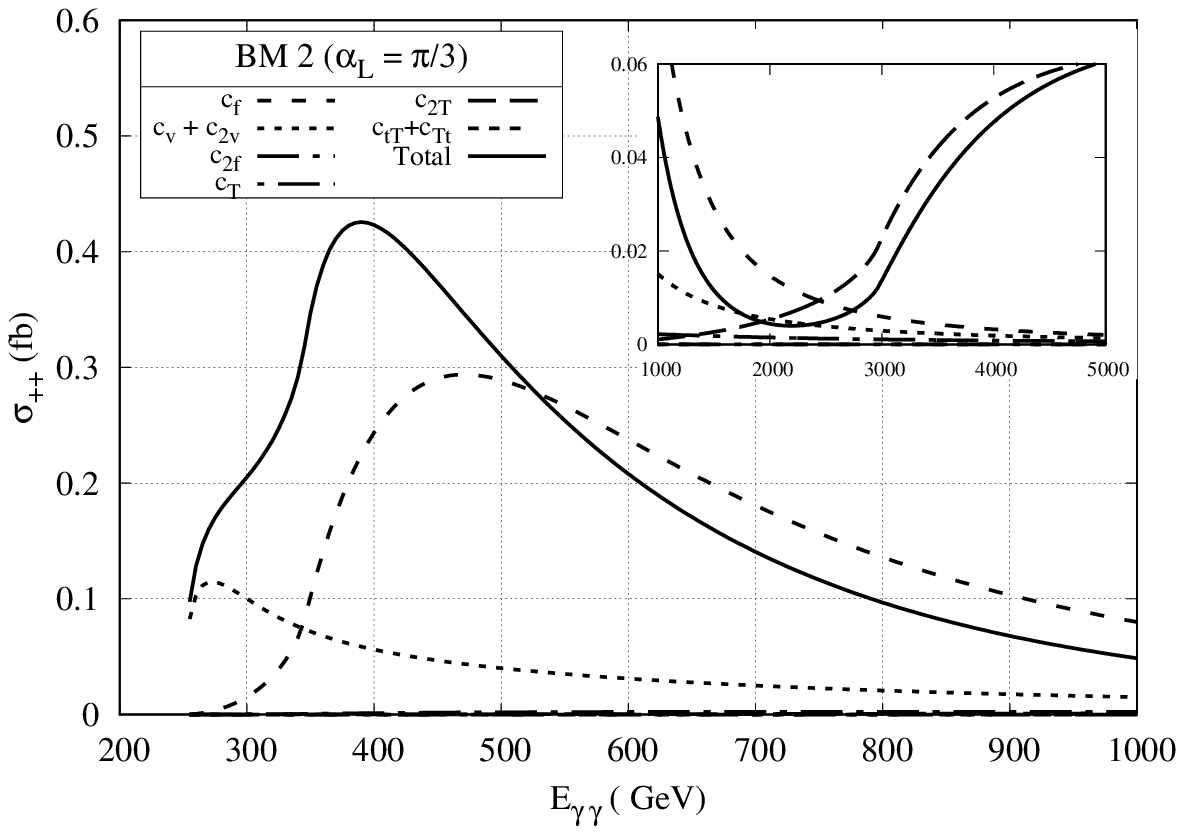,width=0.52\textwidth} 
        \epsfig{file=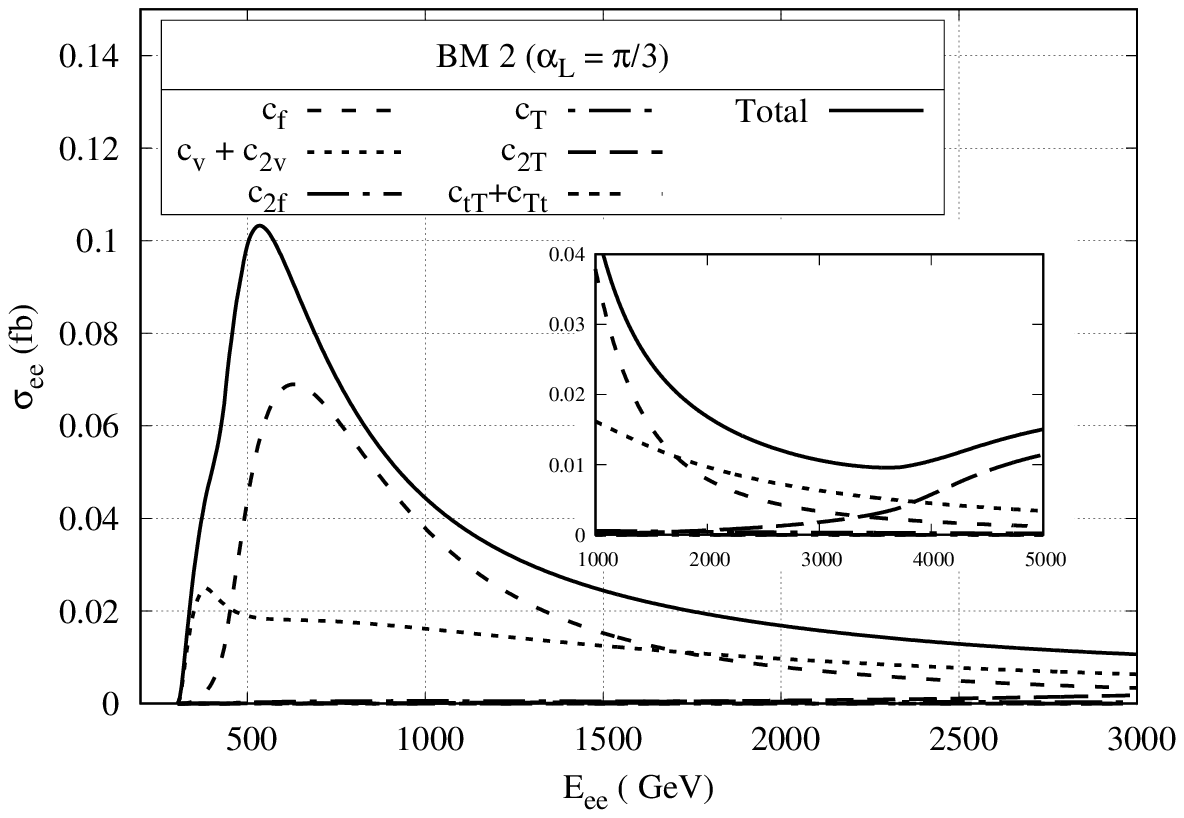,width=0.52\textwidth}
    \caption{Cross-sections for the individual contributions as given in Tables \ref{tab:diagSM} and \ref{tab:diagTP} 
    for $\gamma \gamma \to h h$ same photon helicities (left panel) and
    $e^+ e^- \to hh$ (right panel) in the presence of Doublet Vector Like Top-quark (as given in Section \ref{subsection:5_2}). }  
    \label{fig:appendix_4} 
    \end{center}
\end{figure}

In Fig.\ref{fig:appendix_4} we have shown the results of individual contributions after introducing the vector like quark 
Doublet $(T,B)^T$ (discussed in Sub-section \ref{subsection:5_2}). The results are shown for benchmark point 2 as given in Table
\ref{table:TopPartners}. The results are similar to the ones of the introduction of a singlet vector like quark and once again we witness
the impact of the quartic Higgs-fermionic vertex ($c_{2F})$ after the threshold of the Vector Like Quark mass.


\bibliographystyle{JHEP-2-2}
\bibliography{biblio}

\end{document}